\documentclass[11pt,fleqn,oneside,openany]{book} 
\usepackage[utf8]{inputenc}
\usepackage{amsmath}
\usepackage{mathptmx}
\usepackage[T1]{fontenc}
\usepackage{relsize}
\usepackage{mathrsfs}
\usepackage{siunitx}
\usepackage{wasysym}
\usepackage{bm}
\usepackage{csvsimple}
\usepackage{longtable}
\usepackage{chngcntr}
\usepackage[symbol]{footmisc}

\DeclareUnicodeCharacter{200E}{\textvisiblespace}
\usepackage[top=3cm,bottom=3cm,left=3cm,right=3cm,headsep=10pt,a4paper]{geometry} 

\usepackage{graphicx} 
\graphicspath{{Pictures/}} 

\usepackage{lipsum} 

\usepackage{tikz} 

\usepackage[english]{babel} 

\usepackage{enumitem} 
\setlist{nolistsep} 

\usepackage{booktabs} 

\usepackage{xcolor} 
\definecolor{ocre}{RGB}{243,102,25} 

\usepackage{avant} 
\usepackage{mathptmx} 

\usepackage{microtype} 
\usepackage[utf8]{inputenc} 
\usepackage[T1]{fontenc} 

\usepackage{csquotes}
\usepackage{calc} 
\usepackage{makeidx} 
\makeindex 

\usepackage{titletoc} 

\contentsmargin{0cm} 

\titlecontents{part}[0cm]
{\addvspace{20pt}\centering\large\bfseries}
{}
{}
{}

\titlecontents{chapter}[1.25cm] 
{\addvspace{12pt}\large\sffamily\bfseries} 
{\color{ocre!60}\contentslabel[\Large\thecontentslabel]{1.25cm}\color{ocre}} 
{\color{ocre}}  
{\color{ocre!60}\normalsize\;\titlerule*[.5pc]{.}\;\thecontentspage} 

\titlecontents{section}[1.25cm] 
{\addvspace{3pt}\sffamily\bfseries} 
{\contentslabel[\thecontentslabel]{1.25cm}} 
{}
{\hfill\color{black}\thecontentspage} 
[]

\titlecontents{subsection}[1.25cm] 
{\addvspace{1pt}\sffamily\small} 
{\contentslabel[\thecontentslabel]{1.25cm}} 
{}
{\ \titlerule*[.5pc]{.}\;\thecontentspage} 
[]

\titlecontents{figure}[0em]
{\addvspace{-5pt}\sffamily}
{\thecontentslabel\hspace*{1em}}
{}
{\ \titlerule*[.5pc]{.}\;\thecontentspage}
[]

\titlecontents{table}[0em]
{\addvspace{-5pt}\sffamily}
{\thecontentslabel\hspace*{1em}}
{}
{\ \titlerule*[.5pc]{.}\;\thecontentspage}
[]

\titlecontents{lchapter}[0em] 
{\addvspace{15pt}\large\sffamily\bfseries} 
{\color{ocre}\contentslabel[\Large\thecontentslabel]{1.25cm}\color{ocre}} 
{}  
{\color{ocre}\normalsize\sffamily\bfseries\;\titlerule*[.5pc]{.}\;\thecontentspage} 

\titlecontents{lsection}[0em] 
{\sffamily\small} 
{\contentslabel[\thecontentslabel]{1.25cm}} 
{}
{}

\titlecontents{lsubsection}[.5em] 
{\normalfont\footnotesize\sffamily} 
{}
{}
{}

\usepackage{fancyhdr} 

\fancypagestyle{plain}{\fancyhead{}} 

\makeatletter
\renewcommand{\cleardoublepage}{
\clearpage\ifodd\c@page\else
\hbox{}
\vspace*{\fill}
\thispagestyle{empty}
\newpage
\fi}

\usepackage{amsmath,amsfonts,amssymb,amsthm} 

\newtheoremstyle{ocrenumbox}
{0pt}
{0pt}
{\normalfont}
{}
{\small\bf\sffamily\color{ocre}}
{\;}
{0.25em}
{\small\sffamily\color{ocre}\thmname{#1}\nobreakspace\thmnumber{\@ifnotempty{#1}{}\@upn{#2}}
\thmnote{\nobreakspace\the\thm@notefont\sffamily\bfseries\color{black}---\nobreakspace#3.}} 

\newtheoremstyle{blacknumex}
{5pt}
{5pt}
{\normalfont}
{} 
{\small\bf\sffamily}
{\;}
{0.25em}
{\small\sffamily{\tiny\ensuremath{\blacksquare}}\nobreakspace\thmname{#1}\nobreakspace\thmnumber{\@ifnotempty{#1}{}\@upn{#2}}
\thmnote{\nobreakspace\the\thm@notefont\sffamily\bfseries---\nobreakspace#3.}}

\newtheoremstyle{blacknumbox} 
{0pt}
{0pt}
{\normalfont}
{}
{\small\bf\sffamily}
{\;}
{0.25em}
{\small\sffamily\thmname{#1}\nobreakspace\thmnumber{\@ifnotempty{#1}{}\@upn{#2}}
\thmnote{\nobreakspace\the\thm@notefont\sffamily\bfseries---\nobreakspace#3.}}

\newtheoremstyle{ocrenum}
{5pt}
{5pt}
{\normalfont}
{}
{\small\bf\sffamily\color{ocre}}
{\;}
{0.25em}
{\small\sffamily\color{ocre}\thmname{#1}\nobreakspace\thmnumber{\@ifnotempty{#1}{}\@upn{#2}}
\thmnote{\nobreakspace\the\thm@notefont\sffamily\bfseries\color{black}---\nobreakspace#3.}} 
\makeatother

\newcounter{dummy} 
\numberwithin{dummy}{section}
\theoremstyle{ocrenumbox}
\newtheorem{theoremeT}{Box}
\newtheorem{questionT}{Question}
\theoremstyle{blacknumex}
\newtheorem{exampleT}{Example}[chapter]
\theoremstyle{blacknumbox}

\newtheorem{definitionT}{Definition}[section]
\newtheorem{corollaryT}[dummy]{Corollary}
\theoremstyle{ocrenum}

\RequirePackage[framemethod=default]{mdframed} 

\newmdenv[skipabove=7pt,
skipbelow=7pt,
backgroundcolor=black!5,
linecolor=ocre,
innerleftmargin=15pt,
innerrightmargin=15pt,
innertopmargin=15pt,
leftmargin=0cm,
rightmargin=0cm,
innerbottommargin=15pt]{tBox}

\newmdenv[skipabove=7pt,
skipbelow=7pt,
rightline=false,
leftline=true,
topline=false,
bottomline=false,
backgroundcolor=ocre!10,
linecolor=ocre,
innerleftmargin=5pt,
innerrightmargin=5pt,
innertopmargin=5pt,
innerbottommargin=5pt,
leftmargin=0cm,
rightmargin=0cm,
linewidth=4pt]{eBox}	

\newmdenv[skipabove=7pt,
skipbelow=7pt,
rightline=false,
leftline=true,
topline=false,
bottomline=false,
linecolor=ocre,
innerleftmargin=5pt,
innerrightmargin=5pt,
innertopmargin=0pt,
leftmargin=0cm,
rightmargin=0cm,
linewidth=4pt,
innerbottommargin=0pt]{dBox}	

\newmdenv[skipabove=7pt,
skipbelow=7pt,
rightline=false,
leftline=true,
topline=false,
bottomline=false,
linecolor=gray,
backgroundcolor=black!5,
innerleftmargin=5pt,
innerrightmargin=5pt,
innertopmargin=5pt,
leftmargin=0cm,
rightmargin=0cm,
linewidth=4pt,
innerbottommargin=5pt]{cBox}

\newenvironment{theorem}{\begin{tBox}\begin{theoremeT}}{\end{theoremeT}\end{tBox}}

\makeatletter
\renewcommand{\@seccntformat}[1]{\llap{\textcolor{ocre}{\csname the#1\endcsname}\hspace{1em}}}                    
\renewcommand{\section}{\@startsection{section}{1}{\z@}
{-4ex \@plus -1ex \@minus -.4ex}
{1ex \@plus.2ex }
{\normalfont\large\sffamily\bfseries}}
\renewcommand{\subsection}{\@startsection {subsection}{2}{\z@}
{-3ex \@plus -0.1ex \@minus -.4ex}
{0.5ex \@plus.2ex }
{\normalfont\sffamily\bfseries}}
\renewcommand{\subsubsection}{\@startsection {subsubsection}{3}{\z@}
{-2ex \@plus -0.1ex \@minus -.2ex}
{.2ex \@plus.2ex }
{\normalfont\small\sffamily\bfseries}}                        
\renewcommand\paragraph{\@startsection{paragraph}{4}{\z@}
{-2ex \@plus-.2ex \@minus .2ex}
{.1ex}
{\normalfont\small\sffamily\bfseries}}

\newcommand{\@mypartnumtocformat}[2]{%
\setlength\fboxsep{0pt}%
\noindent\colorbox{ocre!20}{\strut\parbox[c][.7cm]{\ecart}{\color{ocre!70}\Large\sffamily\bfseries\centering#1}}\hskip\esp\colorbox{ocre!40}{\strut\parbox[c][.7cm]{\linewidth-\ecart-\esp}{\Large\sffamily\centering#2}}}%
\newcommand{\@myparttocformat}[1]{%
\setlength\fboxsep{0pt}%
\noindent\colorbox{ocre!40}{\strut\parbox[c][.7cm]{\linewidth}{\Large\sffamily\centering#1}}}%
\newlength\esp
\newlength\ecart
\def\@part[#1]#2{%
\ifnum \c@secnumdepth >-2\relax%
\refstepcounter{part}%
\addcontentsline{toc}{part}{\texorpdfstring{\protect\@mypartnumtocformat{\thepart}{#1}}{\partname~\thepart\ ---\ #1}}
\else%
\addcontentsline{toc}{part}{\texorpdfstring{\protect\@myparttocformat{#1}}{#1}}%
\fi%
\startcontents%
\markboth{}{}%
{\thispagestyle{empty}%
\begin{tikzpicture}[remember picture,overlay]%
\node at (current page.north west){\begin{tikzpicture}[remember picture,overlay]%
\fill[ocre!20](0cm,0cm) rectangle (\paperwidth,-\paperheight);
\node[anchor=north] at (4cm,-3.25cm){\color{ocre!40}\fontsize{220}{100}\sffamily\bfseries\@Roman\c@part}; 
\node[anchor=south east] at (\paperwidth-1cm,-\paperheight+1cm){\parbox[t][][t]{8.5cm}{
\printcontents{l}{0}{\setcounter{tocdepth}{1}}%
}};
\node[anchor=north east] at (\paperwidth-1.5cm,-3.25cm){\parbox[t][][t]{15cm}{\strut\raggedleft\color{white}\fontsize{30}{30}\sffamily\bfseries#2}};
\end{tikzpicture}};
\end{tikzpicture}}%
\@endpart}
\def\@spart#1{%
\startcontents%
\phantomsection
{\thispagestyle{empty}%
\begin{tikzpicture}[remember picture,overlay]%
\node at (current page.north west){\begin{tikzpicture}[remember picture,overlay]%
\fill[ocre!20](0cm,0cm) rectangle (\paperwidth,-\paperheight);
\node[anchor=north east] at (\paperwidth-1.5cm,-3.25cm){\parbox[t][][t]{15cm}{\strut\raggedleft\color{white}\fontsize{30}{30}\sffamily\bfseries#1}};
\end{tikzpicture}};
\end{tikzpicture}}
\addcontentsline{toc}{part}{\texorpdfstring{%
\setlength\fboxsep{0pt}%
\noindent\protect\colorbox{ocre!40}{\strut\protect\parbox[c][.7cm]{\linewidth}{\Large\sffamily\protect\centering #1\quad\mbox{}}}}{#1}}%
\@endpart}
\def\@endpart{\vfil\newpage
\if@twoside
\if@openright
\null
\thispagestyle{empty}%
\newpage
\fi
\fi
\if@tempswa
\twocolumn
\fi}

\newif\ifusechapterimage
\usechapterimagetrue
\newcommand{\thechapterimage}{}%
\newcommand{\chapterimage}[1]{\ifusechapterimage\renewcommand{\thechapterimage}{#1}\fi}%
\def\@makechapterhead#1{%
{\parindent \z@ \raggedright \normalfont
\ifnum \c@secnumdepth >\m@ne
\if@mainmatter
\begin{tikzpicture}[remember picture,overlay]
\node at (current page.north west)
{\begin{tikzpicture}[remember picture,overlay]
\node[anchor=north west,inner sep=0pt] at (0,0) {\ifusechapterimage\includegraphics[width=\paperwidth]{\thechapterimage}\fi};
\draw[anchor=west] (\Gm@lmargin,-4cm) node [line width=2pt,rounded corners=15pt,draw=ocre,fill=white,fill opacity=0.5,inner sep=15pt]{\strut\makebox[22cm]{}};
\draw[anchor=west] (\Gm@lmargin+.3cm,-4cm) node {\huge\sffamily\bfseries\color{black}\thechapter. #1\strut};
\end{tikzpicture}};
\end{tikzpicture}
\else
\begin{tikzpicture}[remember picture,overlay]
\node at (current page.north west)
{\begin{tikzpicture}[remember picture,overlay]
\node[anchor=north west,inner sep=0pt] at (0,0) {\ifusechapterimage\includegraphics[width=\paperwidth]{\thechapterimage}\fi};
\draw[anchor=west] (\Gm@lmargin,-4cm) node [line width=2pt,rounded corners=15pt,draw=ocre,fill=white,fill opacity=0.5,inner sep=15pt]{\strut\makebox[22cm]{}};
\draw[anchor=west] (\Gm@lmargin+.3cm,-4cm) node {\huge\sffamily\bfseries\color{black}#1\strut};
\end{tikzpicture}};
\end{tikzpicture}
\fi\fi\par\vspace*{100\p@}}}

\def\@makeschapterhead#1{%
\begin{tikzpicture}[remember picture,overlay]
\node at (current page.north west)
{\begin{tikzpicture}[remember picture,overlay]
\node[anchor=north west,inner sep=0pt] at (0,0) {\ifusechapterimage\includegraphics[width=\paperwidth]{\thechapterimage}\fi};
\draw[anchor=west] (\Gm@lmargin,-4cm) node [line width=2pt,rounded corners=15pt,draw=ocre,fill=white,fill opacity=0.5,inner sep=15pt]{\strut\makebox[22cm]{}};
\draw[anchor=west] (\Gm@lmargin+.3cm,-4cm) node {\huge\sffamily\bfseries\color{black}#1\strut};
\end{tikzpicture}};
\end{tikzpicture}
\par\vspace*{100\p@}}
\makeatother

\usepackage{hyperref}
\hypersetup{hidelinks,colorlinks=false,breaklinks=true,urlcolor= ocre,bookmarksopen=false,pdftitle={Title},pdfauthor={Author}}
\usepackage{bookmark}
\bookmarksetup{
open,
numbered,
addtohook={%
\ifnum\bookmarkget{level}=0 
\bookmarksetup{bold}%
\fi
\ifnum\bookmarkget{level}=-1 
\bookmarksetup{color=ocre,bold}%
\fi
}
}
\counterwithout{figure}{chapter}
\counterwithout{table}{chapter}
\counterwithout{equation}{chapter}
\usepackage{etoolbox}
\patchcmd{\thebibliography}{\chapter*}{}{}{}

\begin{document}
\begingroup
\thispagestyle{empty}
\begin{tikzpicture}[remember picture,overlay]
\coordinate [below=12cm] (midpoint) at (current page.north);
\node at (current page.north west)
{\begin{tikzpicture}[remember picture,overlay]
\node[anchor=north west,inner sep=0pt] at (0,0) {\includegraphics[width=\paperwidth]{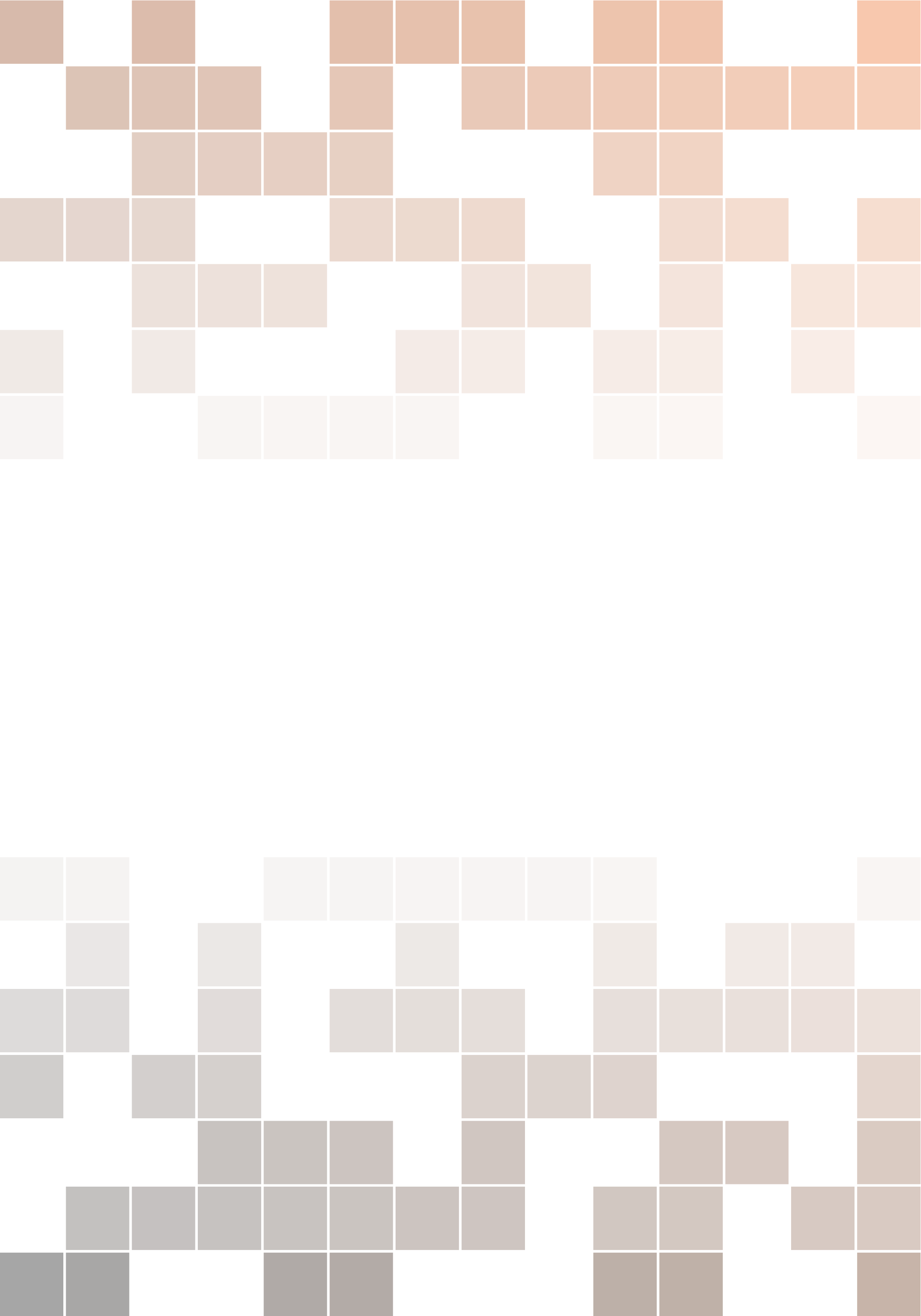}}; 
\draw[anchor=north] (midpoint) node [fill=ocre!30!white,fill opacity=0.6,text opacity=1,inner sep=1cm]{\centering\bfseries\sffamily\parbox[c][][t]{\paperwidth}{\centering {\LARGE 
A global phase fitting approach for the analysis of \\ $^{129}$Xe electric dipole moment measurements \\ \vspace{1cm}} 
{\LARGE Tianhao Liu\\}
\vspace{0.3cm}
{\normalsize Physikalisch-Technische Bundesanstalt, Berlin\\}
}
}; 
\end{tikzpicture}};
\end{tikzpicture}
\vfill
\endgroup

\chapterimage{chapter_head_1.pdf} 

\tableofcontents 

\chapter*{Preface}
\vspace{1cm}
This writeup is a translated excerpt of the author's doctoral thesis in Chinese entitled \textit{"Research on realization and application of \textmu T-level magnetic fields of high homogeneity inside magnetically shielded rooms"}, which will be available at https://i.cnki.net/ in about six months. The author registered at Harbin Institute of Technology under the supervision of Prof.~Liyi Li. The presented work was performed at Physikalisch-Technische Bundesanstalt, Berlin (PTB), where the author joined the group 8.24 (nuclear spin precession of noble gases) first as a guest scientist and then as an employed researcher under the supervision of Dr.~Lutz Trahms. This writeup aims for providing supplemental information for interested readers of the paper published at https://arxiv.org/abs/2008.07975. 
\vspace{0.5cm}

\noindent  The author sincerely thank Dr.~Katharina Rolfs, Dr.~Lutz Trahms, Dr.~Wolfgang Kilian, Dr.~Allard Schnabel, Dr.~Jens Voigt for their vital contributions to the development of the presented EDM data analysis method and their careful proofreading of this writeup.  
\vspace{0.5cm}

\noindent The template for this writeup is "The Legrand Orange Book" and was freely downloaded from http://www.LaTeXTemplates.com.

\vspace{1cm}
{\setlength{\parindent}{0cm} Tianhao Liu \\ Berlin, 14 March 2021 \\ Email: silasliutianhao@gmail.com }

\chapter*{Abstract}

\vspace{1cm}
Measuring the size of permanent electric dipole moments (EDM) of a particle or system provides a powerful tool to test Beyond-the-Standard-Model physics. The diamagnetic $^{129}$Xe atom is one of the promising candidates for EDM experiments due to its obtainable high nuclear polarization and its long spin-coherence time in a homogeneous magnetic field. By measuring the spin precession frequencies of polarized $^{129}$Xe and $^{3}$He, a new upper limit on the $^{129}$Xe atomic EDM $d_\mathrm{A}(^{129}\mathrm{Xe})$ was reported \cite{Sachdeva2019}. This writeup proposes a new evaluation method based on global phase fitting (GPF) for analyzing the continuous phase development of the $^{3}$He-$^{129}$Xe comagnetometer signal. The Cramer-Rao Lower Bound on the $^{129}$Xe EDM for the GPF method is theoretically derived and shows the benefit of achieving high statistical sensitivity without bringing new systematic uncertainties. The robustness of the GPF method is verified with Monte-Carlo studies. By optimizing the analysis parameters and adding few more data that could not be analyzed with the former method used in Ref.~\cite{Sachdeva2019}, a result of
\[ {  d_\mathrm{A} (^{129}\mathrm{Xe})=(1.1 \pm 3.6_\mathrm{(stat)} \pm 2.0_\mathrm{(syst)})\times 10 ^{-28} e~\mathrm{cm}}, \]
is obtained and is used to derive the upper limit of $^{129}$Xe permanent EDM at 95\% C.L.
\[ {\large |d_\text{A}(^{129}\text{Xe})| <  8.3 \times 10^{-28}~e~\mathrm{cm}}. \]
This limit is a factor of 1.7 smaller as compared to the previous result \cite{Sachdeva2019} and a factor of 8.0 as compared to the result in 2001 \cite{Rosenberry2001}.

\vspace{1cm}

\centerline{\includegraphics[width=.6\columnwidth]{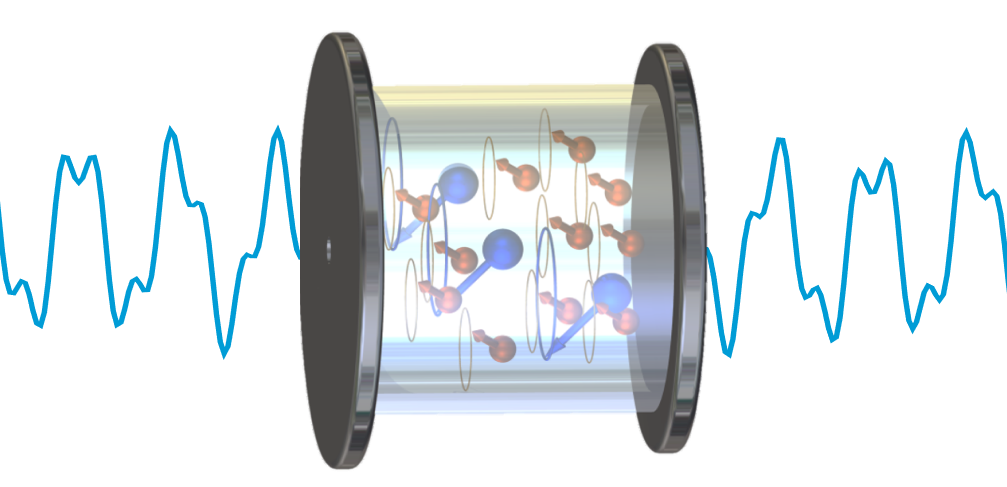}}

\chapterimage{chapter_head_22.pdf}

\chapter{Data collection}

A collaboration between the University of Michigan, the Technical University of Munich and the Physikalisch-Technische Bundesanstalt(PTB) has conducted two measurement campaigns of $^{129}$Xe permanent electric dipole moment (EDM) measurements in June 2017 and July 2018 with similar procedures. The experiments took place at the Berlin Magnetically Shielded Room (BMSR-2) facility at PTB Berlin. Detailed information on the experimental setup of the 2017 campaign can be found in the doctoral thesis of Natasha Sachdeva\cite{Sachdeva2019a}. In this chapter, the overall experimental setup is concisely introduced together with the achieved typical parameters, as a basis for the discussion of data analysis method. 

\section{Measuring principle and procedure}

\begin{figure}[hb]
    \centerline{\includegraphics[width=.45\columnwidth]{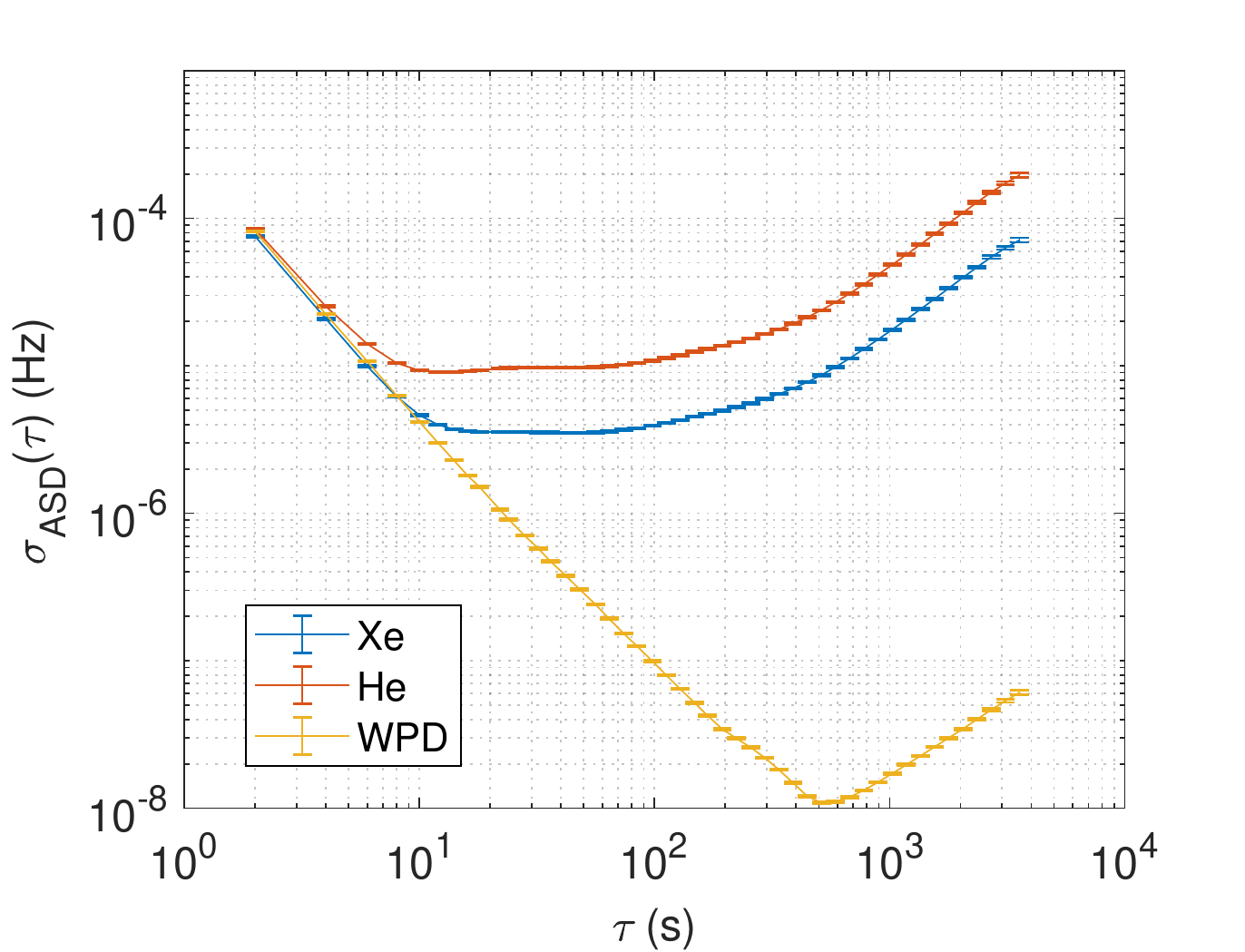}
    \includegraphics[width=.44\columnwidth]{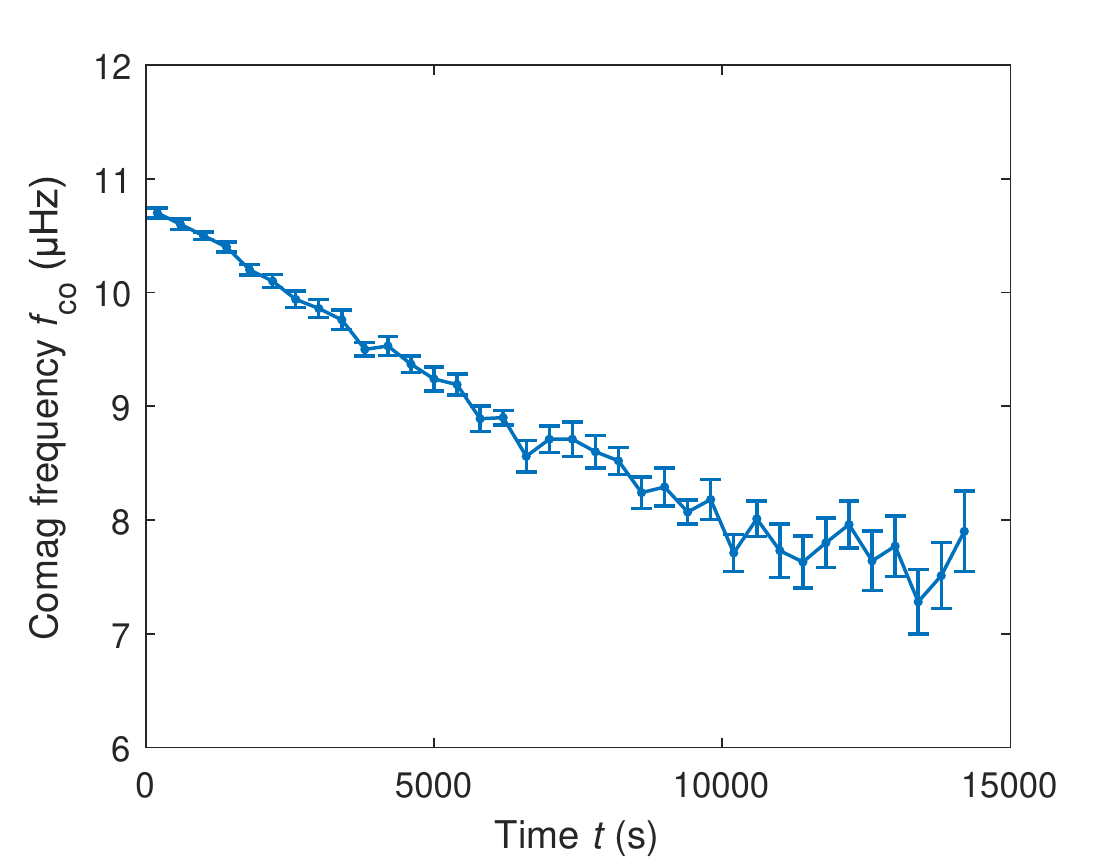}}
    \centerline{(a) \hspace{6cm} (b)}
    \caption{(a) Modified Allan deviation of $^{129}$Xe, $^{3}$He and $\omega_{\text{co}}$ (called Weighted Phase Difference, WPD) signals from the B881 sub-run. (b) Observed anomalous frequency drift from the A84 sub-run. The frequency drift from 0~s to 14000~s is of about 4 \textmu Hz. The constant frequency $f_\text{co} \approx 7.8$~\textmu Hz at the end is caused by well-known sources, including the Earth's rotation and chemical shift.}
    \label{fig:Ano_phase_drift}
\end{figure}

An atom with a total angular momentum $\mathbf{F=I+J}$ has a magnetic moment 
\begin{equation} 
	\label{eqn:Mu_moment}
    \bm{\mu}=\gamma \mathbf{F},
\end{equation}
where $\gamma$ is the gyromagnetic ratio and $\lVert \mathbf{F} \rVert =\sqrt{F(F+1)} \hslash$. For both ground state atoms $^{129}$Xe and $^{3}$He, the total electronic angular momentum $\mathbf{J=0}$ and the nuclear spin quantum number $I=1/2$, leading to the total angular momentum quantum number $F=1/2$ \cite{Sachdeva2019a}. A fundamental or composed particle can also have a permanent EDM $\mathbf{d}$, which must be aligned parallel to its axial spin vector, as it is the only vector for an eigenstate of the isolated particle \cite{Chupp2019}. Otherwise an additional quantum number would be needed, being in disagreement with the Pauli principle. For both atoms $^{129}$Xe and $^{3}$He, 
\begin{equation} 
	\label{eqn:d_EDM}
         \textbf{d} = d \frac{\mathbf{F}}{F},
\end{equation}
where the scalar $d$ is the magnitude of $\mathbf{d}$. Henceforth, EDM denotes $d$ instead of $\mathbf{d}$. The Hamiltonian that describes the impact of the magnetic field $\mathbf{B}$ and the electric field $\mathbf{E}$ on atoms, is written as 
\begin{equation} 
	\label{eqn:Hamiltonian}
      \mathcal{H} = -\bm{\mu} \cdot \textbf{B} - \textbf{d} \cdot \textbf{E}.
\end{equation}
Assuming the magnetic field $\mathbf{B}$ is parallel to the electric field $\mathbf{E}$, the splitting in energy levels of atoms between two states ($\Delta m_F=1$)  becomes 
\begin{equation} 
	\label{eqn:Energy_split}
         \Delta E = \hslash \omega = |\mu B \pm d E |/F,
\end{equation}
where $\omega$ is the spin precession angular frequency and the "+/-" sign means \textbf{B} has the same/opposite direction of \textbf{E}. In principle one can derive EDM $d$ from the measured spin precession frequency $\omega$ with a known electric field amplitude $E$. However, the magnetic field $B$ in Eq.~(\ref{eqn:Energy_split}) becomes an interference term when doing this. To overcome the experimental difficulties on controlling and measuring $\mathbf{B}$, comagnetometry was introduced with two collated species measured at the same time \cite{Gemmel2010}. $^{3}$He is an ideal candidate for comagnetometry due to its ability to be efficiently hyper-polarized and its negligible EDM compared to $d_\text{Xe}$\footnote[2]{In some cases for accuracy, we denote the Xe EDM $d_\text{Xe}$ as $d_\text{A}(^{129}\text{Xe})$}. The Xe-He comagnetometry has been widely used in the Xe EDM measurements \cite{Sachdeva2019,Rosenberry2001,Allmendinger2019}. The weighted angular frequency difference between $^{129}$Xe and $^{3}$He atoms is defined as
\begin{equation} 
	\label{eqn:Co_omega}
    \omega_{\text{co}}= \omega_{\text{Xe}}-\frac{\gamma_{\text{Xe}}}{\gamma_{\text{He}}}\omega_{\text{He}},
\end{equation}
where the spin precession frequency of the $^{3}$He atoms is $\omega_{\text{He}}=|\gamma_{\text{He}}B|$. Substituting this together with Eq.~(\ref{eqn:Energy_split}) into Eq.~(\ref{eqn:Co_omega}), the comagnetometer angular frequency $\omega_{\text{co}}$ is simplified to 
\begin{equation} 
	\label{eqn:C0_omega2}
    \omega_{\text{co}}= -(\hat{\textbf{E}} \cdot \hat{\textbf{B}})\frac{d_{\text{Xe}}}{\hslash F}|E|,
\end{equation}
which is independent of the magnitude $|B|$ of the background magnetic field. In order to quantify the magnetic field stability, Fig.~\ref{fig:Ano_phase_drift}(a) shows the modified Allan deviation of $^{129}$Xe and $^{3}$He frequencies obtained in one measurement (see detailed explanation of the Allan deviation in Sec.~\ref{sec:Fit quality}). Both deviations of $^{129}$Xe and $^{3}$He are flattened at the integration time $\tau \approx 15$~s, indicating the background magnetic field drift of our passively shielded environment. The standard deviation of the comagnetometer angular frequency $\omega_{\text{co}}$ reaches the minimum at $\tau$=550~s as a result of cancelling the external field dependent impact. However, the standard deviation increases alongside $\tau$ after the minimum, implying the existence of 1/f drifts in the comagnetometer frequency. Multiple physical models to describe the observed comagnetometer drift were proposed \cite{Terrano2019,Limes2019,Romalis2014,Allmendinger2014}. The dominant term thereby varies in different models and is still of controversy. From here on, we name it anomalous comagnetometer frequency(phase) drift.  

Fig.~\ref{fig:Ano_phase_drift}(b) shows the observed anomalous frequency drift of $\omega_{\text{co}}$ over a measurement run. This unfavorable frequency drift becomes a potential systematic error source since its value, which is on the \textmu Hz level, exceeds the current sensitivity of $f_{\text{co}}$ per run, which is on the sub-nHz level, by far. Due to the ambiguity of physical mechanism for the observed drift, correcting the frequency drift with a deterministic model is challenging (an example is given in Ref.~\cite{Allmendinger2019}).    

One common approach to mitigate the effect of anomalous comagnetometer drift is repetitively reversing the direction of the electric field \textbf{E} during one run. This allows to separate the impact of $d_{\text{Xe}}$ on $\omega_{\text{co}}$ from other interference terms, like comagnetometer drift. We have applied the E-field modulation method with varied patterns in both campaigns.  

\begin{figure}[!t]
    \centerline{\includegraphics[width=.8\columnwidth]{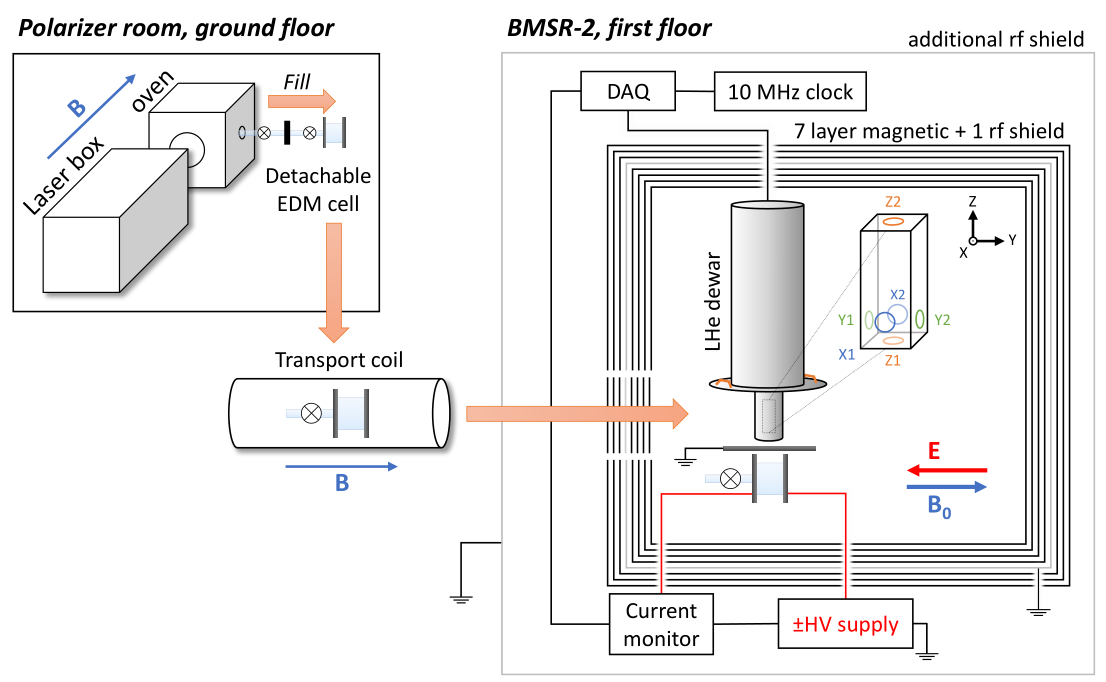}}
    \caption{Conceptual view of the experimental setup (not to scale) for measuring the $^{129}$Xe EDM at BMSR-2 \cite{Sachdeva2019a}.}
    \label{fig:EDM_Setup}
\end{figure}

The schematic view of the experimental setup is shown in Fig.~\ref{fig:EDM_Setup} and the routine procedure is listed as follows. Before the start of routine measurements, a handful of preparation work had been done, such as testing the high voltage supply, cleaning EDM cells, and configuring the superconducting quantum interference device (SQUID) system. 
\par\vspace{10pt}
\begin{list}{}{  \leftmargin=35pt 
                 \rightmargin=25pt} 
    \item\ignorespaces 
    \makebox[-2.5pt]{\begin{tikzpicture}[overlay]
    \node[draw=ocre!60,line width=1pt,circle,fill=ocre!25,font=\sffamily\bfseries,inner sep=2pt,outer sep=0pt] at (-15pt,0pt){\textcolor{ocre}{1}};\end{tikzpicture}} 
    \advance\baselineskip -1pt 
    Fill the Dewar, turn on the $B_0$ coil, demagnetize BMSR-2 with installed degaussing coils and the EDM cell with a degausser. 
   
    \item\ignorespaces 
    \makebox[-2.5pt]{\begin{tikzpicture}[overlay]
    \node[draw=ocre!60,line width=1pt,circle,fill=ocre!25,font=\sffamily\bfseries,inner sep=2pt,outer sep=0pt] at (-15pt,0pt){\textcolor{ocre}{2}};\end{tikzpicture}} 
    \advance\baselineskip -1pt 
    Mix $^{129}$Xe and $^{3}$He atoms in a refillable optical pumping cell (OPC), and polarize it by spin-exchange optical pumpig at a laser room close to BMSR-2.

    \item\ignorespaces 
    \makebox[-2.5pt]{\begin{tikzpicture}[overlay]
    \node[draw=ocre!60,line width=1pt,circle,fill=ocre!25,font=\sffamily\bfseries,inner sep=2pt,outer sep=0pt] at (-15pt,0pt){\textcolor{ocre}{3}};\end{tikzpicture}} 
    \advance\baselineskip -1pt 
    Expand the spin-polarized gas from the valved OPC to the evacuated EDM measurement cell and then transport it into BMSR-2 using a battery-powered 400 \textmu T solenoid.

    \item\ignorespaces 
    \makebox[-2.5pt]{\begin{tikzpicture}[overlay]
    \node[draw=ocre!60,line width=1pt,circle,fill=ocre!25,font=\sffamily\bfseries,inner sep=2pt,outer sep=0pt] at (-15pt,0pt){\textcolor{ocre}{4}};\end{tikzpicture}} 
    \advance\baselineskip -1pt 
    Induce a $\pi$/2 flip of the $^{129}$Xe and $^{3}$He spins and record the spin precession by a SQUID system. Turn on the electric field and modulate it with a pre-designed pattern until the amplitude of the spin precession signal is lower than that of noise.   
\end{list} \vskip5pt 

\section{Scientific instruments}
\subsection{Chamber and fields}

\begin{figure}[htb]
    \centering
    \includegraphics[width=.2\columnwidth]{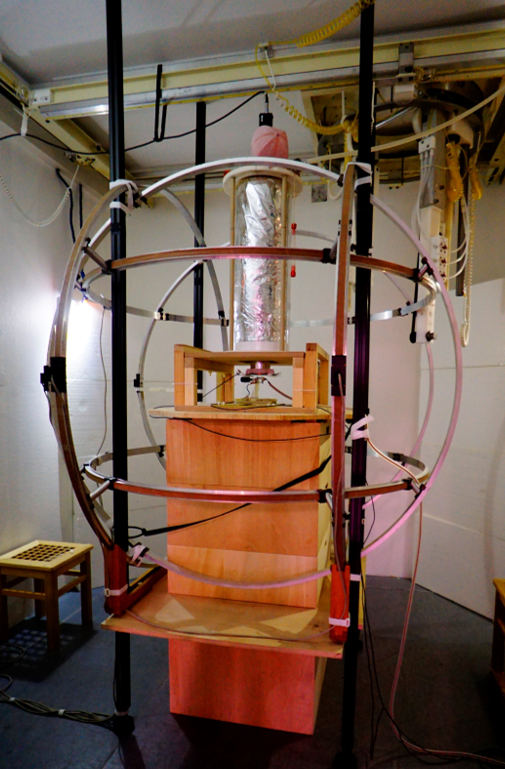}
    \hspace{0.5cm}
    \includegraphics[width=.32\columnwidth]{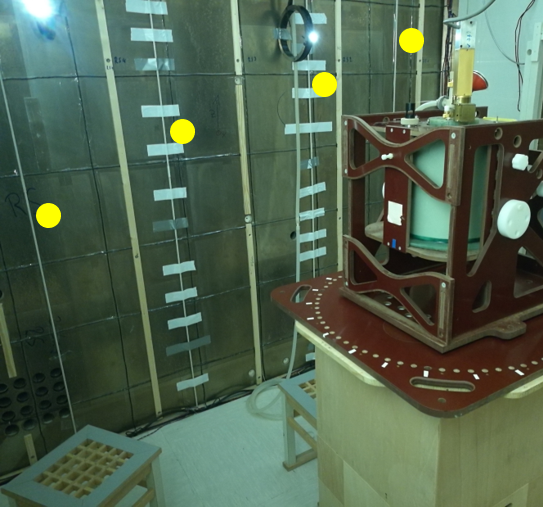}
    \hspace{0.5cm}
    \includegraphics[width=.38\columnwidth]{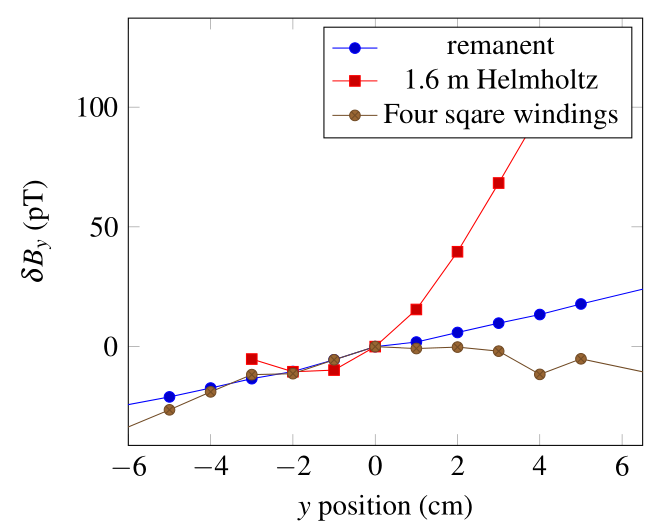}  
    \leftline{ \hspace{1.3cm} (a) \hspace{3.5cm} (b) \hspace{6cm} (c)}
    \caption{(a) The movable 3-axis Helmholtz coil and (b) a new built-in coil set of four attached square windings marked with four yellow dots. (c) The measured field changes for those two kinds of coils at $B_0$=3 \textmu T. The field was corrected with the chamber's remanent magnetic field.}
    \label{fig:Coil_uniformity}
\end{figure}

Commonly, for EDM experiments, a stable \textmu T-level magnetic field of high homogeneity is required to achieve longer relaxation time and smaller systematic error. Table~\ref{tab:Demands_field_uniformity} lists the demands and the realized values of field uniformity from several ongoing EDM experiments. The common solution is to (i) enclose the experiment inside a magnetic shield to suppress external field disturbances, and (ii) construct internal coil sets that generate the desired homogeneous field \cite{Chupp2019}. For our setup, we placed a \diameter 1.6~m Helmholtz coil inside BMSR-2 to generate the required field, as shown in Fig.~\ref{fig:Coil_uniformity}(a). BMSR-2 is an 8-layer magnetically shielded room comprised of 7 layers of Permalloy and a 10-mm-thick aluminum rf-shielding layer enclosing a space of 2.9 m $\times$ 2.9 m $\times$ 2.8 m, which is considered as the most powerful magnetic shield with an effective area over 10~m$^3$. The measured magnetic field generated by the Helmholtz coil is shown in Fig.~\ref{fig:Coil_uniformity}(c) and the average gradient in the central $\pm 5$~cm is around 12~pT/cm at a field strength of $B_0 = 3$~\textmu T, while the residual magnetic field gradient in the center region is around 3~pT/cm.

\begin{table}[h]
    \centering
    \caption{Used coil system and required field homogeneity along the main axis in various EDM experiments. Note that the specifications in most EDM experiments are manifold. Here we solely listed the relevant parameters and more details can be found in the cited literature.}
    \begin{tabular}{l l p{15mm} c  p{15mm} c}
    \toprule
    \textbf{Experiment} & \textbf{Coil type} & \textbf{Range} & \textbf{Gradient} & \textbf{Field} & \textbf{Literature} \\
	& & \textbf{(cm)}  &  \textbf{(ppm/cm)} & \textbf{(\textmu T)} & \\
    \midrule
    cyroEDM, SNS    & Cos$\theta$ coil        & $\pm 7.5$  & <3.3  & 3   & \cite{PerezGalvan2011}\\
    n2EDM, PSI      & Dedicated coils         & $\pm 16$   & <3.1  & 1   & \cite{Abel2019b}\\
    XeEDM, Japan    & Gapped solenoid         & $\pm 1$    & <166  & 3   & \cite{Sakamoto2015} \\    
    MuonEDM, J-PARC & Superconductor          & $\pm 5$    & <0.02 & $3 \times 10^6$ & \cite{Abe2018} \\   
    nEDM, LANL      & Gapped solenoid         & $\pm 14$   & <3    & 1   & \cite{Dadisman2018}   \\   
    XeEDM, PTB      & Square coil          & $\pm 5$    & <1  & 3   & \cite{Liu2021} \\
    \bottomrule
    \end{tabular}
    \label{tab:Demands_field_uniformity}
\end{table}

The present limit for residual field gradients is around 1~pT/cm in a central cube of 0.5~m side length in a temporary 3-layer MSR \cite{Sun2020}. Since it was not possible to modify solely the degaussing coil of the 20-year-old BMSR-2 \cite{Bork2001}, it was decided to upgrade the chamber after the two EDM campaigns. An additional inner layer covering some problematic pieces from the original construction with strong residual magnetization was installed and equipped with an up-to-date degaussing coil system. As another part of my PhD thesis, a built-in coil set with four square windings was designed and directly attached to the Permalloy walls to reach an ultra-high field homogeneity in the central region, as shown in Fig.~\ref{fig:Coil_uniformity}(b) \cite{Liu2021}. Compared to the previously used Helmholtz coil, the field homogeneity increases by a factor of 10 to around 1~pT/cm in the central $\pm 5$~cm. The built-in coil set inside the upgraded BMSR-2 will serve as the $B_0$ field in forthcoming experiments. 

\begin{figure}[ht]
    \centerline{\includegraphics[width=.5\columnwidth]{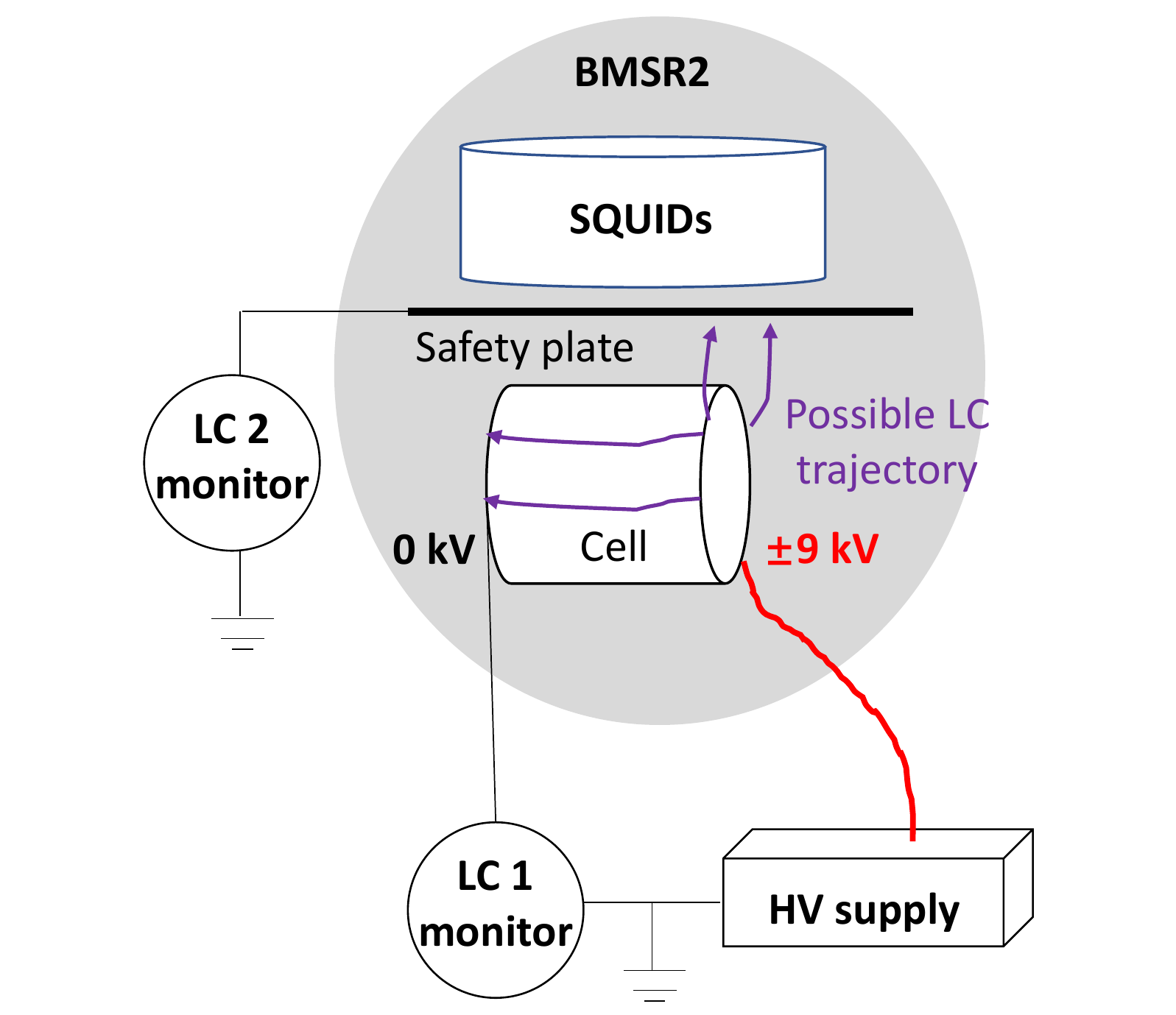}}
    \caption{The sketch for the high voltage power supply and two leakage current monitors.}
    \label{fig:E_field_Setup}
\end{figure}

The electric field was generated across the whole cell volume by applying the double-polarity DC high voltage (generated by a Trek 610-E supply placed outside the BMSR-2) directly to one of the silicone wavers attached to each side of the EDM cell whereas the other side was kept at ground level. The output voltage within the range of $\pm$ 10~kV was controlled by an analog signal from a programmed DAQ-card. For both campaigns, a current monitor to measure the leakage current across the cell was installed in order to estimate false EDM effects which could potentially be induced by flowing currents. In the 2018 campaign, another current monitor was added to record the leakage current to the safety ground, which was below the Dewar to protect the SQUIDs. The schematic of current monitors is showed in Fig.~\ref{fig:E_field_Setup}. The current monitors are composed of a current to voltage converter and a diode box for overload protection. Its converting factor is 10~mV/pA and their measurement range is $\pm$ 10~nA, with an overall noise of approximately 10~pA. 

\subsection{Measurement Cells, spin polarization and excitation}

In the two campaigns, we used three valved EDM cells, named PP1, PP2 and PP3 and one refillable optical pumping cell(OPC). These cells are cylindrical and their dimensions are listed in Table~\ref{tab:Cell}. The valved EDM cells were formed by attaching two silicon wafers to a Duran cylinder on both sides by diffusion bonding.

\begin{table}[h]
    \centering
    \caption{Parameters of used cells in EDM experiments.}
    \begin{tabular}{l l l}
    \toprule
    \textbf{Cell name} & \textbf{Type} & \textbf{Description}\\
    \midrule
    PP1 & Valved EDM & Duran/Pyrex $l=18.5$ mm, $d=20.5$ mm \\
    PP2 & Valved EDM & Duran/Pyrex $l=21.8$ mm, $d=20.4$ mm \\
    PP3 & Valved EDM & Duran/Pyrex $l=21.8$ mm, $d=20.4$ mm \\
    OPC & refillable optical pumping cell & Duran/Pyrex $l=90 $ mm, $d=49 $ mm\\    
    \bottomrule
    \end{tabular}
    \label{tab:Cell}
\end{table}

\begin{figure}[htbp]
    \centerline{\includegraphics[width=.54\columnwidth]{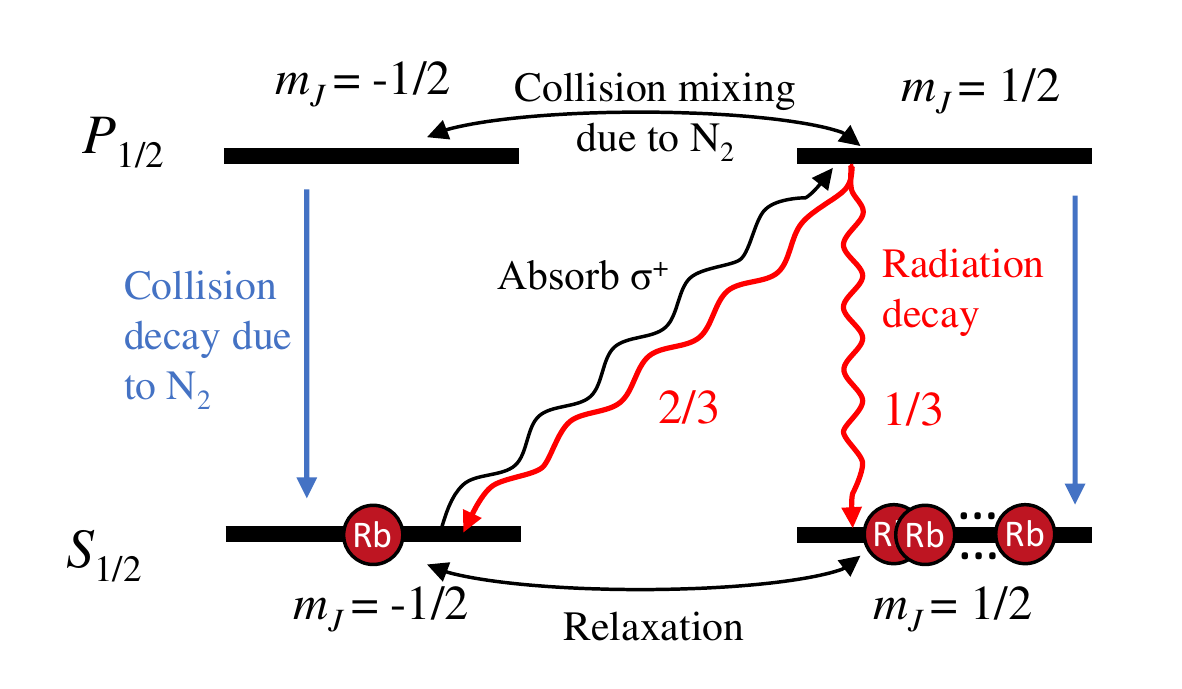}
    \includegraphics[width=.41\columnwidth]{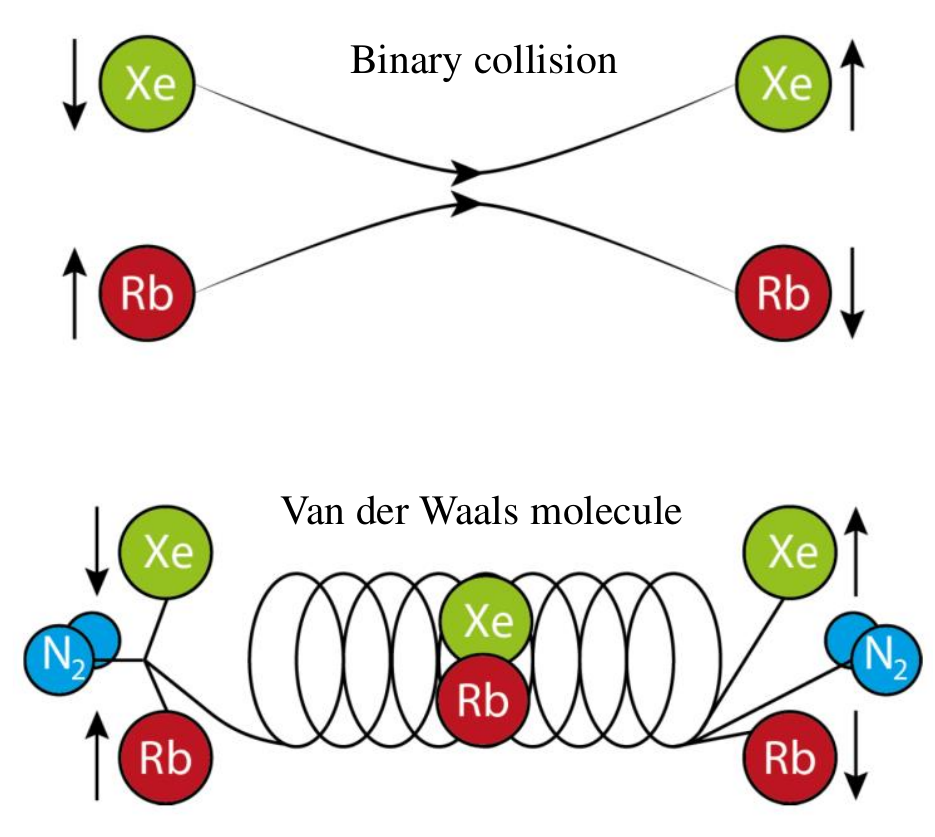}}
    \leftline{ \hspace{4cm} (a)  \hspace{6cm} (b) \cite{Stefan2018}}
    \caption{Illustration of the spin-exchange optical pumping. (a) Step 1: Optical pumping of Rubidium vapor (b) Step 2: Spin-exchange of Rb-Xe/He (for $^3$He only binary collisions occurs) \cite{Walker1997}.}
    \label{fig:SEOP}
\end{figure}

The EDM measurement sensitivity nearly scales with the polarization of the noble gases $^{3}$He and $^{129}$Xe. Spin-exchange Optical Pumping (SEOP) technology is able to hyperpolarize noble gases in the order of one bar, where electronic polarization produced in alkali-metal atoms by optical pumping (see Fig.~\ref{fig:SEOP}(a)) is transferred to noble gas nucleus during collisions via the hyperfine interaction (see Fig.~\ref{fig:SEOP}(b)) \cite{Gentile2017}. In the $^{129}$Xe EDM experiment, we used a 1-2 bar gas mixture of Xe ( enriched to $90 \pm 2 \%$   $^{129}$Xe), $^{3}$He, and N$_2$ in an OPC containing a Rb droplet and the polarization was achieved via SEOP with polarized Rb at PTB in a mobile polarizer like a previous one \cite{Korchak2013} but established as a batch mode system. N$_2$ is included as a buffer gas to suppress light-trapping from radiative decay of the excited state \cite{Walker1997}. The partial pressure of atoms are listed in Table~\ref{tab:partial pressure}.  Normally, the polarized gas from the OPC was transferred into the EDM cell in two batches, with around 1 bar for the first run and around 0.5~bar for the second run. 

An important parameter describing the state of atoms inside the measurement cell is the gas diffusion. In our case, the gases are in diffusion mode due to high pressure (>0.5~bar) and room temperature. The diffusion time is $\tau_\text{d}=R^2/D$ \cite{Cates1988} for a spherical cell of radius $R$, where $D$ is the diffusion coefficient and can be calculated  by applying the formulas and parameters given in Ref.~\cite{Allmendinger2017}. For our setup, the estimated diffusion times are listed in Table~\ref{tab:partial pressure}.  Since the spin Lamour period $\tau_\text{L} = 1/ \gamma B_0$ is much smaller than the diffusion time $\tau_d$, the species are in adiabatic mode (also called high pressure regime), meaning that the magnetic moments see the local magnetic field and do not average out gradients. 

\begin{table}[h]
    \centering
    \caption{The percentage of gas mixing and typical time of the gas species. The Lamor period is calculated for $B_{0}=2.7$~\textmu T. The diffusion time describes the effective speed of atoms in moving one cycle around the cell.}
    \begin{tabular}{c c c c c}
    \toprule
    \textbf{Species} & \textbf{2017} & \textbf{2018}& \textbf{Lamor period} & \textbf{Diffusion time}\\
    \midrule
    Xe        & 18 $\%$  & 15 $\%$ & 0.03~s & 0.50~s\\  
    $^{3}$He  & 73 $\%$  & 70 $\%$ & 0.10~s & 1.43~s\\  
    N$_2$      & 9 $\%$  & 15 $\%$ &  & \\
    \bottomrule
    \end{tabular}
    \label{tab:partial pressure}
\end{table}

\begin{figure}[ht]
    \centerline{\includegraphics[width=.5\columnwidth]{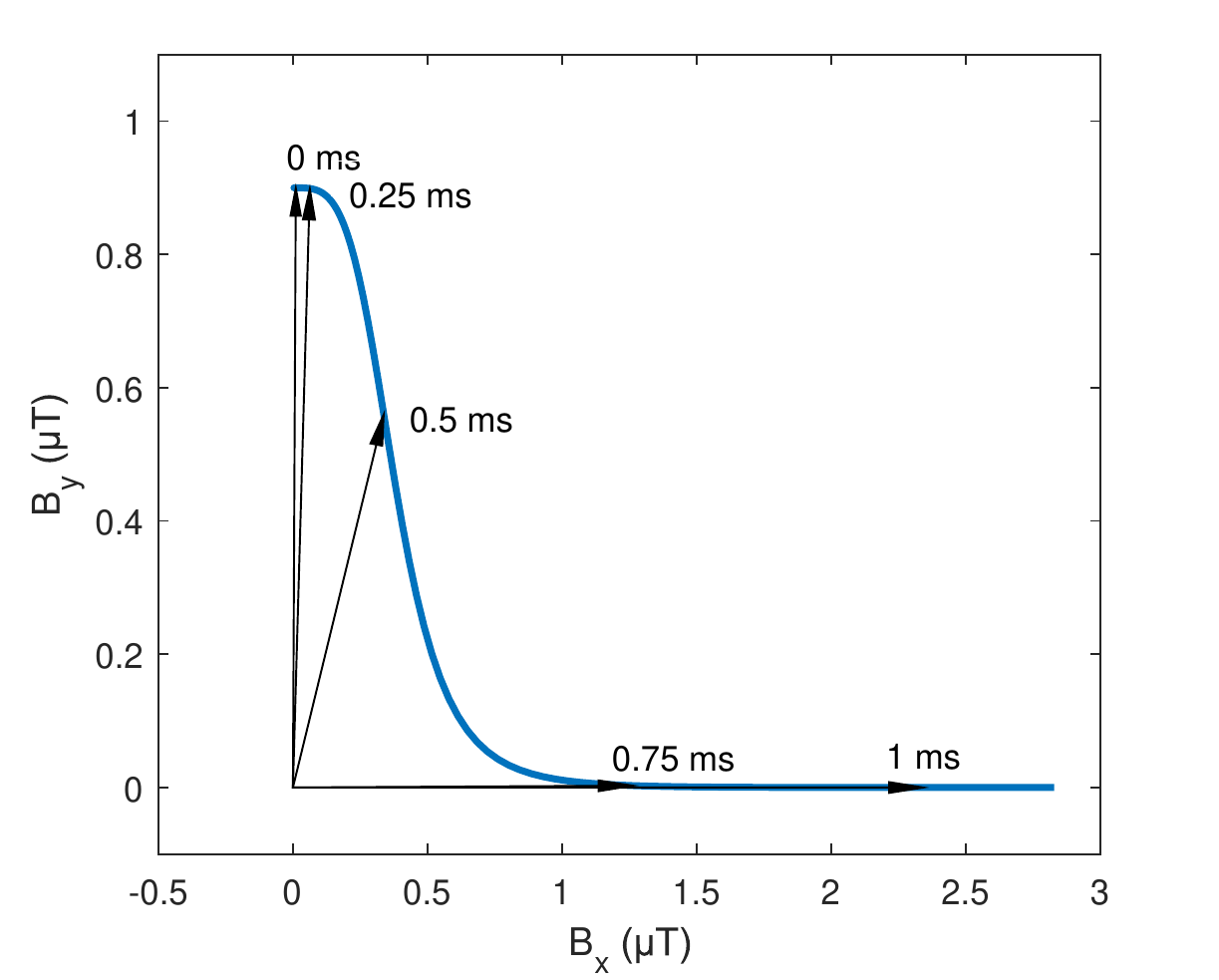}} 
    \caption{The measured field during a DC field switch used in the 2018 campaign by a triaxial Fluxgate.}
    \label{fig:Dc_Flip}
\end{figure}

The polarised gas was expanded into the evacuated EDM cell, which was then transported into BMSR-2 with a 400~\textmu T magnetic field. Later on, the 400 \textmu T transport field was adiabatically ramped down and the EDM cell was placed underneath the dewar before the door of BMSR-2 was closed. When the field drift after closing the door was stable, taking normally several minutes, the $\pi/2$ flip was performed. In the 2017 campaign, we applied a pulsed AC magnetic field resonant with the $^{129}$Xe and $^{3}$He precession frequencies. The $\pi$/2 pulse was generated by a \diameter 1.5~m Helmholtz coil in the $x$-axis powered by an Agilent programmable function generator. In the 2018 campaign, we diabatically switched the background field from the $y$ axis to the $x$ axis. The measured field change during the field switch is shown in Fig.~\ref{fig:Dc_Flip}. The field switch was completed within 2~ms, being much shorter than the Lamour frequency of both atoms as listed in Table~\ref{tab:partial pressure}.     

After the \ang{90} spin flip, the transverse magnetization starts to decay with the transverse relaxation time $T_2^*$, which is given by
\begin{equation} 
	\label{eqn:T2star}
    \frac{1}{T_2^*}=\frac{1}{T_{2\text{b}}}+\frac{1}{T_{2,\text{grad}}},
\end{equation}
where $T_{2\text{b}}$ is a constant background rate independent of the gradient, and the second term is the loss of phase coherence between the spins of the sample due to magnetic field gradients \cite{Cates1988}. Consequently, the longest $T_2^*$ time is achieved at the lowest gradient, showing the importance of a homogeneous background field.

\subsection{SQUID Gradiometer and noise performance}

A SQUID (for superconducting quantum interference device) is one of the most sensitive magnetic field detectors for our signal frequency around 100~Hz \cite{Storm2017}. Our two Xe EDM campaigns used different SQUID systems (MRX-I for 2017 and MRX-III for 2018) as detectors for the precessing spin signals. Both had two channels ($Z1$,$Z2$) with $Z2$ sensitive to the same $B$-field direction as $Z1$ but 12~cm further away from the EDM cell. MRX-I(MRX-III) used PTB-fabricated W9L(W9M) SQUID magnetometers on a $7 \times 7$ ($3 \times 3$)~mm$^2$ chip with a typical field noise level as low as 1.3(3.6)~$\text{fT}/\sqrt{\text{Hz}}$ at 100~Hz. The cold to warm distance for MRX-I(MRX-III) was 12(6)~mm.

\begin{figure}[htpb]
    \centerline{\includegraphics[width=.7\columnwidth]{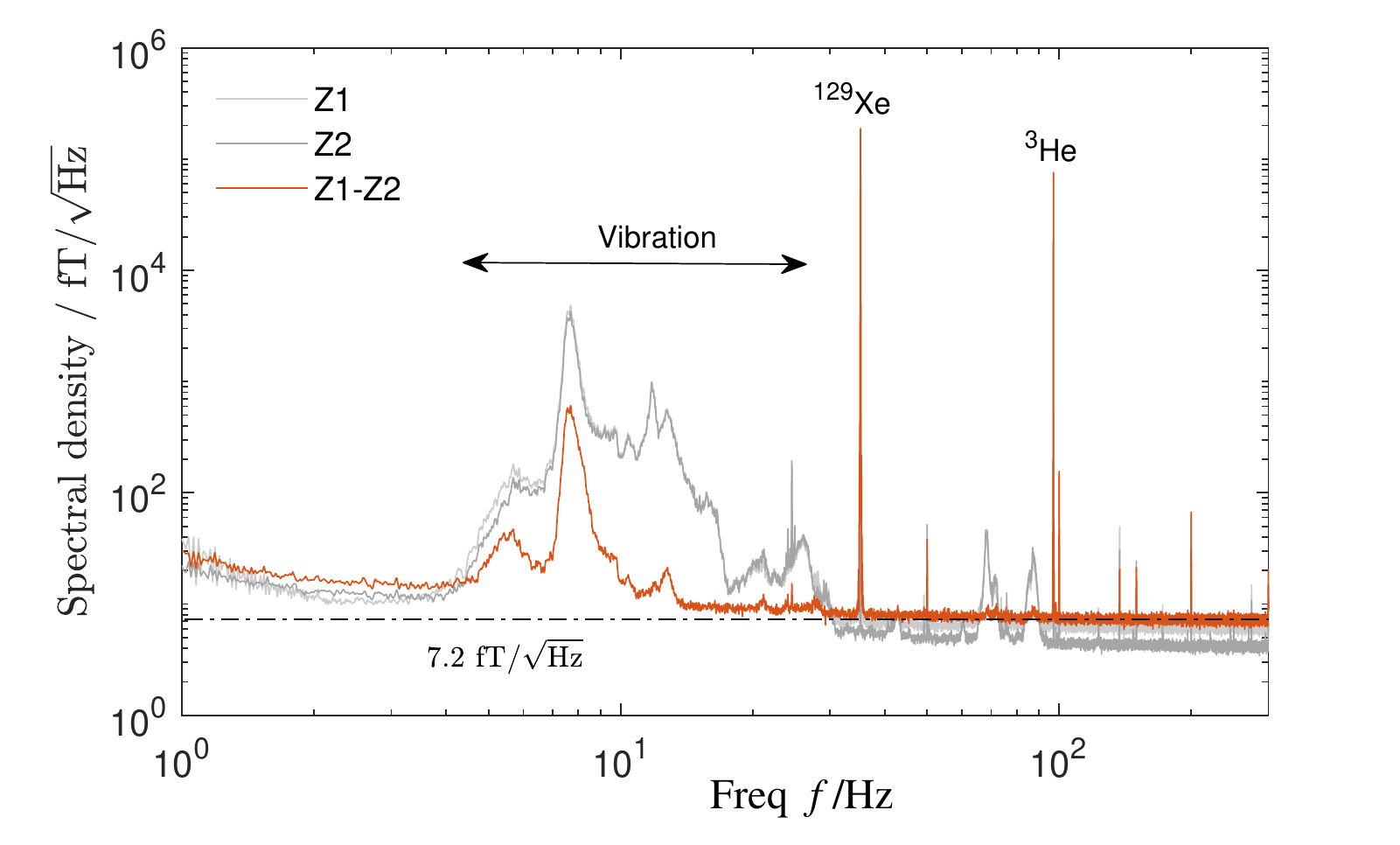}}
    \caption{ The amplitude spectral density of data lasting 100~s from the starting of the first sub-run for two magnetometer channels ($Z1$ and $Z2$) and one software gradiometer ($Z1-Z2$). The white noise level of the gradiometer is $\rho_\mathrm{\omega} \approx 7.2~\mathrm{fT}/\sqrt{\mathrm{Hz}}$. The variance of the white noise is $\sigma^2_\mathrm{\omega}=f_\mathrm{s}\rho_\mathrm{\omega}^2/2=(154~\mathrm{fT})^2$.}
    \label{fig:Noise}
\end{figure}

The measurement data were recorded with a custom-designed data acquisition system. The sampling rate $f_\mathrm{s}$, nominally 915.5245 Hz, was derived from an Oscilloquartz BVA8607 external clock stable up to $10^{-11}$ Hz over time scales relevant for the experiment. Fig.~{\ref{fig:Noise}} shows the power spectral density of the SQUID signal from the B88 run of the 2018 campaign. The two peaks at 35.2~Hz and 97~Hz were the signal of $^{129}$Xe and $^{3}$He, respectively. The white noise level of the SQUID magnetometer for $Z1$($Z2$) were around 6.2(4.2)~$\text{fT}/\sqrt{\text{Hz}}$. The signal in the frequency range of $4$~Hz – $25$~Hz was caused by the vibration of magnetometers in a field gradient. In the analysis performed in this work, in order to suppress the impact of vibrational noise, a software SQUID gradiometer ($Z1-Z2$) was used in the analysis performed in this work. As shown by the orange curve in Fig.~{\ref{fig:Noise}}, the close by vibrational noise at the $^{129}$Xe and $^{3}$He frequencies were essentially eliminated and mainly the intrinsic BMSR-2 vibrational mode at 7.8~Hz still existed. On the other side, the white noise level also increased to 7.2~$\mathrm{fT}/\sqrt{\mathrm{Hz}}$ and the signal amplitude reduced by less than 2\%.

\section{Experimental setup and parameters}

\begin{figure}[hb]
    \centerline{\includegraphics[width=.97\columnwidth]{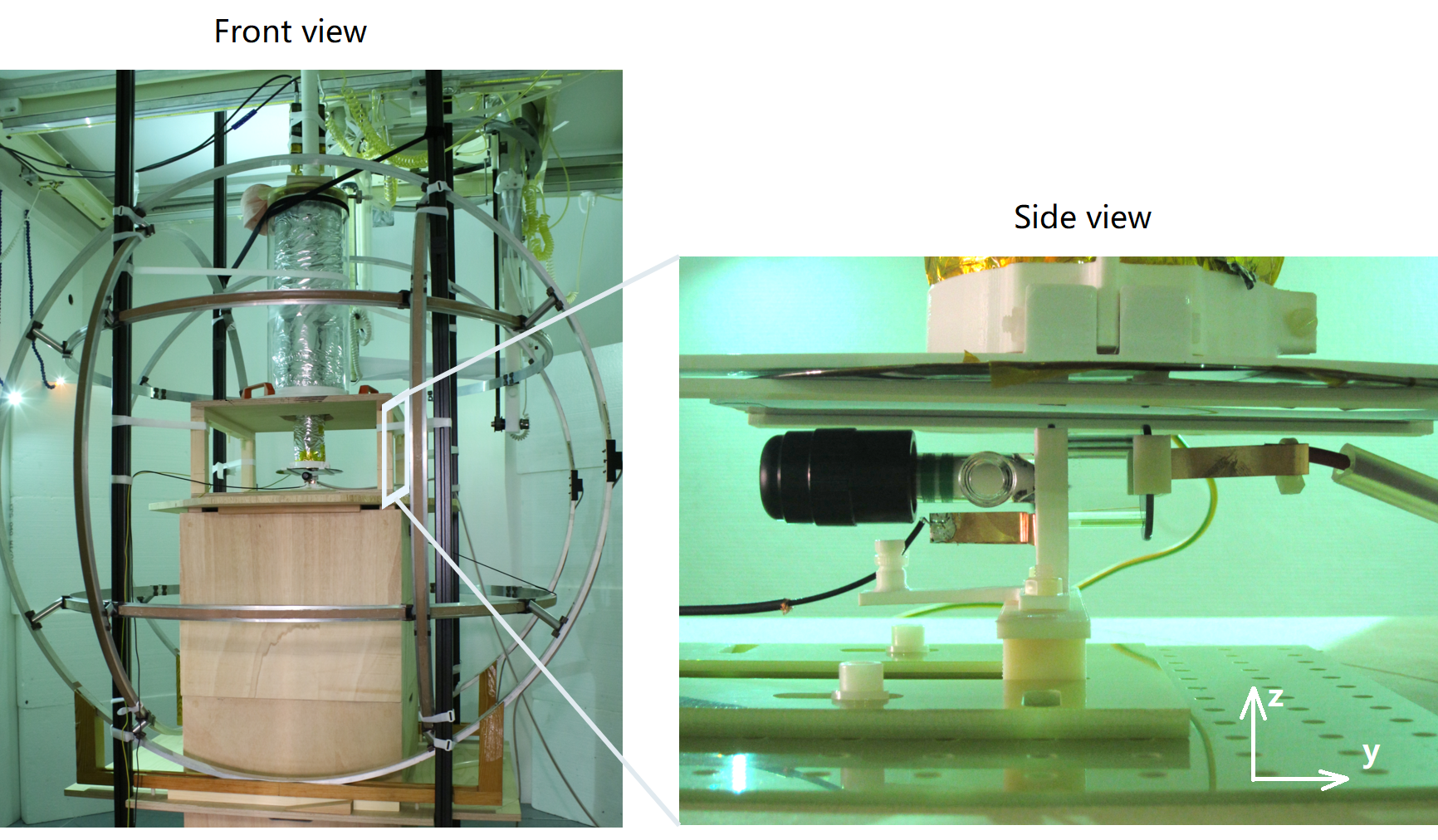}}
    \centerline{(a) \hspace{7cm}  (b)}
    \centerline{\includegraphics[width=\columnwidth]{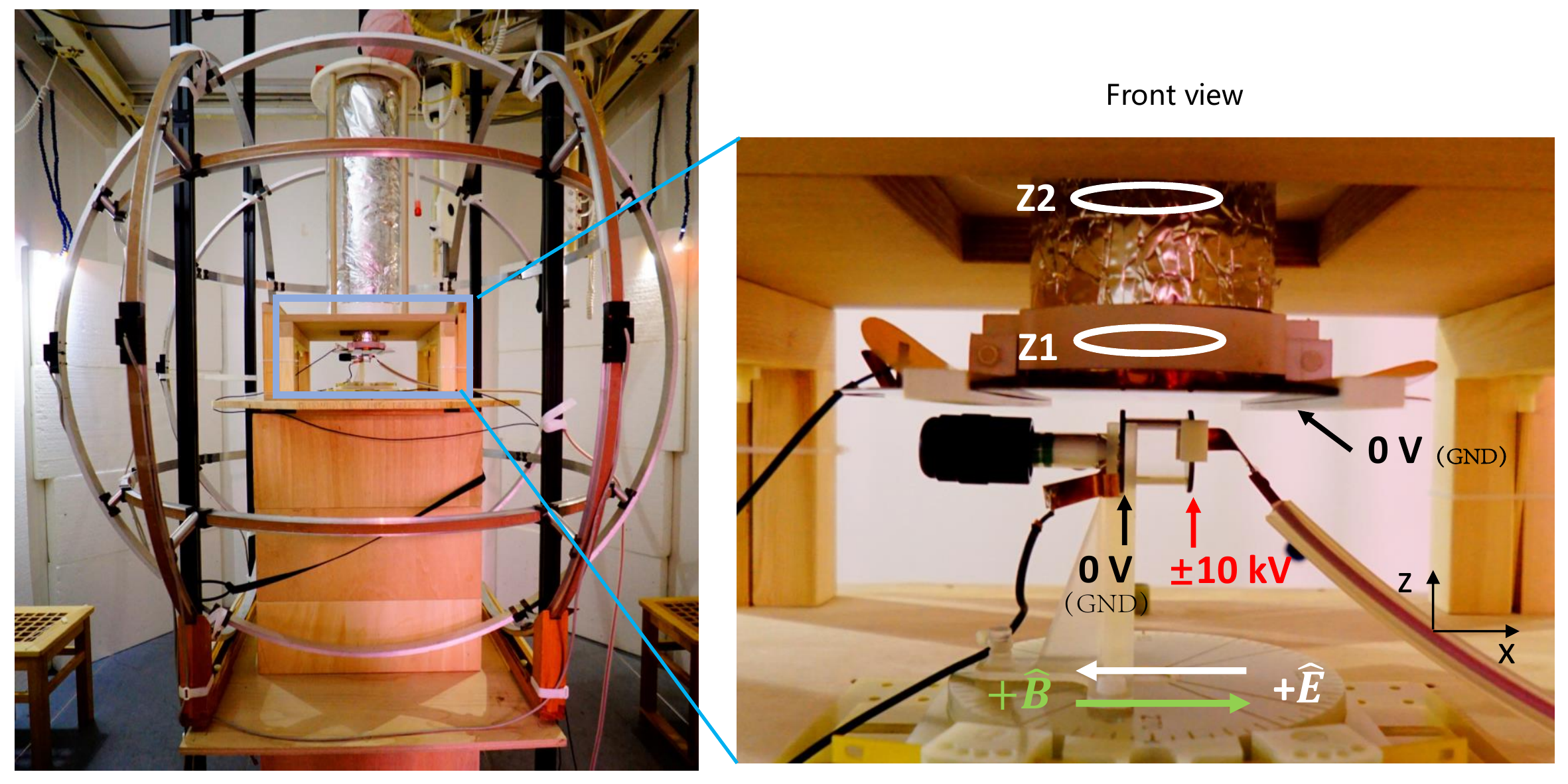}}
    \centerline{(c) \hspace{7cm}  (d)}
    \caption{The hardware setup of 2017 (a and b) and 2018 (c and d) XeEDM campaign at PTB. The background field $B_0$ was generated by the Helmholtz of the greatest diameter in both campaigns while the direction changed from the $y$ axis in 2017 to the $x$ axis in 2018.}
    \label{fig:Setup_2018}
\end{figure}

The hardware setup for the 2018 campaign is shown in Fig.~\ref{fig:Setup_2018}. The typical parameters used in the two campaigns are listed in Table~\ref{tab:exp_para}. The main improvement for the 2018 campaign was the increase of polarization by a factor of 1.5 for $^{129}$Xe and 10 for $^{3}$He, improving the SNR by these factors. The increased signal amplitudes, however, lead to a large anomalous comagnetometer frequency drift, thus limiting the segment length of electric field patterns. In the last week of the 2018 campaign, we intendedly reduced the polarization of $^{3}$He but further increased the $^{129}$Xe polarization, resulting in a much smaller comagnetometer frequency drift, allowing the use of a longer segment length. Other upgrades included an increase of the maximum electric field from 6~kV to 9~kV and an slight improvement of $T_2^*$ time. All of these factors together lead to an improvement of EDM sensitivity per run by a factor of 5 compared to the start of the 2018 campaign and the full 2017 campaign. 

\begin{table}[h]
    \centering
    \caption{The main parameters realized in both measurement campaigns.}
    \begin{tabular}{c c c c }
    \toprule
    \textbf{Parameters} & \textbf{2017} & \textbf{2018, first two weeks}& \textbf{2018, last week} \\
    \midrule
    Run Start amplitude He/Xe(pT) &  6/40  & 60/60 &   28/78  \\
    $T_2^*$ He/Xe(s) &     7000/8000  &  9000/9000  &  9000/9000 \\
    $B_0$ (\textmu T)       & 2.6    & 3.0  & 3.0 \\  
    White gradiometer noise ($\text{fT}/\sqrt{\text{Hz}}$) & 6.7  & 7.2 & 7.2 \\  
    Distance to SQUID (mm)      & 50  & 36 &  36 \\
    E field (kV/cm) & 2.75 & 4.13  & 4.13 \\
    Segment length (s) & 400,800 &100,200,400  & 300,600 \\
    Valid run/sub-run number &  17/17 & 19/52 & 9/18\\
    \bottomrule
    \end{tabular}
    \label{tab:exp_para}
\end{table}

\chapter{Data analysis method}

\section{Overall process} 

\begin{figure}[ht]
    \centerline{\includegraphics[width=0.8\columnwidth]{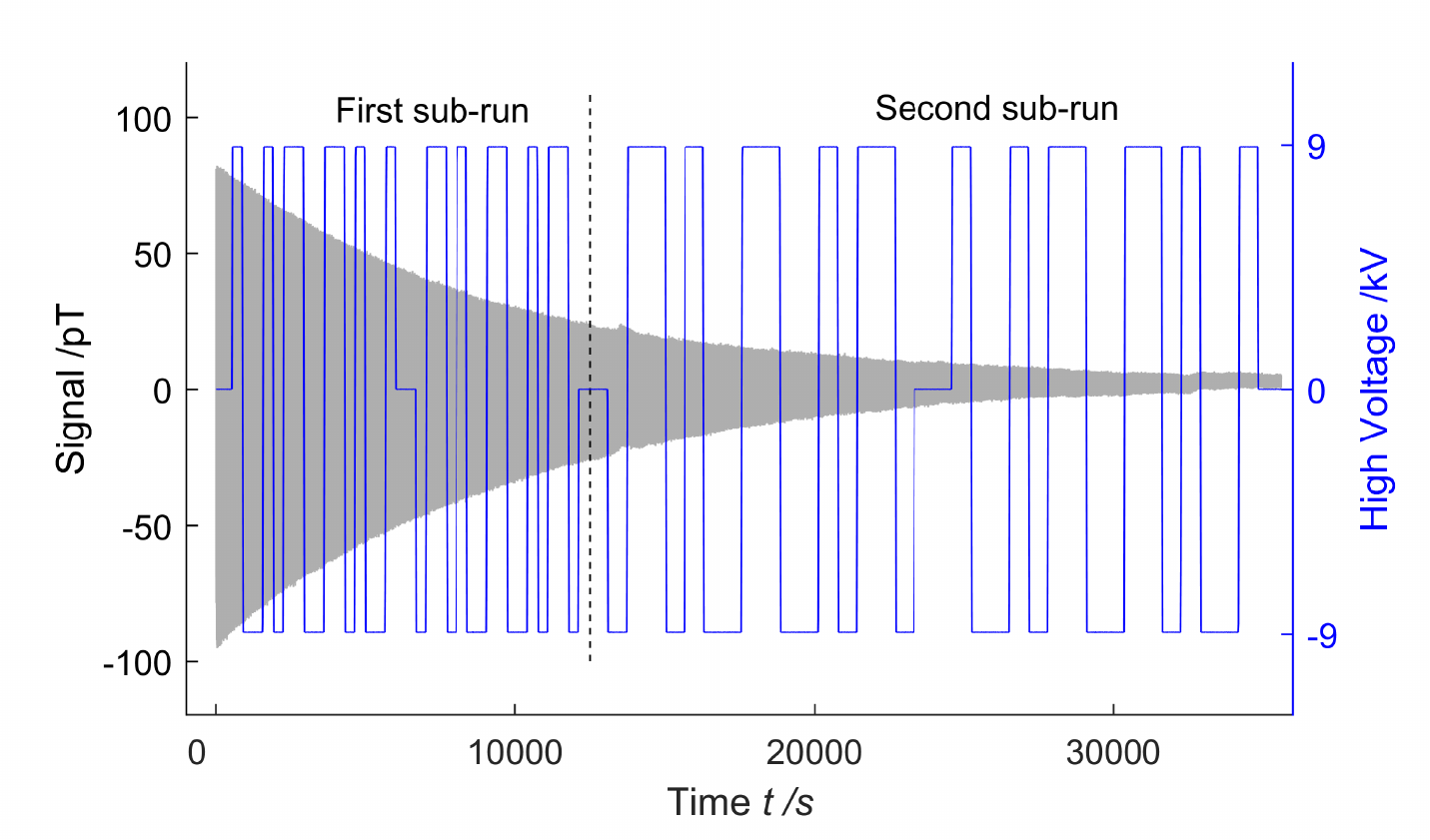}}
    \caption{The recorded gradiometer $Z1-Z2$ signal (gray curve) and the modulated high voltage signal (blue line) of the run B88 from the 2018 campaign.}
    \label{fig:Grad_HV} 
\end{figure}

Fig.~\ref{fig:Grad_HV} shows the recorded SQUID gradiometer signal (gray curve) on the left $y$-axis and the modulated high voltage signal (blue line) on the right $y$-axis of the B88 run from the 2018 campaign lasting 35000~s exemplarily.

Before going into details about the data analysis procedure, we first introduce several important terms regarding the data analysis as follows:
\begin{dBox}
    \begin{tabular}{c c l}
    \textbf{Run}&:&A run starts after the \ang{90} spin flip and lasts for 4-8 hours until the signal vanished \\
    \textbf{Sub-run}&:&A sub-run consists of a planned sequence of electric field states lasting 1-4~h\\
    \textbf{Segment}&:&The applied high voltage is kept constant in a segment lasting 100-800~s(5000~s)\\
    \textbf{Block}&:&A block of 1-20~s measured data is used to derive the precession phases\\ 
    \end{tabular}
\end{dBox}

The electric field pattern with repetitively reversed signs was applied to mitigate the effect of the comagnetometer frequency drift. The common $E$ pattern for one sub-run consisted of 36 segments with an equal time interval $t_\mathrm{s}$, and the sign of $\mathbf{E}$ changed according to the following sequence $\pm$[0 + - - + - + + - - + + - + - - + 0, 0 - + + - + - - + + - - + - + + - 0]. The segments of zero voltage were added to allow for systematic error studies.

In the 2017 campaign, each run consisted of only one sub-run. Contrarily a run in the 2018 campaign normally covered 2-4 sub-runs. This is the result of increased starting amplitudes of the two spin precession signals and slightly longer $T_2^*$ times, permitting a longer measurement time. The typical number of segments in one sub-run was 36, accounting for over 90\% cases. There were 2 sub-runs with 3 segments only, and 4 sub-runs with 18 segments. For the two sub-runs shown in Fig.~\ref{fig:Grad_HV}, the segments last 300~s and 600~s. The first sub-run ranging from 50~s to 12400~s is used as an example in the data analysis section.  

\begin{figure}[ht]
    \centerline{\includegraphics[width=\columnwidth]{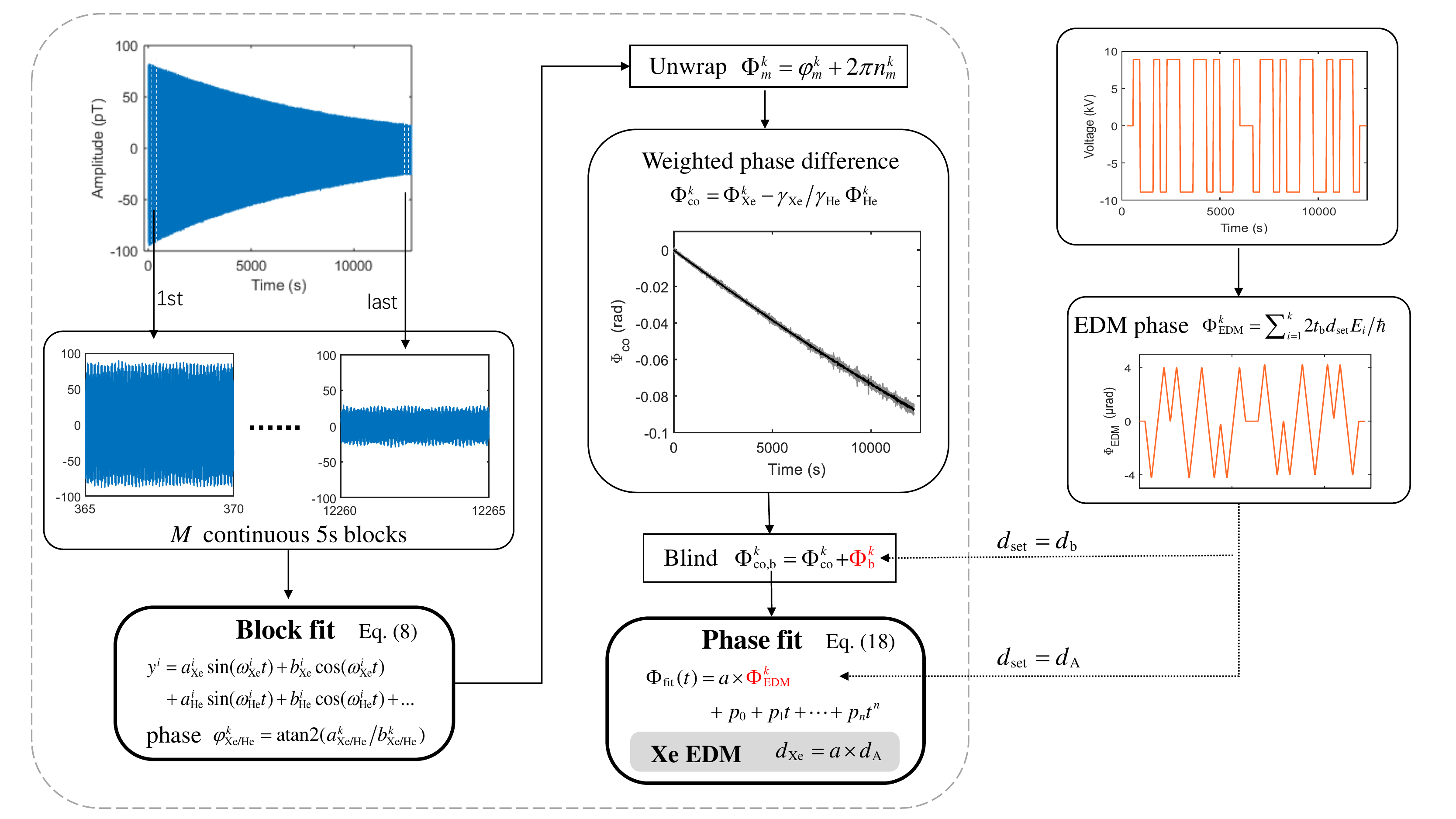}}
    \caption{Schematic process of the GPF method.}
    \label{fig:Analysis Procedure}
\end{figure}

The sub-run data is cut into continuous blocks of the same length $t_b$ \footnote[2]{One could also think about using varied block length $t_b$ in the analysis of one sub-run, to account for the decrease in SNR due to signal decay}. The analysis first extracts the phases of the precession of $^{129}$Xe and $^{3}$He for the center of each block, then unwraps them into continuous precessing phases over the complete sub-run, after which the comagnetometer phase, also called weighted phase difference (WPD) $\Phi_{\mathrm{co}}$, is calculated. At this point a blinding function can be added being generated from the measured $E$-field time evolution before the complete comagnetometer phase $\Phi_{\mathrm{co,b}}$ is fitted to a sum of a constructed EDM function and a polynomial. From the fitting algorithum one obtains an EDM value with its uncertainty. Fig.\ref{fig:Analysis Procedure} outlines the process flow and the detailed procedure is described in the following sections. 

In the following, each sub-run is named with four characters. The first letter A/B represents the 2017/2018 campaign, and the following two characters denote the number of the corresponding run, while the last character specifies the sub-run number within the run. For example, the first sub-run of the 88 run in the 2018 campaign is denoted as 'B881'.

\section{Phase estimation}

For a holding magnetic field of 3~\textmu T, the signal recorded by the SQUID gradiometer is a beat-note between the helium and xenon signals, at 35.2~Hz and 96.7~Hz, respectively. In addition to the precession signal, various noise was present in the data, consisting mainly of pink noise (1/f noise), noise from power lines at 50~Hz frequency plus higher order harmonics and remaining vibrational noise. Since the duration of one block $t_b$ was orders of magnitude smaller than the $T_2^*$ times of the detected signals, the amplitudes of these two sinusoidal signals in one block can be set to constants. The $^{129}$Xe and $^{3}$He phases for each block were determined by a time-domain fit with the function
\begin{equation} 
    \label{eqn:sincos_fit}
    \begin{aligned}
    y=& a_{\mathrm{Xe}}\sin(\omega_{\mathrm{Xe}}t)+b_{\mathrm{Xe}}\cos(\omega_{\mathrm{Xe}}t)+a_{\mathrm{He}}\sin(\omega_{\mathrm{He}}t)+b_{\mathrm{He}}\cos(\omega_{\mathrm{He}}t)+ \\
    & a_{i}\sin(\omega_{i}t)+b_{i}\cos(\omega_{i}t)+c+d \cdot t,
    \end{aligned}
\end{equation}
where $a_{\mathrm{Xe/He}/i},b_{\mathrm{Xe/He}/i},\omega_{\mathrm{Xe/He}},c$,~and~$d$ are the fit parameters and $\omega_{i=1,2,3,4}=2\pi \times 50i$~s$^{-1}$ represent the power frequency and its harmonics. The constant and linear terms $c$ and $d\cdot t$ describe the background magnetic field and its small drift as seen by the SQUID. These nonlinear fits were performed using the separable non-linear least squares method, also called Variable Projection(VP) method \cite{Golub2003}, where the nonlinear parameters $\omega_{\text{Xe/He}}$ were estimated separately from the linear parameters (see the detailed steps in Box.\ref{box:Variable Projection}). 

\begin{theorem}
    \label{box:Variable Projection}
    \hspace{1cm} \textbf{Variable Projection Method} \newline

    The inputs are observations $y$, time $t$ and the initial guesses of the frequencies $\omega_{\text{Xe/He}}$. \newline

    \textbf{Step 1:} To minimize the correlation between the fit terms in Eq.~(\ref{eqn:sincos_fit}), the time of each block is assigned to be symmetrical around zero.  \newline

    \textbf{Step 2:}  Using the frequencies $\omega_{\text{Xe/He}}$ to create a design matrix 
    \begin{equation*} 
        \begingroup 
        \setlength\arraycolsep{2pt}
        \mathbf{D}(\omega_{\text{Xe}}, \omega_{\text{He}})=
        \begin{bmatrix} 
            \sin(\omega_{\text{Xe}}t),&\cos(\omega_{\text{Xe}}t),
            &\sin(\omega_{\text{He}}t),&\cos(\omega_{\text{He}}t),
            &\sin(\omega_{i}t),&\cos(\omega_{i}t),
            &\textbf{1},&t 
        \end{bmatrix}.
        \endgroup
    \end{equation*}

    \textbf{Step 3:}  Calculate the parameter vector $P=\mathbf{D}(\omega_{\text{Xe}}, \omega_{\text{He}})^{-1}y^\intercal$.     \newline
    
    \textbf{Step 4:}  Applying the Levenberg-Marquardt least-squares minimization to find the frequencies $\omega_{\text{Xe/He}}$ that minimize $\epsilon= \begin{Vmatrix}  y - \mathbf{D}(\omega_{\text{Xe}}, \omega_{\text{He}})P \end{Vmatrix}$.     \newline

    \textbf{Step 5:} Repeat the steps 2-4 using the resulting frequencies $\omega_{\text{Xe/He}}$ until the change of the sum of residual square $\epsilon$ is less than a given threshold. \newline

    \textbf{Step 6:} Calculate the covariance matrix of the parameter vector $P$ with  $\text{Cov} =\epsilon/\nu(\mathbf{D}^\intercal\mathbf{D})^{-1}$, where $\nu$ is the number of degrees of freedom. \newline

    The outputs are $\omega_{\text{Xe/He}}$, $P=[a_{\mathrm{Xe/He}/i},b_{\mathrm{Xe/He}/i},c,d]$ and its covariance matrix \text{Cov}. 
\end{theorem}

The accuracy of the Variable Projection method depends on the precision of the initial guess of the frequencies, even so for a high signal to noise ratio due to the multiple local minimum points. A precise initial guess can also decrease computation time. The initial guess of $\omega_{\text{Xe/He}}$ was determined from the frequency with the highest magnitude in the FFT spectrum of the data $y$ over a sub-run within the expected region around Lamour frequencies.  

The uncertainties of the fit parameters were determined from the covariance matrix with 
\begin{equation} 
    \label{eqn:para_std}
    \delta P_i=\sqrt{(\text{Cov}(i,i))}, i=1,2,...,16
\end{equation}
The wrapped phase $\phi_{\text{Xe/He}}$ in the range of $(-\pi,\pi)$ was calculated from the parameters $P$ as 
\begin{equation} 
    \label{eqn:phase}
    \phi_{\text{Xe/He}}=\text{Arg}(a_{\text{Xe/He}}+j \cdot b_{\text{Xe/He}}),
\end{equation}
where the Arg() function gets the principle argument of a complex number and $j$ is the imaginary unit. Note that the phases derived here for block $k$ corresponding to the time at the middle of blocks, i.e. $t=(k-0.5)*t_b$, due to the shift of the time data in the VP fit. The phases uncertainties were calculated according to the law of error propagation as
\begin{equation} 
    \label{eqn:phase_std}
    \delta\phi_{\text{Xe/He}}=\frac{\sqrt{(a_{\text{Xe/He}} \cdot \delta b_{\text{Xe/He}})^2+(b_{\text{Xe/He}} \cdot \delta a_{\text{Xe/He}})^2 - 2a_{\text{Xe/He}}b_{\text{Xe/He}}\text{Cov}(a_{\text{Xe/He}},b_{\text{Xe/He}})}}{a_{\text{Xe/He}}^2+b_{\text{Xe/He}}^2}.
\end{equation}

\subsection{Residual and vibrational noise}

\begin{figure}[ht]
    \centerline{\includegraphics[width=.48\columnwidth]{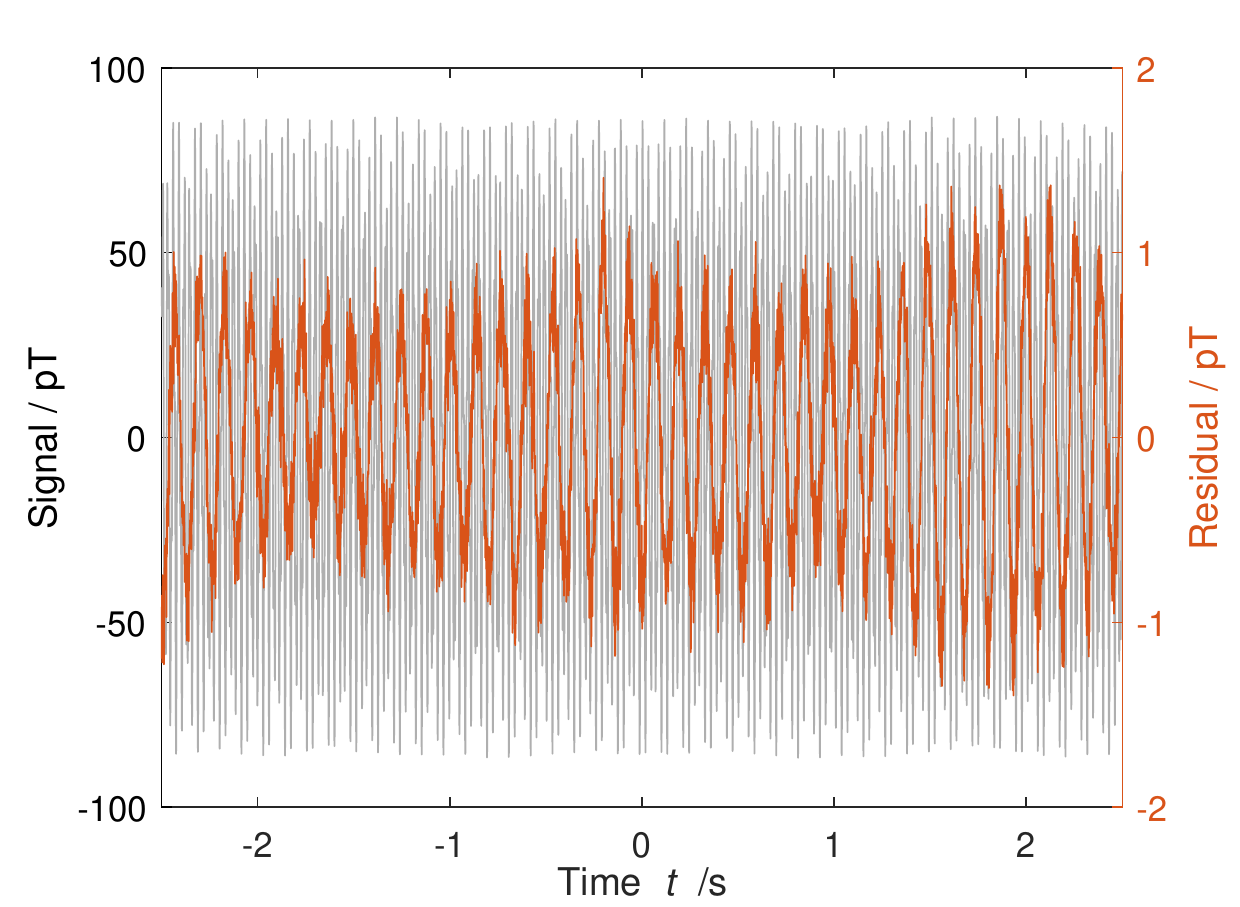} \hspace{0.5cm}
    \includegraphics[width=.48\columnwidth]{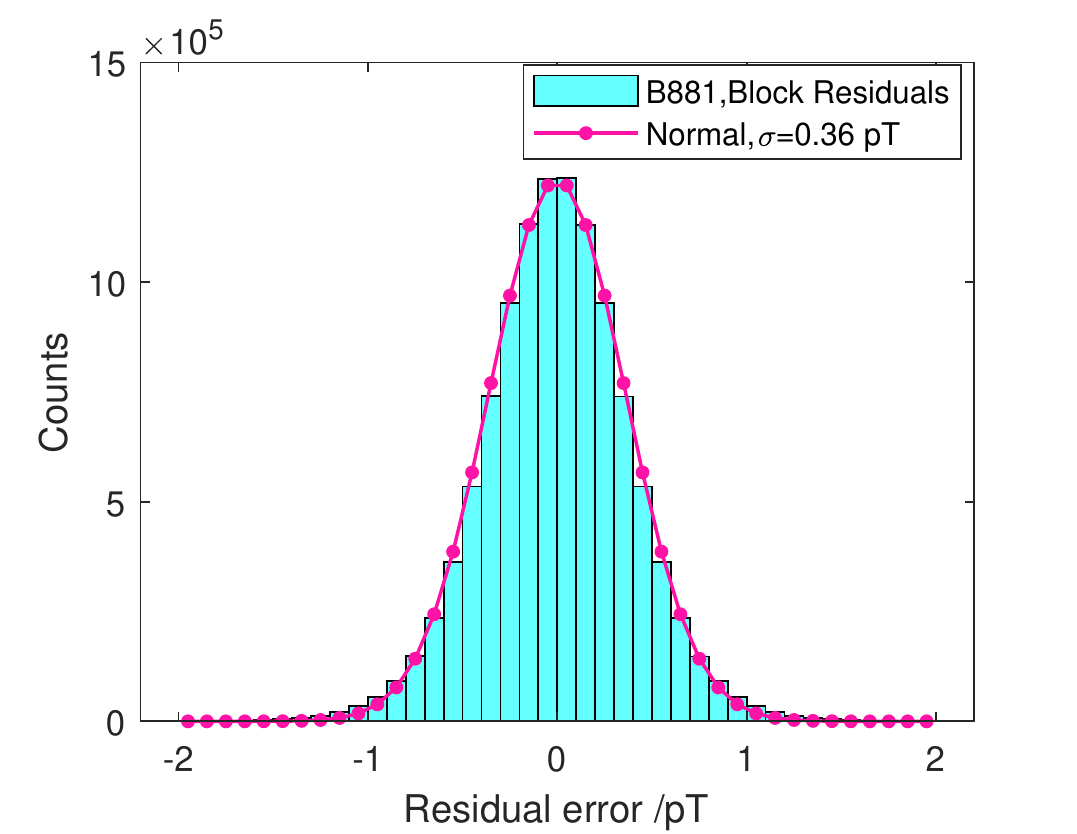}}
    \leftline{ \hspace{3cm} (a)  \hspace{7.5cm} (b)}
    \caption{(a) Signal (gray) and residual (red) of the first block from the sub-run B881. (b) Histogram of the residual error from all blocks of the sub-run B881 (in total 2436 blocks and 11150790 sample points). The purple line with dots shows a Gaussian fit to the histogram of the residuals.}
    \label{fig:fit_quality}
\end{figure}

\begin{figure}[htbp]
    \centerline{\includegraphics[width=.45\columnwidth]{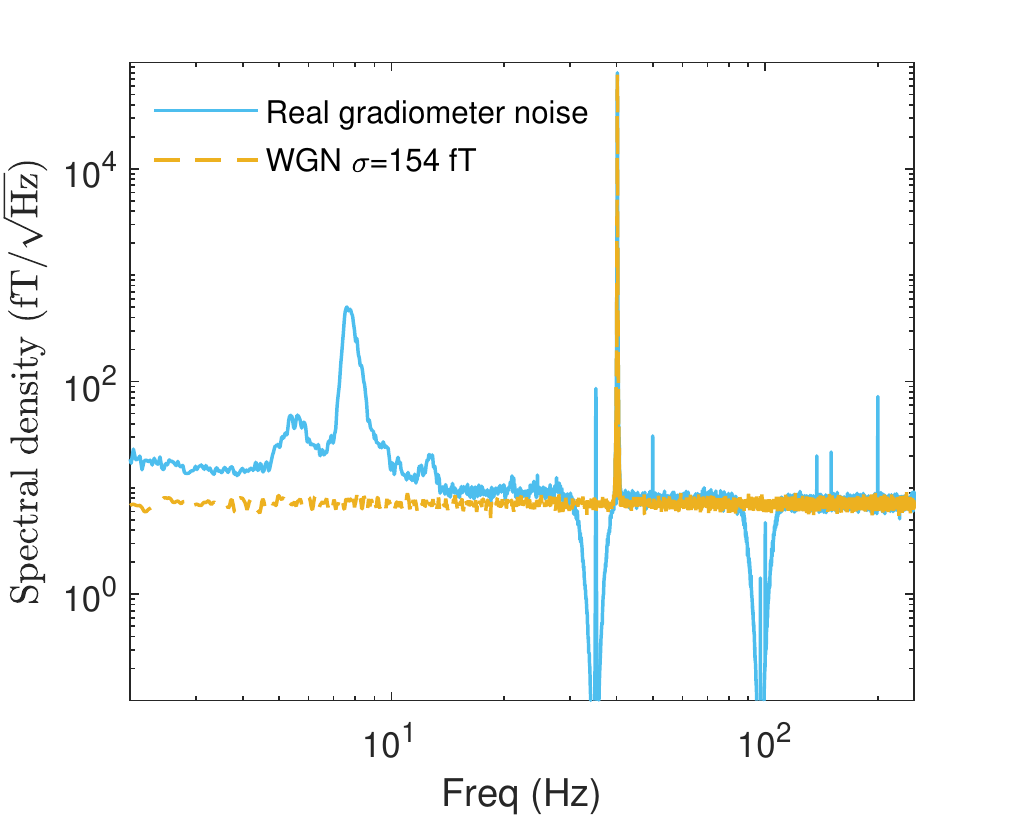} \hspace{1cm}\includegraphics[width=.55\columnwidth]{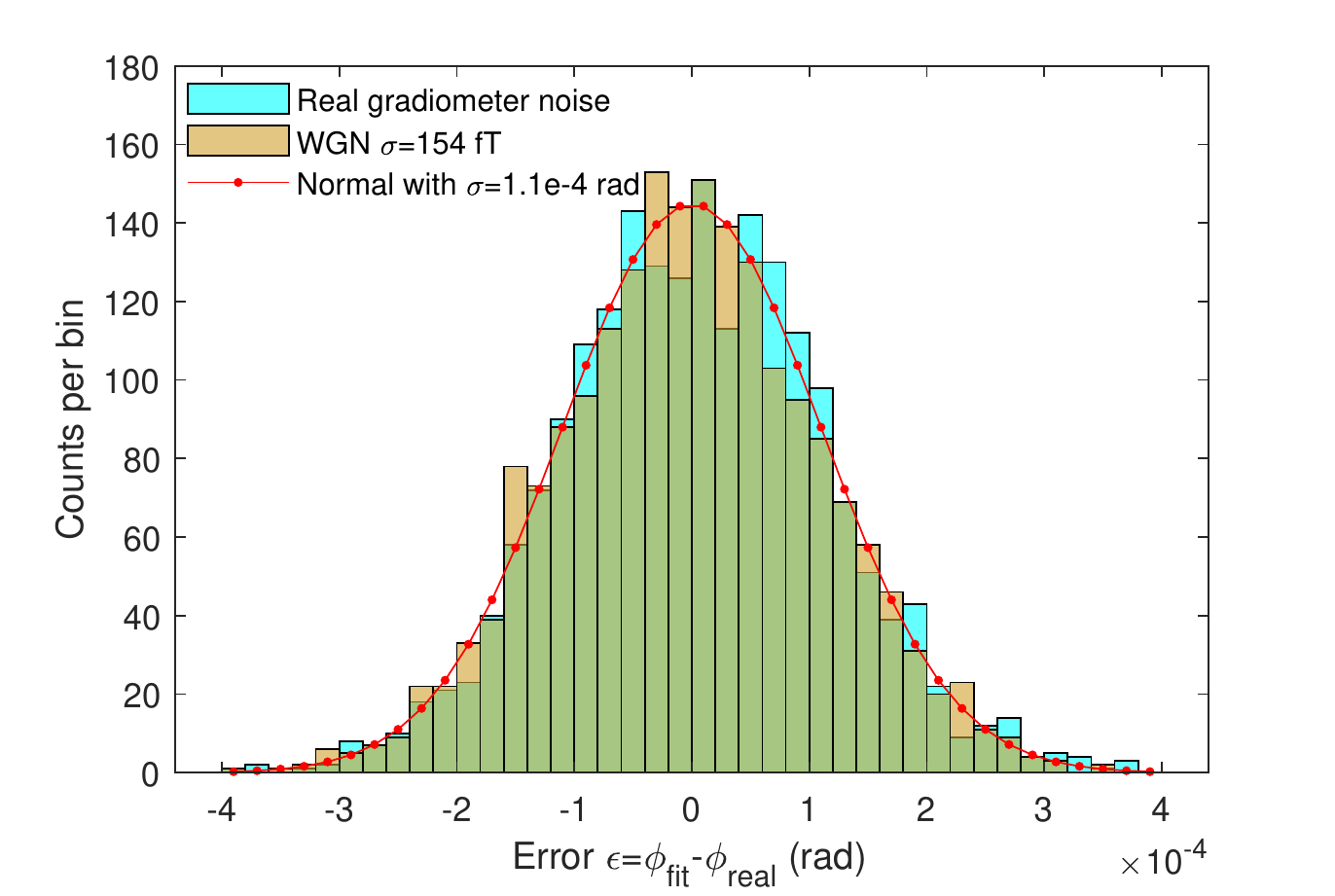}}
    \leftline{ \hspace{3.5cm}(a) \hspace{7cm} (b)}
    \caption{(a)Power spectral density curves of the synthetic data with two kinds of noise. The two holes in real noise data at 35~Hz and 97~Hz are the result of two bandstop filters. (b) Histograms of the phase error from the synthetic data with real gradiometer noise (light blue) and white noise (yellow). The data lasts for 10000 s and consists of 2000 blocks.}
    \label{fig:syn_phase}
\end{figure}

Fig.~\ref{fig:fit_quality}(a) shows the SQUID gradiometer data (gray) of the first 5~s block from the exemplary sub-run B881 and the residual (red) of the fit to this data (be aware of the different $y$-axis scaling). The residual is dominated by mechanical vibration in the frequency range of 4~Hz - 25~Hz, which is not fitted but present in the measured signal as shown in Fig.~\ref{fig:Noise}. The power of the vibrational noise, which was estimated via integrating the spectral density curve, is around (0.40~pT)$^2$, being a factor of 6.8 larger than the white noise power $\sigma^2_\mathrm{\omega}=f_\mathrm{s}\rho_\mathrm{\omega}^2/2=(0.15~\mathrm{pT})^2$\footnote[2]{The power of vibrational noise of this sub-run is the third highest among all 87 sub-runs with the recorded maximum vibrational noise power of (0.46~pT)$^2$.}. Such a large residual makes the goodness-of-fit $ \chi^2$ poor, and leads to an overestimation of the phase uncertainty with respect to the statistical dispersion of the phase error (see the Monte-Carlo simulation in Fig.~\ref{fig:syn_phase}). The vibrational noise as seen from the histogram of all block residuals within a sub-run, however, satisfied Gaussian distribution well, as shown in Fig.~\ref{fig:fit_quality}(b).

\begin{figure}[htbp]
    \centerline{\includegraphics[width=.48\columnwidth]{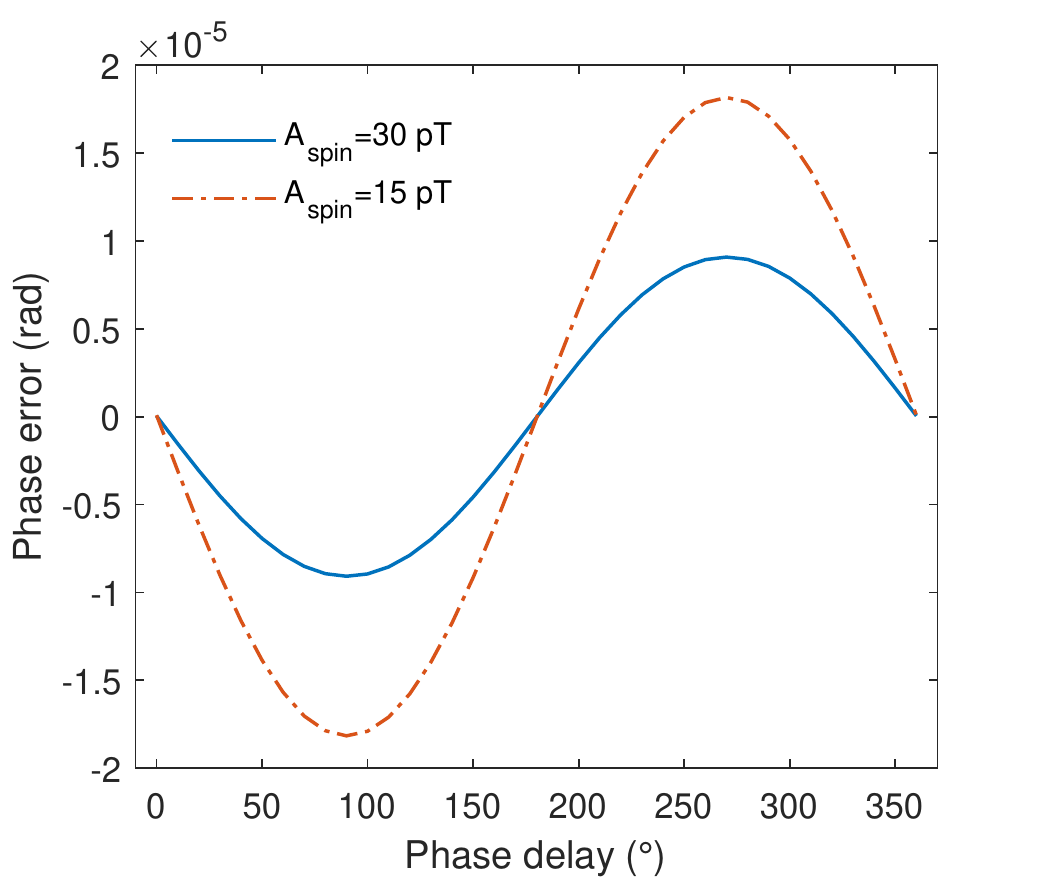} \hspace{1cm}\includegraphics[width=.48\columnwidth]{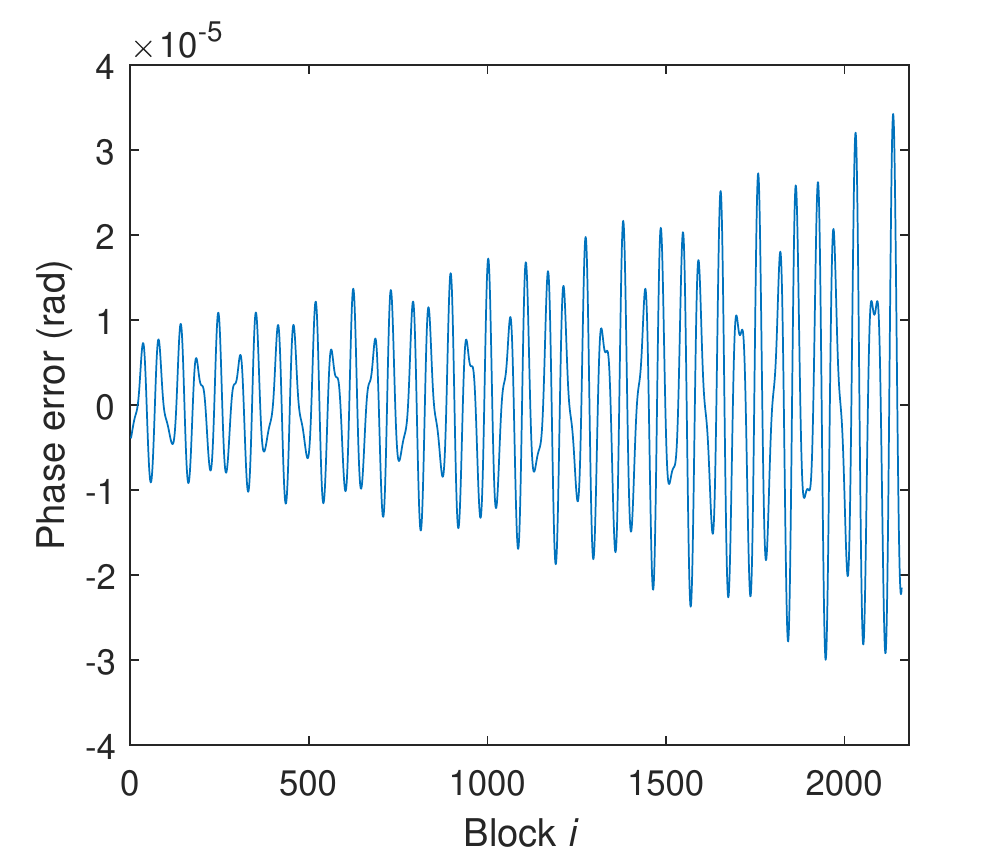}}
    \leftline{ \hspace{3.5cm}(a) \hspace{7cm} (b)}
    \caption{(a) Bias of the estimated phase of the precession signal. The amplitude for Xe signal at 35~Hz is 30~pT or 15~pT. (b) Bias of the estimated phase over one sub-run. The synthetic data has 2160 blocks of length $t_b=5$~s. The initial amplitude for Xe signal at 35~Hz is 30~pT and its decay constant $T_2^*$ is 8000~s. For both simulations, the amplitude of the vibration signal at 7.8~Hz is 1~pT. }
    \label{fig:syn_phase_bias}
\end{figure}

The phase bias caused by the vibrational noise, however, also depends on the orthogonality between the noise and the spin precession signal. For an ideal case where the vibration signal is orthogonal to the spin precession signal, it does not cause any error on the phase estimator even though it leads to a large residual. For our case, the precession frequencies of the $^{129}$Xe and $^{3}$He atoms were far above the frequency range of the mechanical vibration, resulting in approximate orthogonality between them at a block-time interval of 5~s (see the analysis in Fig.~\ref{fig:syn_phase_bias}).

To quantitatively analyse the effect of the vibrational noise on the estimated phase, we applied Monte-Carlo simulations. Synthetic spin precession signal was generated by a single sinusoidal function with a constant amplitude $A=30$~pT. Two kinds of noise, white noise and real noise, were added separately to the synthetic signal. The white noise with $\sigma =154$~fT (the standard deviation of the white noise in real gradiometer data) was generated in MATLAB, while the real noise was obtained by filtering out the two precession signals from measurement data with two 2-Hz bandwidth bandstop filters. The frequency of simulated precession signal was set to 40.2~Hz, which is 6~Hz higher than the $^{129}$Xe precession frequency in real data in order to avoid the artificial correlation between the real noise and the synthetic precession signal. The total time was 10000~s with 2000 blocks. Fig.~\ref{fig:syn_phase}(a) shows the power spectral density of the two sets of data. 

The error of the fitted phase for block $i$ is defined as $\epsilon_i =\phi_\mathrm{fit,i}-\phi_\mathrm{real,i}$. Here $\phi_\mathrm{real,i}$ was known and $\phi_\mathrm{fit,i}$ was obtained from the fit to block $i$. The histograms of $\epsilon_i$ for these two synthetic data sets are plotted in Fig.~\ref{fig:syn_phase}(b). The error for the white noise data is in good agreement with the normal distribution with $\sigma = 1.11\times 10^{-4} $ rad. \footnote[2]{In Sec.~\ref{sec:CRLB} it will be proven that this value is very close to the CRLB on phase uncertainty, which corroborates the high sensitivity of the used estimator.}  The error for the real noise data also satisfies the Gaussian distribution, with a similar result as the white noise data. This implies that the vibrational noise did not cause evident phase uncertainty, although the standard deviation of it is greater than the white noise. 

The above analysis shows that the phase bias caused by the vibrational noise is too small to be observed, compared to the random deviation caused by white noise. To further cross-check the systematic impact of the vibrational noise, we generated another set of synthetic data without adding any noise. The used signal model was $y=A_\text{spin}\sin(2\pi*35*t)+\sin(2\pi*7.8*t+\phi_0)$, including two sinusoidal signals, namely the precession signal at 35~Hz and the vibrational signal at 7.8~Hz with a constant amplitude of 1~pT. The data length was set to 5~s, the used block length. Fig.~\ref{fig:syn_phase_bias}(a) shows the bias in the estimated phase of the precession signal at 35~Hz using a single sinusoidal fit, as a function of the initial phase $\phi_0$ of the vibrational signal.
The phase bias is a sinusoidal function of the vibrational phase delay. The maximum phase error is inversely proportional to the precession amplitude. For $A_\text{spin}=30$~pT, the maximum phase error is 9~\textmu rad, being more than 10 times smaller than the uncertainty caused by the white noise, as shown in Fig.~\ref{fig:syn_phase}(b). We also found that the phase bias is almost irrelevant to the precession frequency as long as the frequency difference between the vibrational signal and the precession signal is larger than 0.2~Hz.  

For a whole sub-run, the phase bias varies in different blocks due to the changed phase delay between the vibrational signal and the precession signal. Fig.~\ref{fig:syn_phase_bias}(b) shows phase biases for a synthetic sub-run with 36 segments of 300~s length. The initial amplitude for Xe was 30~pT and its decay constant $T_2^*$ was set to 8000~s. The amplitude of the vibration signal was again 1~pT. The profile of the phase bias increases with time due to the signal decay. Although the magnitude of phase bias is greater than the phase amplitude caused by a potential EDM (see Fig.~\ref{fig:Voltage&Phase}(b)), the resultant EDM bias shall be small due to the low correlation between it and the electric field pattern. There is no solid physical reason to assume that the vibrational noise is correlated with the electric field direction. For the synthetic data shown in Fig.~\ref{fig:Voltage&Phase}(b), the EDM bias caused by the vibrational noise is only 4.9 $\times 10^{-31}~e~\mathrm{cm}$, being roughly three orders of magnitude smaller than our overall EDM statistical uncertainty. Important to note that in real data, the phase of vibrational noise at different blocks cannot be fully correlated as the correlation time for the vibrational signal is just in the minute regime and not a single frequency is generated by the vibrational disturbance. Thus, their impact on the derived EDM value has to be even smaller. In Sec.~\ref{sec:Real_noise}, the EDM bias caused by a large amount of real noise data will be further analyzed. 

In conclusion, the existence of the vibrational noise resulted in a large overestimation of the phase uncertainty. However, the residual error of each data point is Gaussian distributed. Owing to the almost-perfect orthogonality between the vibrational noise and the spin precession signal, the phase bias caused by the vibrational noise is more than 10 times smaller than the phase error caused by the white noise in our data, as validated by the theoretical analysis as well as Monte-Carlo simulations. Due to the negligible correlation between the phase bias caused by the vibrational noise and the potential EDM-induced phase, the bias on the derived EDM caused by the vibrational noise is proven to be ignorable at the current statistical $^{129}$Xe EDM sensitivity.

\subsection{Precession decay}

The amplitudes of the measured precession signals are expected to decay in time exponentially withhttps://www.overleaf.com/project/60622eaf971b8fd9ebf3369f a time constant $T_2^*$. The precession amplitudes for $^{129}$Xe and $^{3}$He at each block can be derived from the Variable Projection(VP) fit as 
\begin{equation} 
    \label{eqn:Amplitude}
    A_{\text{Xe/He}}=\sqrt{a_{\text{Xe/He}}^2+ b_{\text{Xe/He}}^2}.
\end{equation}

\begin{figure}[ht]
    \centerline{\includegraphics[width=.48\columnwidth]{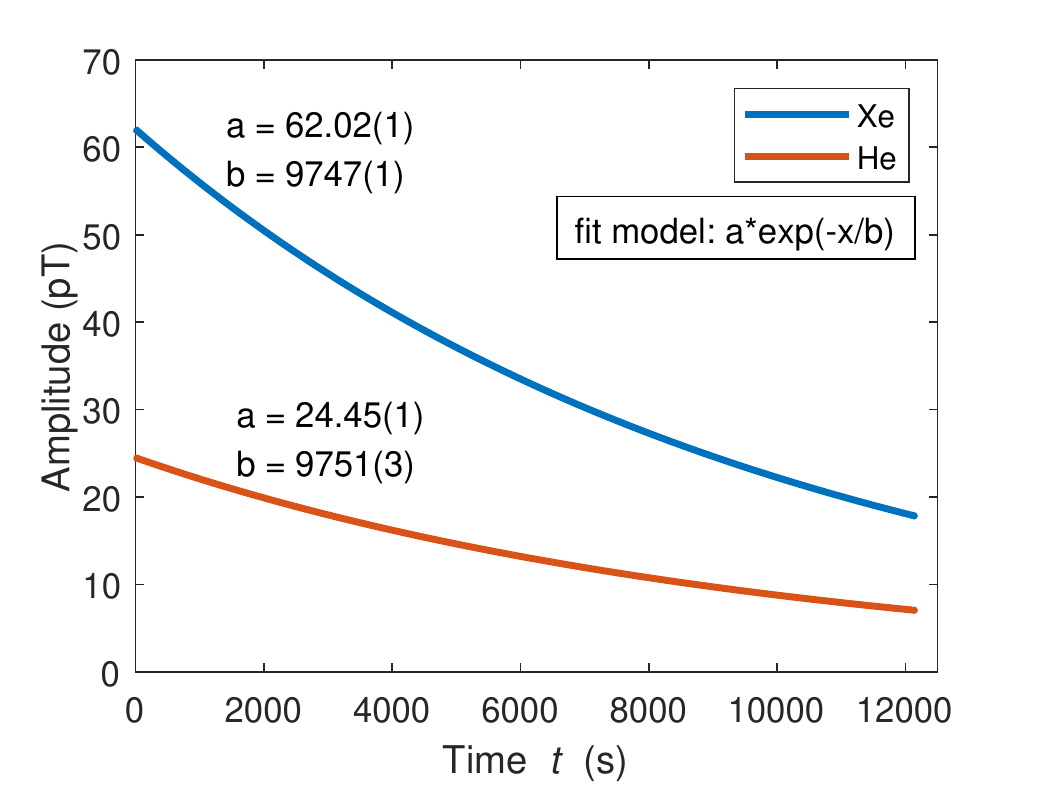}  \hspace{1cm} \includegraphics[width=.5\columnwidth]{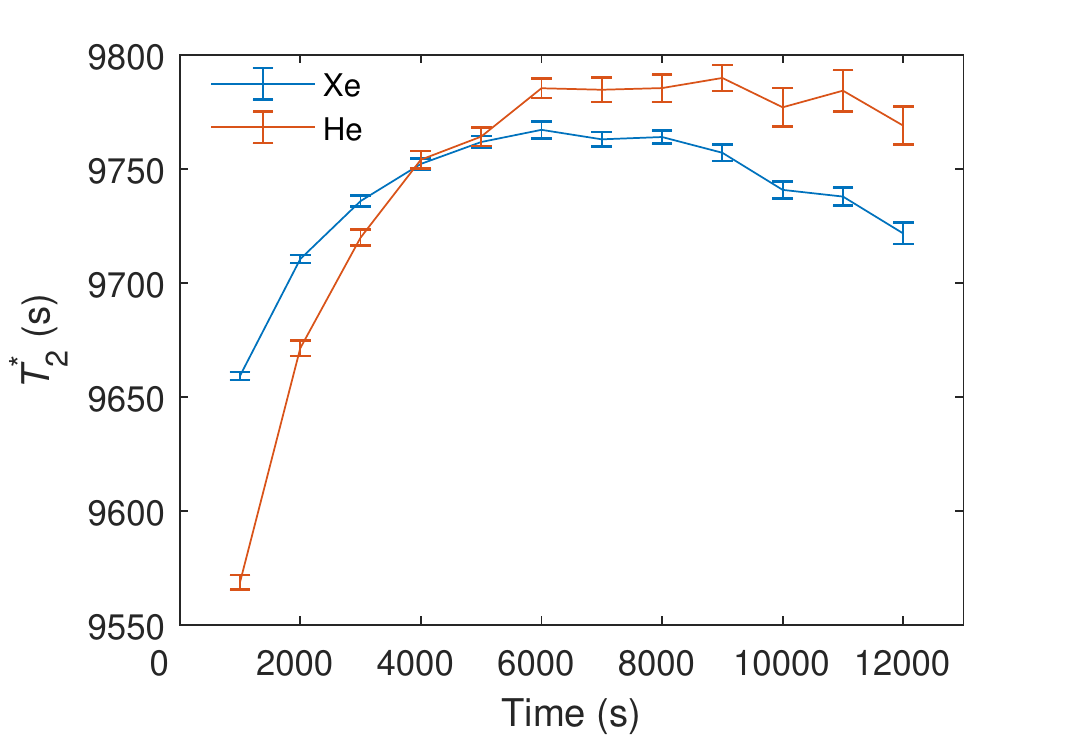}}
    \leftline{ \hspace{3cm} (a)  \hspace{8cm} (b)}
    \caption{(a) Block precession amplitudes of $^{129}$Xe and $^{3}$He in the sub-run B881. The parameters close to the lines are the result of fits with function $y=a e^{-t/b}$. (b) Variance of $T_{2}^*$ with time for 1000~s time windows .}
    \label{fig:Amplitude}
\end{figure}

Fig.~\ref{fig:Amplitude}(a) shows the exponential decay of the precession amplitudes. The starting amplitudes are $A_\text{Xe,0}=62.0$~pT and $A_\text{He,0}=24.5$~pT. By fitting an exponential function $y=A_0 e^{-t/T_2^*}$ to the data of whole sub-run, the transverse relaxation time is estimated to be $T_{2,\text{Xe}}^*=9747$ and $T_{2,\text{He}}^*=9751$. However, $T_{2}^*$, if fitted with a shorter time span, varies with time, as shown in Fig.~\ref{fig:Amplitude}(b). This is expected to be caused by the interaction between spins inside the cell \cite{Limes2019}. The signal decay happening in one block contributes to the residual error in the VP fit, since it violates the constant amplitude assumption. For a sinusoidal signal with amplitude $A = 100$~pT and $ T_2^* = 8000$~s, the residual error caused by the signal decay for a block with $t_\text{b}$=20~s is equal to that generated by the white noise with a deviation of $\sigma=47$~fT. This is roughly one third of the recorded white noise. Note that this residual caused by signal decay does not affect the estimated phase, but leads to an overestimate of the phase uncertainty, similar to the vibrational noise.      

\subsection{Phase unwrapping}
The phases $\phi_{\text{Xe/He}}$ derived in Eq.~(\ref{eqn:phase}) are wrapped in the range $[0,2\pi)$. The accumulated phase $\Phi$ in the block $k$ of the continuously precessing atoms is the sum of the wrapped phase and a multiple of $2\pi$
\begin{equation} 
    \label{eqn:Phase}
    \Phi_{\text{Xe/He}}^{k}=\phi_{\text{Xe/He}}^{k}+2 \pi n_{\text{Xe/He}}^{k},
\end{equation}
where the cycle of the first block is set as $n_{\text{Xe/He}}^{1}=0$. The subsequent cycle numbers were determined as
\begin{equation} 
    \label{eqn:integer_wrap}
    n_{\text{Xe/He}}^{k}=\text{Mod}(\Phi_{\text{Xe/He,f}}^{k-1}+\omega_{\text{Xe/He}}^{k-1}t_b,2\pi),
\end{equation}
where $t_b$ is the block length and the Mod() function returns the integer part of a number, $\omega_{\text{Xe/He}}^{k-1}$ is the averaged frequency from the middle of block $k-1$ to the middle of block $k$. Since the spin precession frequencies are so stable that we used the averaged frequency in the block $k-1$ as a substitute because it was already obtained together with the phase in the VP fit. If $\Phi_{\text{Xe/He}}^{k}-(\Phi_{\text{Xe/He}}^{k-1}+\omega_{\text{Xe/He}}^{k-1}t_b)$ is either $> \pi$ or $<-\pi$,  $n_{\text{Xe/He}}^{k}$ is  incremented or decremented by one, respectively, to ensure a continuous phase evaluation. Since the frequency of spins in one block is sufficiently stable to prevent error in determining the integer $n_{\text{Xe/He}}$, the uncertainty of the unwrapped phase $\delta\Phi_{\text{Xe/He}}$ is identical to $\delta\phi_{\text{Xe/He}}$. 

\begin{figure}[ht]
    \centerline{\includegraphics[width=.48\columnwidth]{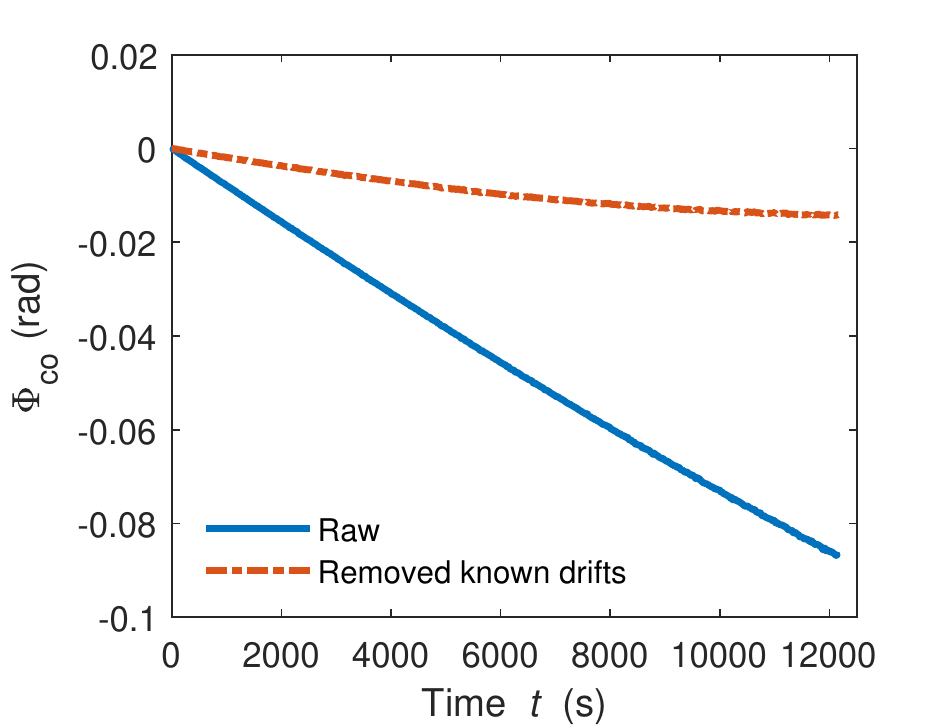} \hspace{1cm}\includegraphics[width=.48\columnwidth]{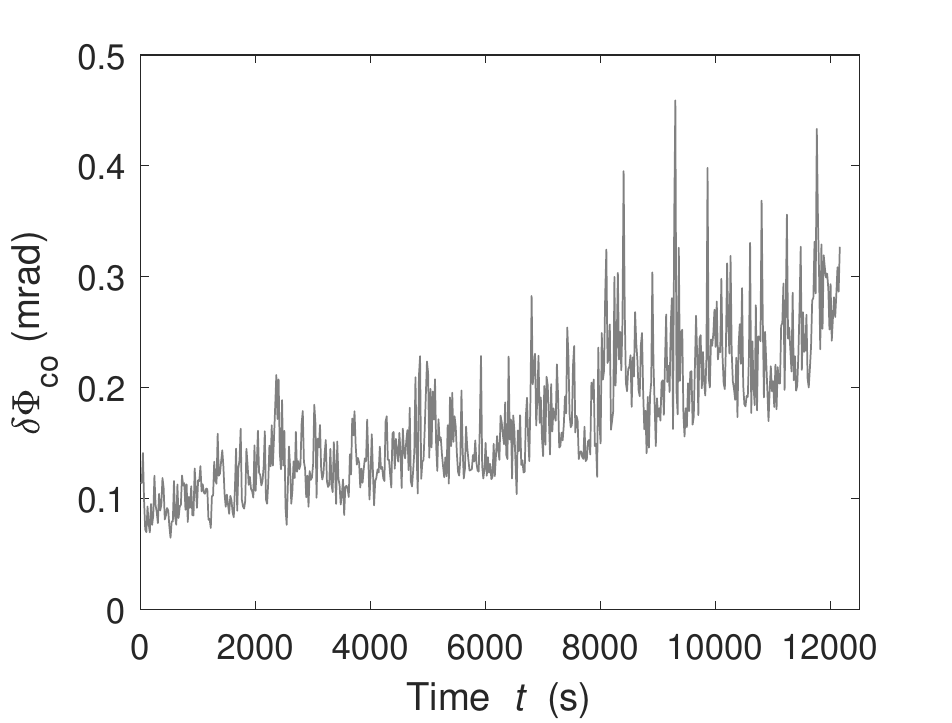}}
    \leftline{ \hspace{3.5cm}(a) \hspace{7cm} (b)}
    \caption{(a) The obtained comagnetometer phase $\Phi_{\text{co}}$ (blue line) and the anomalous phase drift (red dashed line), which was calculated by removing the known phase drifts from raw $\Phi_{\text{co}}$ with a deterministic model. Compared to the phase drift, the phase uncertainty  $\delta \Phi_{\text{co}}$ is much smaller, thus being separately plotted in (b).}
    \label{fig:Phi_co}
\end{figure}

The comagnetometer phase for each block $k$ is finally determined by 
\begin{equation} 
    \label{eqn:co_phase}
    \Phi_{\text{co}}^{k}=\Phi_{\text{Xe}}^{k}-\frac{\gamma_{\text{Xe}}}{\gamma_\text{He}}\Phi_{\text{He}}^{k}.
\end{equation}
The used gyromagnetic ratio between $^{3}$He and $^{129}$Xe atoms was $\gamma_{\text{He}} / \gamma_{\text{Xe}}=2.75408160$ \cite{Fan2016}. The uncertainty of the comagnetometer phase was propagated from $\delta\phi_{\text{Xe/He}}$. Fig.~\ref{fig:Phi_co}(a) shows the obtained comagnetometer phase $\Phi_{\text{co}}$ for sub-run B881.
The averaged frequency over this sub-run is calculated with the start phase and the end phase to be $\approx 12$~\textmu Hz, which is much smaller than the Lamor frequency  $\approx 35$~Hz, validating the usage of comagnetometer. The linear character of the $\Phi_{\text{co}}$ curve is the result of several known mechanisms, mainly the chemical shift and the frequency shift caused by the Earth's rotation. By removing these drifts with deterministic models from the raw $\Phi_{\text{co}}$ as done in Ref.~\cite{Allmendinger2019}, the remaining phase change is plotted as the dashed red line, which is the so-called "anomalous comagnetometer phase drift". The phase uncertainty $\delta \Phi_{\text{co}}$ as derived from the block fit is separately plotted in Fig.~\ref{fig:Phi_co}(b) because it is much smaller compared to the amplitude of phase drift. Even though this uncertainty is heavily dominated by the vibrational noise it exponentially increases with time $t$ as a result of decaying precession amplitudes.  

\section{EDM estimation}

\begin{figure}[ht]
    \centerline{\includegraphics[width=.48\columnwidth]{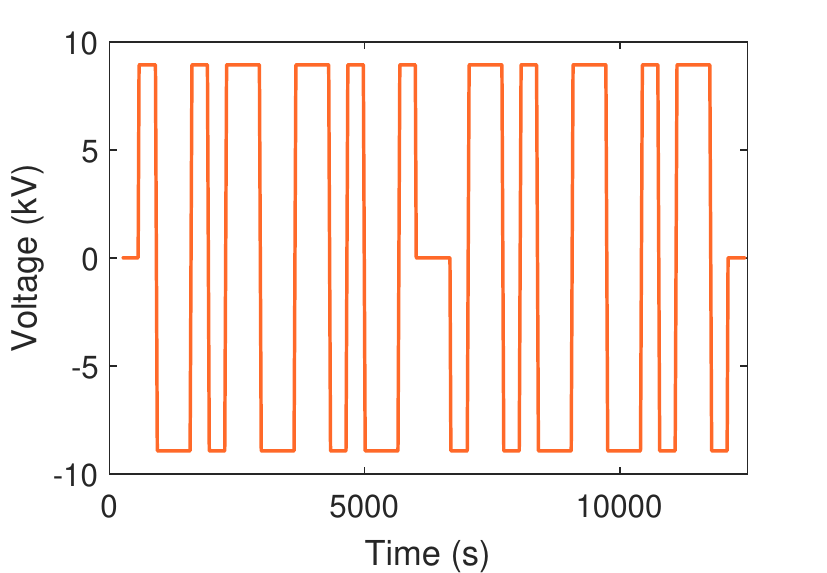} \hspace{1cm}\includegraphics[width=.49\columnwidth]{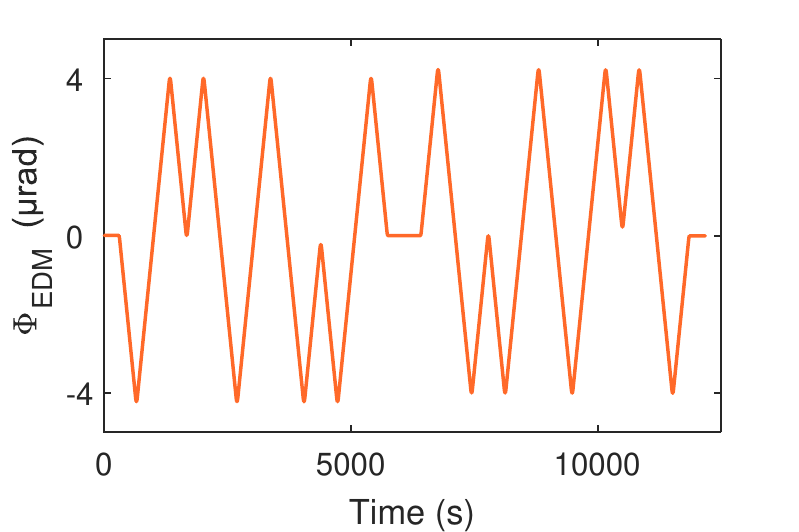}}
    \leftline{ \hspace{3cm}(a) \hspace{8cm} (b)}
    \caption{(a) Measured high voltage $U$ and (b) constructed EDM phase function $ \Phi_{\text{EDM}}$ of the sub-run B881. $\Phi_{\text{EDM}}$ is calculated for $d_\text{set}=1 \times 10^{-27} e$~cm.}
    \label{fig:Voltage&Phase}
\end{figure}

According to Eq.~(\ref{eqn:C0_omega2}), the accumulated phase for a hypothetical $^{129}$Xe EDM $d_{\text{set}}$ at the block $k$ is 
\begin{equation}
    \label{eqn:Co_phase3}
    \Phi_{\text{EDM}}^k= t_b \frac{2d_\text{set}}{\hslash}\sum_{i=1}^{k}({\textbf{E}_i} \cdot \hat{\textbf{B}}),
\end{equation}
where ${\textbf{E}_i}$ is the average electric field within the block $i$. Important to note that the extracted phase in Eq.~(\ref{eqn:Phase}) refers to the middle time of each block. So here the time interval of the block $i$ is shifted to $((i-1.5)t_b,(i-0.5)t_b)$. For an electric field pattern as shown in Fig.~\ref{fig:Voltage&Phase}(a) with an almost square-wave pattern, the accumulated phase is a quasi-triangular wave, as shown in Fig.~\ref{fig:Voltage&Phase}(b). To calculate $\Phi_{\text{EDM}}^k$ with the measured voltage from experiment, the offset of the input voltage signal has to be removed by subtracting the measured voltage before the high voltage was switched on.

By replacing $d_\text{set}$ with a computer-generated pseudorandom EDM value $d_\text{b}$, the bias phase $\Phi_{\text{b}}$ is calculated and then used to blind the comagnetometer phase with
\begin{equation}
    \label{eqn:blind}
    \Phi_{\text{co,b}}^k= \Phi_{\text{co}}^k+\Phi_{\text{b}}^k.
\end{equation}
The value of $d_{\text{b}}$ is saved in an independent file in a binary format and $\Phi_{\text{co,b}}^k$ is used for deriving the $^{129}$Xe EDM.  

The measured phase $\Phi_{\mathrm{co}}^k$ originated not only from the potential $^{129}$Xe EDM, but also from other sources such as chemical shift and the aforementioned anomalous comagnetometer frequency drift \cite{Limes2019,Sachdeva2019a}. These contributions can be phenomenologically parametrized by a polynomial of $g$th order. Hence, the comagnetometer phase is fitted with the function 
\begin{equation}
    \label{eqn:GPF}
     \Phi_{\mathrm{fit}}^k= a \Phi_{\mathrm{EDM}}^k + p_0 + p_1 \tilde{P}_1(t_k) + p_2\tilde{P}_2(t_k) + \cdots + p_g \tilde{P}_g(t_k) ,
\end{equation}
where $a, p_0, p_1, p_2, …, p_g$ are the global fit parameters. Here the time series $t_k$ are normalized to the interval [0,1] and shifted Legendre polynomials $\tilde{P}_n(t_k)$ are applied to decrease the correlation between polynomial coefficients \cite{Refaat2009}. The fit was conducted by using the iterative least squares estimation method with the built-in function $nlinfit$ in MATLAB. Thereby the inverse values of the phase variances $(\delta\Phi_{\mathrm{co}}^k)^2$ are used as weights, in order to account for the signal decay over a sub-run. Although the phase uncertainty $\delta\Phi_{\mathrm{co}}^k$ is overestimated due to vibrational noise, the relative ratio of it between blocks is still correct, as shown in Fig.~\ref{fig:Phi_co}(b). The covariance matrix of the fit is calculated with 
\begin{equation}
    \label{eqn:std_GPF}
    \text{Cov}_\text{p} =\frac{\mathbf{r}^{\intercal} \cdot \mathbf{r}}{\mathbf{J}^{\intercal} \cdot \mathbf{W}^{-2} \cdot \mathbf{J} \cdot \nu},
\end{equation}
where $r_k=(\Phi_{\text{co,b}}^{k}-\Phi_{\text{fit}}^k)/\delta\Phi_{\text{co,b}}^{k} $ is the weighted residual of the polynomial fit, $\mathbf{J}$ is the Jacobian matrix, $\mathbf{W}$ is the diagonal phase uncertainty matrix, and $ \nu$ is degrees of freedom. The obtained $^{129}$Xe EDM value is 
\begin{equation}
    \label{eqn:d_Xe}
        d_\mathrm{A} (^{129}\mathrm{Xe})=a \cdot d_\mathrm{set}, 
\end{equation}
and its uncertainty is calculated as $\delta d_\mathrm{A} (^{129}\mathrm{Xe}) =  \text{Cov}_\text{p}(1,1)^{0.5} \cdot d_\text{set}$.  

Once the systematic error is calculated, data unblinding can be done by repeating the same analysis but with the raw phase data $\Phi_{\text{co}}$.   

\subsection{Polynomial order}
\begin{figure}[ht]
    \centerline{\includegraphics[width=.7\columnwidth]{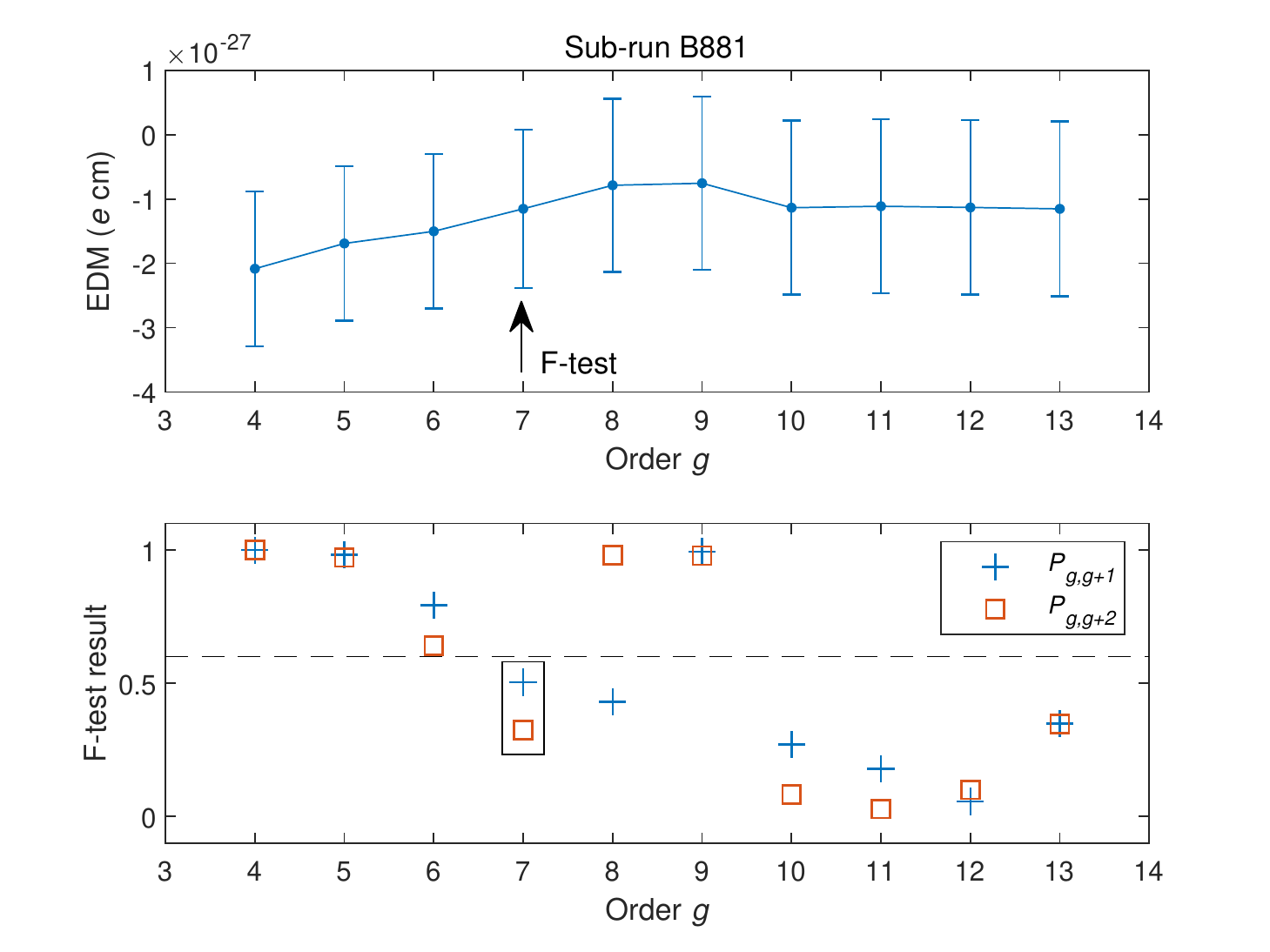} }
    \caption{ The fitted EDM result in dependence of the polynomial order $g$ in the fit model to the blinded data set of the sub-run B881. The arrow in the upper plot indicates the order determined by $F$-test. The bottom plot shows the $F$-test probability $P_{g,g+1}$ and $P_{g,g+2}$ with the dashed line implying the chosen threshold $P_{\text{min}}=0.6$. }
    \label{fig:F_test}
\end{figure}

One open question is the choice of the polynomial order in Eq.~(\ref{eqn:GPF}) needed to fully describe the comagnetometer phase drift. To determine it, we applied an $F$-test where the significance of adding $q$ term(s) to the polynomial function with $g$th order was evaluated by the integral probability
\begin{equation}
    \label{eqn:P}
         P_{g,g+q}=\int_{0}^{F_{g,g+q}}P_\mathrm{F}(F;q,N-g-q)dF,
\end{equation}
where $P_\mathrm{F}$ is the probability density function of the $F$-distribution and $N$ is the number of data points \cite{Bevington1992}. The upper bound of the integral is
\begin{equation}
    \label{eqn:F}
        F_{g,g+q}=\frac{(N-g-q)(\chi_g^2-\chi_{g+q}^2)}{q \cdot \chi_{g+q}^2}.
\end{equation}
The order of the fit was defined sufficient when $P_{g,g+1}$  and $P_{g,g+2}$ are both smaller than a chosen threshold of $P_{\mathrm{min}}$. Fig.~\ref{fig:F_test} shows the fitted EDM results and $F$-test probabilities as a function of the used order for the sub-run B881. The 7th order was the smallest order which satisfied $P_{g,g+1}<0.6$ and $P_{g,g+2}<0.6$.

\subsection{Correlation study}
Note that the uncertainty as deduced from the fit algorithm is "correlated uncertainty" as it accounts for the impact of the correlation between EDM parameter $a$ and other polynomial parameters. The correlation between fit parameters depends on the model function. The constructed phase function of $M$ nonzero high voltage segments as shown in Fig.~\ref{fig:Voltage&Phase}(b) was designed to be orthogonal to the polynomial function of the order up to $\log_2 M-2$, that is 
\begin{equation}
    \label{eqn:Orth}
             \sum\limits_{k=1}^{JM} \Phi_{\text{EDM}}^k t_k^n = 0, \hspace{1cm} n \leq \log_2 M-2,
\end{equation}
where $J$ is the number of blocks in one segment. Due to this orthogonality, the correlation between the EDM parameter $a$ and the polynomial parameters are generally small. The correlation matrix for the sub-run B881 (see Fig. 1) is given in Table~\ref{tab:corr_GPF_fit}. The correlations between the EDM parameter $a$ and the polynomial coefficients are significantly smaller than 1, but nonzero, since the polynomials higher than 3rd order are not orthogonal to this sub-run of 32 non-zero high voltage segments. Another reason for the nonzero correlation is the increasing phase uncertainty with time, making the weighted sum of the EDM function and the polynomial to be nonzero.

\begin{table}[h]
    \centering
    \caption{Correlation matrix of the first sub-run for the fit with a 7th order polynomial and the block length $t_b$=5~s.}
    \begin{tabular}{c c c c c c c c c c}
    \hline\noalign{\smallskip}
          & $a$ & $p_0$ & $p_1$  & $p_2$  & $p_3$ & $p_4$ & $p_5$ & $p_6$  & $p_7$ \\
    \noalign{\smallskip}\hline\noalign{\smallskip}
    $a$   & \textbf{1.0} & \textbf{0.0} & \textbf{0.0} & \textbf{0.0} & \textbf{0.0} & \textbf{0.2} & \textbf{0.0} & \textbf{0.2} & \textbf{0.2} \\
    $p_0$ & \textbf{0.0} & 1.0 & 0.5 & 0.2 & 0.1 & 0.0 & 0.1 & 0.1 & 0.0\\
    $p_1$ & \textbf{0.0} & 0.5 & 1.0 & 0.5 & 0.2 & 0.1 & 0.1 & 0.1 & 0.0 \\
    $p_2$ & \textbf{0.0} & 0.2 & 0.5 & 1.0 & 0.5 & 0.2 & 0.1 & 0.0 & 0.0 \\
    $p_3$ & \textbf{0.0} & 0.1 & 0.2 & 0.5 & 1.0 & 0.5 & 0.2  & 0.1 & 0.0 \\
    $p_4$ & \textbf{0.2} & 0.0 & 0.1 & 0.2 & 0.5 & 1.0 & 0.5  & 0.2 & 0.1 \\
    $p_5$ & \textbf{0.0} & 0.1 & 0.1 & 0.1 & 0.2 & 0.5 & 1.0 & 0.5 & 0.1 \\
    $p_6$ & \textbf{0.2} & 0.1 & 0.1 & 0.0 & 0.1 & 0.2  & 0.5 & 1.0 & 0.5 \\
    $p_7$ & \textbf{0.2} & 0.0 & 0.0 & 0.0 & 0.0 & 0.1 & 0.1 & 0.5 & 1.0 \\
    \noalign{\smallskip}\hline
    \end{tabular}
    \label{tab:corr_GPF_fit}
\end{table}

\subsection{Fit quality}
\label{sec:Fit quality}
\begin{figure}[ht]
    \includegraphics[width=\columnwidth]{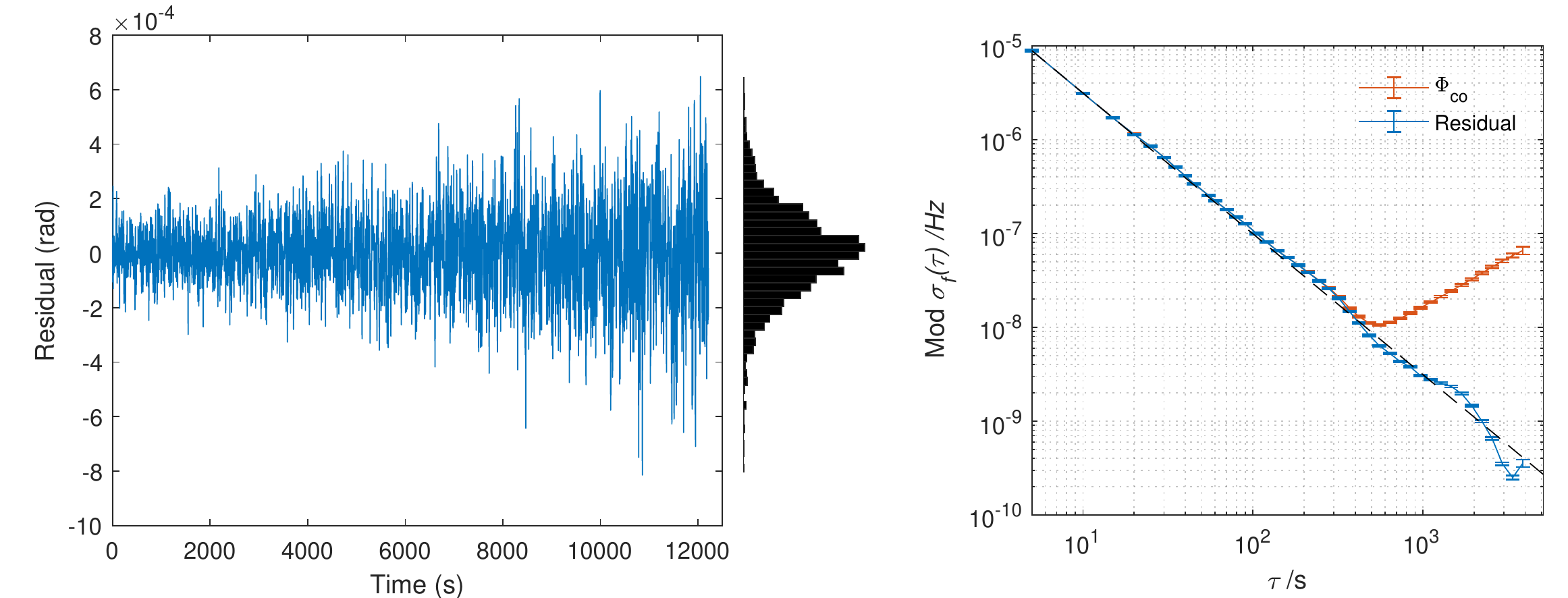} 
    \leftline{ \hspace{3cm}(a) \hspace{8cm} (b)}
    \caption{The result of the sub-run B881. (a) Residual phase for the fit with a 7th order polynomial and the histogram of it. (b) The modified Allan deviation and its error bar of the accumulated comagnetometer phase and the residual phases. To fulfill the MAD statistics criteria \cite{Allan1981}, only data are shown for integration time $\tau < 4000$~s.}
    \label{fig:Residual_phase}
\end{figure}

The modified Allan deviation (MAD) is an established tool to evaluate the low-frequency drift of a time series of phases $\Phi$, which is defined as  
\begin{equation}
    \label{eqn:Allan}
        \text{MAD} \sigma_f(\tau)=\frac{1}{2\pi}\sqrt{\frac{\sum\limits^{P-3n-1}_{j=1}(\sum\limits_{k=j}^{j+n-1}\Phi^{k+2n}-2\Phi^{k+n}+\Phi^{n})^2}{2n^2\tau^2(P-3n+1)}},
\end{equation}
where the integration time $\tau$ is the product of the count $n$ and the block length $t_b$, and the total measurement time $T$ is subdivided into $P$ time intervals of equal length $\tau$, such that $P\tau \approx T$. The residual phase of the sub-run B881 is shown in Fig.~\ref{fig:Residual_phase}(a) with the adjacent histogram showing a Gaussian-like distribution. The residual generally increases with time due to the signal decay. The MAD of the exemplary sub-run is plotted in Fig.~\ref{fig:Residual_phase}(b). $\sigma_f$ of $\Phi_{\mathrm{co}}^k$ reaches the minimum at the integration time $\tau$ of 550~s and then increases due to the comagnetometer frequency drift. For the residual phase $\Phi_{\mathrm{co}}^k-\Phi_{\mathrm{fit}}^k$ of this sub-run, the MAD decreases with increasing integration time according to $\sigma_f \propto \tau^{-3/2}$ (dashed line in Fig.~\ref{fig:Residual_phase}) over the considered range, down to 0.4~nHz. This behavior is an indicator that the comagnetometer phase $\Phi_{\mathrm{co}}^k$  is adequately described by the fit model of Eq.~(\ref{eqn:GPF}) with the chosen number of polynomials by $F$-test criteria, given that the residual is dominated by white phase noise. For all the other sub-runs, the modified Allan deviation showed a similar behavior as this presented sub-run. 

To cross-check the derived uncertainty for $d_\text{Xe}$ using the covariance, we applied the log profile likelihood method, which is an established tool to assess parameter identifiability in nonlinear regression \cite{Venzon1988}. It generalises Fisher information-based confidence intervals to the non-linear setting, resulting in appropriate confidence regions. In short, the parameter of interest $d_\text{Xe}$ is profiled by scanning along its axis and re-optimising all other parameters $p _{j =0,1,...,g}$ in Eq.~(\ref{eqn:GPF}) for each value of $d_\text{Xe}$. Thus, the profile likelihood is defined as $\text{PL}\left( d_\text{Xe} \right) = \mathop {\max }\limits_{{p_{j =0,1,...,g}}} L\left( {p_j; d_\text{Xe}} \right)$, with $L$ being the likelihood function. Let $\chi _{\alpha ,1}^2$ denote the $\alpha$ quantile of the $\chi^2$ distribution with one degree of freedom, the region for which the inequality 
\begin{equation}
    \label{PLD}
        L \left( {\hat \theta } \right) - {\rm{PL}}\left( d_\text{Xe}  \right) \le \chi _{\alpha ,1}^2
\end{equation}
is satisfied yields the confidence interval of the parameter $d_\text{Xe}$ to a given confidence level $\alpha$. $\hat \theta$ is the best estimate of all parameters. For a confidence level of $\alpha=95.5$\%, $\chi _{\alpha ,1}^2 \approx 2$. 

\begin{figure}[h]
    \centerline{\includegraphics[width=0.6\columnwidth]{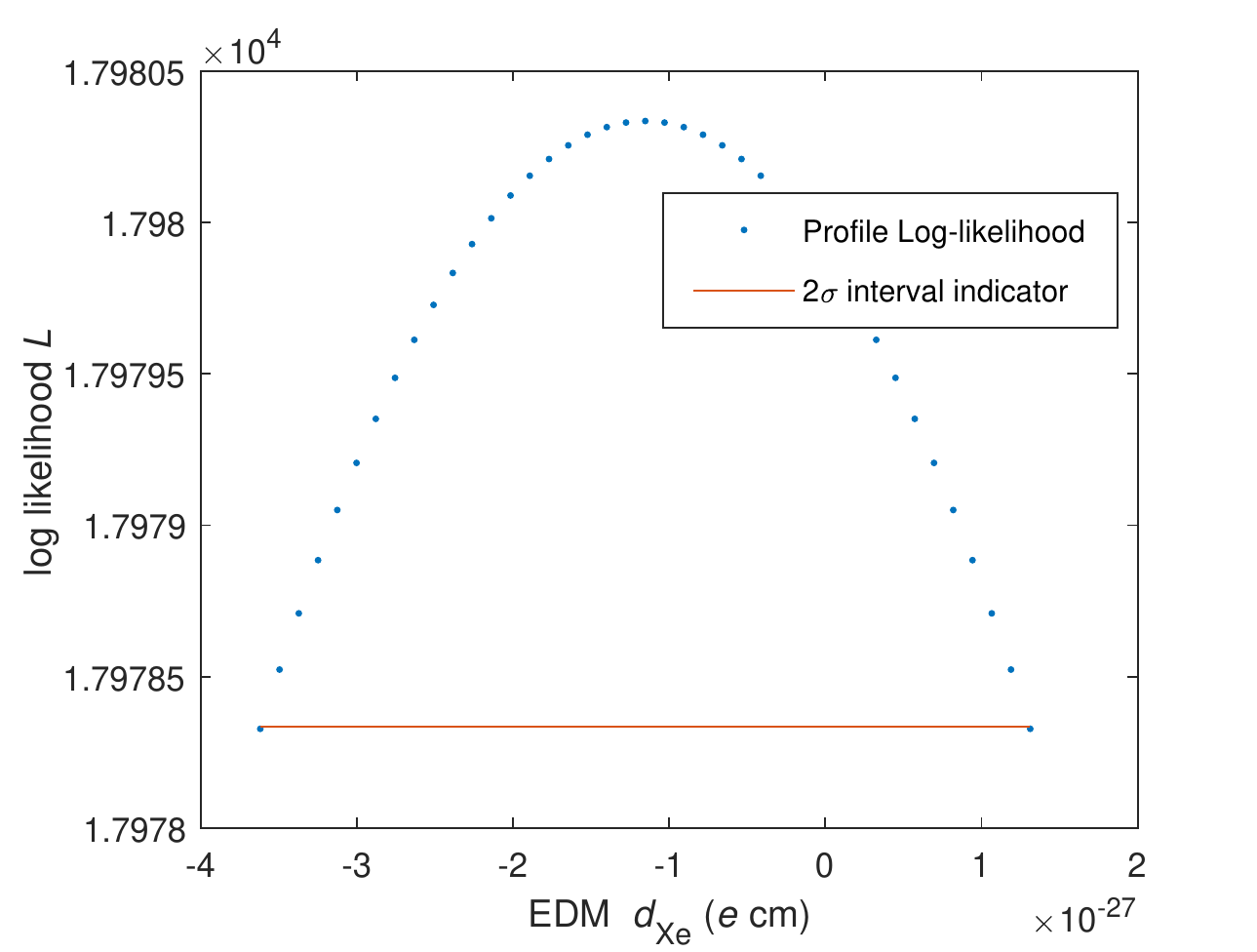}}
    \caption{The log profile maximum likelihood for the exemplary sub-run B881. The varied range for the EDM $d_\text{Xe}$ is the $2\sigma$ confidence interval derived with the covariance. The intersection points of the blue curve and the red line represent the $2\sigma$ confidence interval according to the log profile maximum likelihood method.}
    \label{fig:Log_Profile}
\end{figure}

The blue dots in Fig.~\ref{fig:Log_Profile} show the calculated log profile maximum likelihood values for the sub-run B881 over the $2\sigma$ confidence interval of $d_\text{Xe}$ obtained with the GPF method, that is $[-1.15-2 \times 1.23, -1.15+2 \times 1.23]\times 10^{-27}~e~\mathrm{cm}$. The region between the two inter intersection points of the red line ($y = L\left( {\hat \theta } \right) - 2$) and the profile log-likelihood curve is the $2\sigma$ confidence interval of $d_\text{Xe}$, which is [-3.618, 1.312]$\times 10^{-27}~e~\mathrm{cm}$ and in perfect agreement with the covariance-based GPF result. The profile likelihood method was applied to all sub-runs and achieved almost identical results with the covariance-based GPF method. 

\section{Theoretical sensitivity}
\label{sec:CRLB}
The theoretical limit of the $^{129}$Xe EDM uncertainty $\delta d_\mathrm{A} (^{129}\mathrm{Xe})$ can be derived as the Cramer-Rao Lower Bound (CRLB), which also provides insights into optimizing experimental parameters. For the sake of simplicity, only the spin-precession signal of Xe is considered. For the GPF method, $d_\mathrm{A} (^{129}\mathrm{Xe})$ is estimated in two steps: VP fits to SQUID data to obtain the phases of each block, and a single fit to the cumulative phase to estimate the EDM value. The CRLB is also calculated with these two steps.

\subsection{Phase uncertainty}
The real-valued data are assumed to be 
\begin{equation}
    \label{eqn:y_phase}
        y[n]=A\cos{(2\pi f_0 n \Delta t + \phi_0)} + w [n], \hspace{1cm} n=-\tfrac{N-1}{2}, -\tfrac{N-1}{2}+1, ..., \tfrac{N-1}{2},
\end{equation}
where $w [n]$ is white Gaussian noise (WGN) with the variance $\sigma_{\text{w}}^2$ , $N$ is the length of data set and assumed to be an odd number, the time interval between two sampling point is $\Delta t =1/f_{\text{s}}$, and $\theta=[A,f_0,\phi_0]$ are fit parameters. If the spectral density of the white noise $\rho_{\text{w}}$ is given, the variance is $\sigma_{\text{w}}^2=f_{\text{s}} \rho_{\text{w}} / 2$. Assuming that the data length $N$ is large enough, the Fisher information matrix becomes 
\begin{equation}
    \label{eqn:Fisher_n}
        I[\theta]=\frac{1}{\sigma_{\text{w}}^2} \begin{bmatrix} 
            \frac{N}{2} & 0 & 0 & \\
            0 & 2A^2\pi^2\mathlarger{‎‎\sum}\limits_{n=-(N-1)/2}^{(N-1)/2}n^2 &  \pi A^2\mathlarger{‎‎\sum}\limits_{n=-(N-1)/2}^{(N-1)/2}n\\ 
            0 & \pi A^2\mathlarger{‎‎\sum}\limits_{n=-(N-1)/2}^{(N-1)/2}n & \frac{NA^2}{2} \\ 
        \end{bmatrix} = \frac{1}{\sigma_{\text{w}}^2} \begin{bmatrix} 
            \frac{N}{2} & 0 & 0 & \\
            0 & 2A^2\pi^2 \frac{N^3}{12} & 0\\ 
            0 & 0 & \frac{NA^2}{2} \\ 
        \end{bmatrix}  .
\end{equation}
We have 
\begin{equation}
    \label{eqn:CRLB_phase}
        \delta \phi \geq \sqrt{\frac{2\sigma_{\text{w}}^2}{A^2 N}} ,
\end{equation}
which is a factor of 2 smaller than the result in Ref.~\cite{KAY1993} for the count $n$ starting from 0 to $N-1$. This is because here we normalize the input $x$ data and derive the phase in the center of the block ($n=0$), instead of the starting phase($n=-(N-1)/2$).

\subsection{EDM uncertainty}
The CRLB for the parameters in the fit model Eq.~(\ref{eqn:CRLB_EDM_a}) is the reciprocal of the Fisher information matrix 
\begin{equation}
    \label{eqn:Fisher_GPF}
        \mathbf{I} =      
        \begin{bmatrix} 
            \sum\limits_{k=1}^{JM}\frac{ (\Phi_{\text{EDM}}^k)^2}{\delta \phi_k^2} & \cdots & \cdots & \sum\limits_{k=1}^{JM}\frac{ \Phi_{\text{EDM}}^k t_k^g}{\delta \phi_k^2} \\
            \sum\limits_{k=1}^{JM}\frac{\Phi_{\text{EDM}}^k}{\delta \phi_k^2} & \sum\limits_{k=1}^{JM}\frac{1}{\delta \phi_k^2} & \cdots & \vdots \\ 
            \vdots & \vdots & \ddots & \vdots \\ 
            \sum\limits_{k=1}^{JM}\frac{\Phi_{\text{EDM}}^k t_k^g}{\delta \phi_k^2} & \sum\limits_{k=1}^{JM}\frac{t_k^g}{\delta \phi_k^2} & \cdots &\sum\limits_{k=1}^{JM}\frac{t_k^g t_k^g}{\delta \phi_k^2} \\ 
        \end{bmatrix} .
\end{equation}

Here $M$ is the number of segments in one sub-run, $J$ is the number of blocks in one segment and the product $JM$ is the total number of blocks in this sub-run. For the sake of simplicity, the standard polynomial is used instead of the shifted Legendre polynomial. The variance of the fit parameters $Q=[a,p_0,p_1,...,p_n]$ in Eq.~(\ref{eqn:GPF}) of the GPF is 
\begin{equation}
    \label{eqn:CRLB_EDM_a}
        (\delta Q)^2 \geq \text{diag}(\mathbf{I}^{-1}).
\end{equation}

In order to simplify Eq.~(\ref{eqn:CRLB_EDM_a}), we further assume the phase uncertainty $\delta\phi$ is a constant , meaning that the precession amplitude is a constant over the whole sub-run, and $\sum\limits_{k=1}^{JM} \Phi_{\text{EDM}}^k t_k^i = 0 $ for $i$ going from 0 to $g$, which can be satisfied if the pattern has $2^{g+2}$ nonzero high voltage segments. Under these two assumptions, the Fisher matrix becomes 
\begin{equation}
    \label{eqn:Fisher_GPF_diagonal}
        \mathbf{I} =      
        \begin{bmatrix} 
            \sum\limits_{k=1}^{JM}\frac{ (\Phi_{\text{EDM}}^k)^2}{\delta \phi_k^2} & 0 & \cdots & 0 \\
            0 & \sum\limits_{k=1}^{JM}\frac{1}{\delta \phi_k^2} & 0 & 0\\ 
            \vdots & \vdots & \ddots & \vdots \\ 
            0 & 0 & \cdots &\sum\limits_{k=1}^{JM}\frac{t_k^g t_k^g}{\delta \phi_k^2} \\ 
        \end{bmatrix}.
\end{equation}
The considered CRLB now is simplified to the so called ideal or uncorrelated CRLB, leading the CRLB on the EDM parameter as 
\begin{equation}
    \label{eqn:CRLB_EDM}
        (\delta d_\text{Xe})^2 \geq  \frac{1}{\sum\limits_{k=1}^{JM}\frac{ (\Phi_{\text{EDM}}^k)^2}{\delta \phi_k^2}}.
\end{equation}
The square of the constructed EDM phase function $\Phi_{\text{EDM}}^2$ is a periodic function with a cycle of $2M$ for the applied high voltage pattern shown in Fig.~\ref{fig:Voltage&Phase}(b). The sum of the EDM phase function over one sub-run can be simplified to the product of the number of segments with the sum over two segments, i.e., $ \sum\limits_{k=1}^{JM} (\Phi_{\text{EDM}}^k)^2 =\frac{M}{2} \sum\limits_{k=1}^{2J} (\Phi_{\text{EDM}}^k)^2$. Due to the symmetry $ \sum\limits_{k=1}^{J} (\Phi_{\text{EDM}}^k)^2 = \sum\limits_{k=J+1}^{2J} (\Phi_{\text{EDM}}^k)^2$, it can be further simplified to  $\sum\limits_{k=1}^{JM} (\Phi_{\text{EDM}}^k)^2 = M (\frac{2|E|t_b}{\hslash})^2 \sum\limits_{k=1}^{J} k^2$. Substituting it into Eq.~(\ref{eqn:CRLB_EDM}) gives  
\begin{equation}
    \label{eqn:CRLB_EDM2}
    (\delta d_\text{Xe})^2 \geq  \frac{\delta \phi^2}{ M (\frac{2|E|t_b}{\hslash})^2 \sum\limits_{k=1}^{J} k^2}
    = \delta \phi^2 (\frac{\hslash}{2|E|t_b})^2 \frac{6}{MJ(J+1)(2J+1)}.
\end{equation}
Further expanding $\delta \phi$ in Eq.~(\ref{eqn:CRLB_EDM2}) with Eq.~(\ref{eqn:CRLB_phase}) and assuming $J \gg 1$, the overall CRLB for $d_\text{Xe}$ becomes 
\begin{equation}
    \label{eqn:CRLB_EDM_simp}
    \delta d_\text{Xe,GPF} \geq \frac{\sigma_{\text{w}}}{A}\frac{\hslash}{2|E|}\sqrt{\frac{6M^2 \delta t}{T^3}},
\end{equation}
where $T$ is the total measurement time and $\delta t =1 / f_{\text{s}}$ is the sampling interval. Eq.~(\ref{eqn:CRLB_EDM_simp}) implies that a smaller number of segments and longer measurement time lead to a smaller uncertainty.   

For the PC method, the CRLB on the $^{129}$Xe EDM for $ M$ segments is derived in Ref.~\cite{Sachdeva2019a} as 
\begin{equation}
    \label{eqn:CRLB_PC}
    \delta d_\text{Xe,PC}  \geq \frac{\sigma_{\text{w}}}{A}\frac{\hslash}{2|E|}\sqrt{\frac{24 M^2 \delta t}{T^3}},
\end{equation}

The PC method applies linear fits to the comagnetometer phases within one segment to derive the comagnetometer frequency of each segment, which requires the addition of an interception term as a starting phase, increasing the variance by a factor of four compared to a linear fit without interception term. In the GPF method the accumulated comagnetometer phases within one sub-run are analyzed in a single fit, therefore the uncertainty does not increase as the interception term is orthogonal to the EDM function (see Eq.~(\ref{eqn:Fisher_GPF})). 

One has to emphasize that Eq.~(\ref{eqn:CRLB_EDM_simp}) derived here is for ideal data with white noise and a constant precession amplitude. For the real data set, the final expression of the CRLB is more complicated than in Eq.~(\ref{eqn:CRLB_EDM_simp}) and we have to use the more general Fisher information matrix in Eq.~(\ref{eqn:Fisher_GPF}).   

Eq.~(\ref{eqn:CRLB_EDM_simp}) could be used to estimate the uncertainty of a single run in the 2018 campaign with the achieved experimental parameters listed in Table~\ref{tab:exp_para}. For example, for a sub-run with the following parameters: the standard deviation of the SQUID signal is $\sigma_{\text{w}}=154$~fT; the starting amplitude of $^{129}$Xe and $^{3}$He are 70~pT and 25~pT, the $T_2^*$ time of both atoms are 8000~s;  the  electric field  pattern includes 36 segments of length $t_{\text{b}}=400$~s with a maximum value of $E=4$~kV/cm, the derived CRLB on the EDM uncertainty with Eq.~(\ref{eqn:Fisher_GPF}) is $\delta d_{\text{Xe}}=8.2 \times 10 ^{-28} e$~cm. This value is already close to the uncertainty of the current Xe EDM limit derived from the data of 80 sub-runs \cite{Sachdeva2019}. For $T_{2,\text{Xe/He}}^* \rightarrow \infty $, the CRLB given by Eq.~(\ref{eqn:CRLB_EDM_simp}) reduces to $\delta d_{\text{Xe}}=4.2 \times 10 ^{-28} e$~cm, again showing the importance of a homogeneous magnetic field.  

\section{Validation}
To verify the GPF method, Monte-Carlo simulations were applied. We first give details on the generation model of synthetic data and then show the performance of the GPF method under real vibrational noise as well as synthetic white noise. Special attention was paid to the impact of anomalous phase drift on the EDM result.

\subsection{Synthetic data generation}
The parameters used to generate synthetic data are typical for the measurements of the last week of the 2018 campaign, which was chosen because of its highest statistical significance. In the simulation, some parameters were fixed, including the sampling rate $f_{\text{s}} = 915.5245$~Hz, the background magnetic field $B_0=3$ \textmu T, the maximum electric field $E_0=4$~kV/cm, the starting amplitude $A_{\text{Xe}}^0=70$~pT, $A_{\text{He}}^0=25$~pT, and the transverse relaxation time  $T_{2,\text{Xe}}^*=T_{2,\text{He}}^*=8000$~s. 

The frequency of each spin species at the sampling point $j$ was generated as 
\begin{equation}
    \label{eqn:f_syn}
     f_{\text{Xe/He}}(j) = \gamma_{\text{Xe/He}} B_0 + f_{\mathrm{lin}}^{\mathrm{Xe/He}}+ f_{\mathrm{ano}}^{\mathrm{Xe/He}}+ f_{\mathrm{EDM}}^{\mathrm{Xe}}
\end{equation}
where $ f_{\mathrm{lin}}^{\mathrm{Xe/He}}$ is the linear drift caused by the chemical shift and Earth's rotation. $ f_{\mathrm{ano}}^{\mathrm{Xe/He}}$ denotes the anomalous frequency drift, which is modelled as an exponentially decaying function with the characteristic time of $T_1$ \cite{Limes2019,Terrano2019,Thrasher2019}, that is 
\begin{equation}
    \label{eqn:Ano_f}
    f_{\mathrm{lin}}^{\mathrm{Xe/He}} = u^\mathrm{Xe/He}e^{-t_j/T_1^\mathrm{Xe/He}},
\end{equation}
where $t_j$ is the time referring to the sampling point $j$. $u^{\text{Xe/He}}$ and $T_1^{\text{Xe/He}}$ are random parameters uniformly distributed in a reasonable range. Here, the word 'reasonable' means that the synthetic drift with Eq.~(\ref{eqn:Ano_f}) behaves in a similar way as the real measurement. $f_{\mathrm{EDM}}^{\mathrm{Xe}}$ is the frequency caused by a given EDM value $d_{\text{syn}}$ with the HV pattern according to Eq.~(\ref{eqn:C0_omega2}).  

The accumulated phase at each data point is calculated from the frequency with
\begin{equation}
    \label{eqn:Phi_syn}
    \Phi_{\text{Xe/He}}^j = \sum\limits_{i=0}^{j} 2 \pi f_{\text{Xe/He}}(i) t_{\text{s}} ,
\end{equation}
where $t_{\text{s}}=1/f_{\text{s}}$ is the sampling interval. The spin precession signal is 
\begin{equation}
    \label{eqn:SQUID_syn}
    V_{\mathrm{Xe/He}}^j = A_0^{\mathrm{Xe/He}} e^{-t_j/T_2^{\mathrm{Xe/He}}}\sin{\Phi_{\mathrm{Xe/He}}^j}.
\end{equation}

Special attention has to be paid to the rounding error of the synthetic phase. For a spin frequency of 100~Hz and a total measurement time of $T=20000$~s, the phase at the end will reach $1.2 \times 10^7$~rad. The maximum phase caused by the EDM, however, is on the \textmu rad level as shown in Fig.~\ref{fig:Voltage&Phase}(b). To avoid rounding error, the phase has to be wrapped before integration in Eq.~(\ref{eqn:Phi_syn})  as $\Phi_{\text{Xe/He,wrap}}^j=\sum\limits_{i=0}^{j} (2 \pi f_{\text{Xe/He}}(i)t_{\text{s}}-2 \pi \times \text{clc}(i))$. $\text{clc}(i)$ is the cycle number of the phase $\Phi_{\text{Xe/He}}(i)$.

The created synthetic SQUID signal is the sum of the $^{129}$Xe and $^{3}$He precession signals. The drift of the background field and the starting phase are ignored since they do not significantly affect the data analysis. The parameter ranges used in our simulations are listed in Table~{\ref{tab:Syn_para}}. 

\begin{table}[h]
    \centering
    \caption{ Ranges of the parameter values used for generating synthetic spin precession data for the Monte-Carlo simulation.}
    \begin{tabular}{c c | c c}
    \hline\noalign{\smallskip}
     Para. & Range & Para.  & Range \\
    \noalign{\smallskip}\hline\noalign{\smallskip}
    $u^{\mathrm{He}}$    &  3.5-4.5 \textmu Hz   &  $T_1^{\mathrm{He}}$  & 9000-14000 s   \\
    $u^{\mathrm{Xe}}$    &   9-11 \textmu Hz  &  $T_1^{\mathrm{Xe}}$  & 9000-14000 s   \\
    $f_{\mathrm{lin}}^{\mathrm{He}}$  &  4-10 \textmu Hz      & $f_{\mathrm{lin}}^{\mathrm{Xe}}$ & 4-10 \textmu Hz     \\
    \noalign{\smallskip}\hline
    \end{tabular}
    \label{tab:Syn_para}
\end{table}

\begin{figure}[ht]
    \includegraphics[width=\columnwidth]{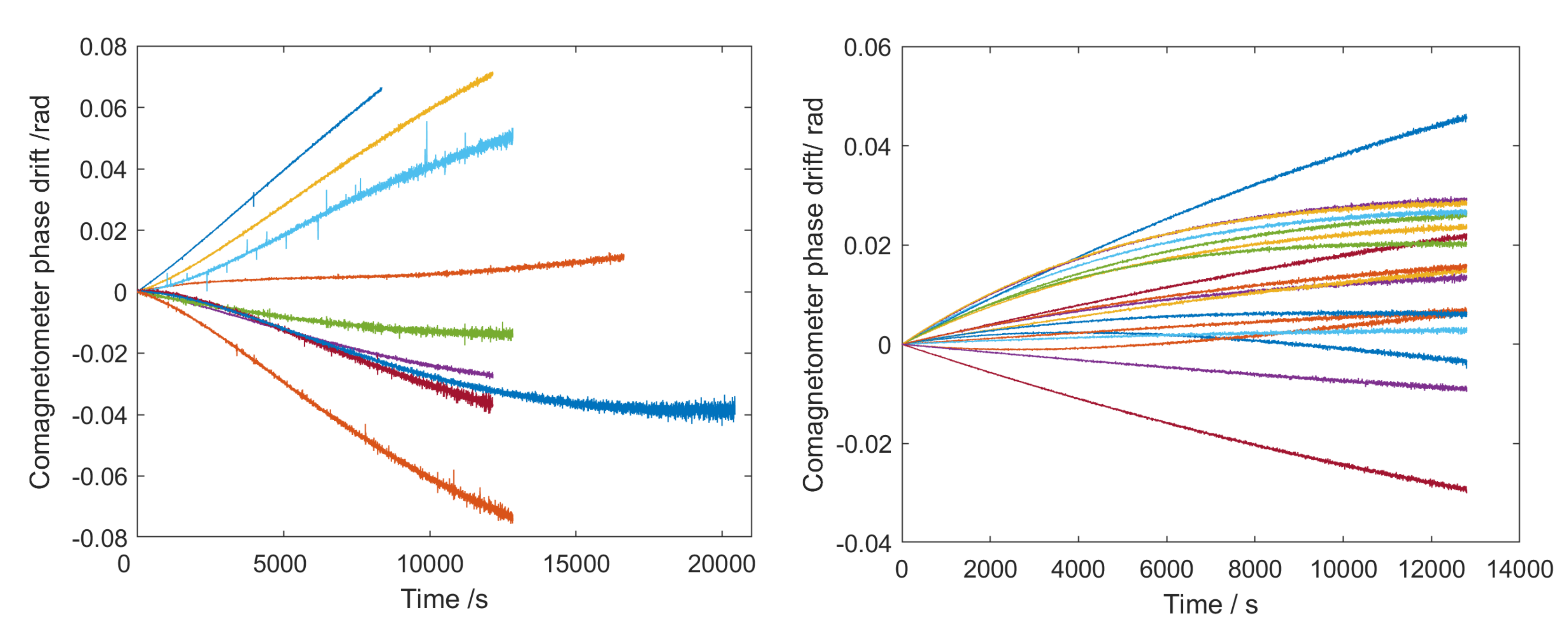} 
    \leftline{ \hspace{3cm}(a) \hspace{8cm} (b)}
    \caption{Measured comagnetometer phase drift $\Phi_{\text{co}}$ of the first sub-runs from 9 runs in the 2018 campaign (a) and from 18 synthetic data set(b).}
    \label{fig:Check_WPD}
\end{figure}

Fig.~\ref{fig:Check_WPD} compares the measured $\Phi_{\text{co}}$ during the last week of the 2018 campaign and the synthetic $\Phi_{\text{co}}$. For the measured result, only the data of the first sub-run are shown. The linear drifts due to the Earth's rotation and chemical shift were previously subtracted. Spikes on the measured data are due to SQUID jumps (see more details in Sec.~\ref{sec:data_ano} and are not relevant for the GPF method as being accounted for by enlarged $\delta \Phi_{\text{co}}$ for blocks with those spikes). The synthetic comagnetometer phase drifts for 18 random sub-runs are presented in Fig.~\ref{fig:Check_WPD}(b), which shows a similar phase drift as the experimental result. 

\subsection{White noise and real noise}
\label{sec:Real_noise}
\begin{figure}[htpb]
    \centerline{\includegraphics[width=0.6\columnwidth]{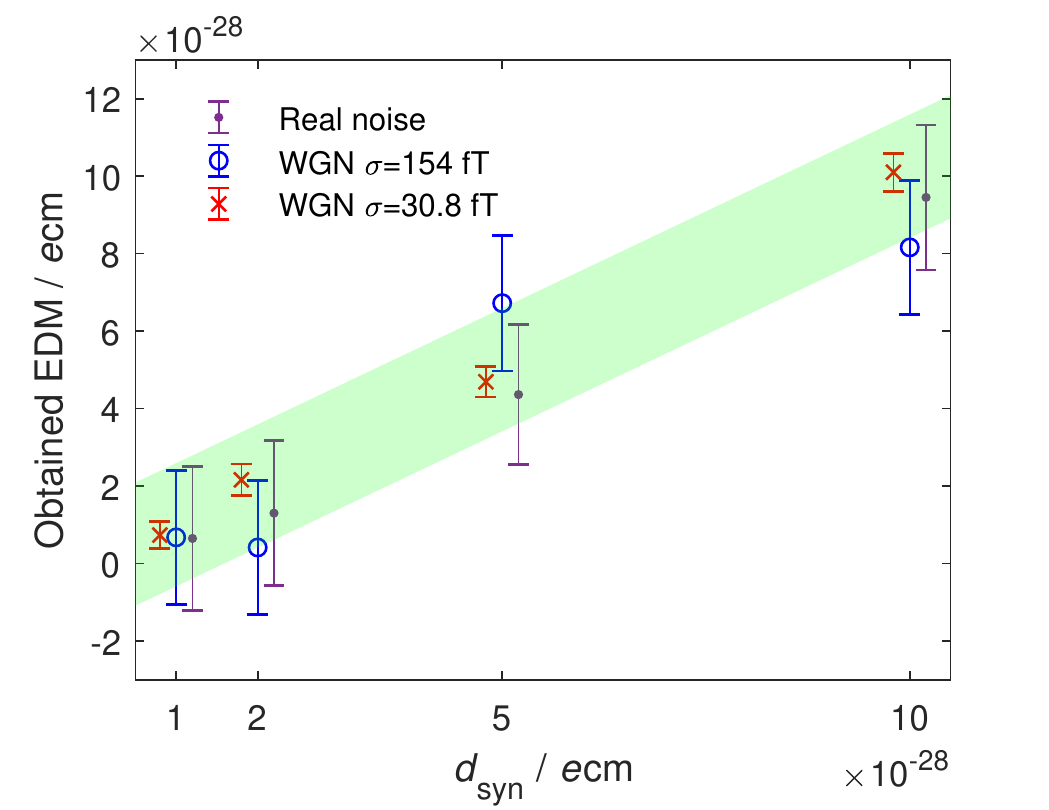}}
    \caption{ \label{fig:MC} The derived EDM values using the synthetic data sets. The horizontal coordinates for the real noise (in red) and WGN with $\sigma = 30.6 $~fT (in purple) are shifted with $2 \times 10^{-29}~e~\mathrm{cm}$ and $-2\times 10^{-29}~e~\mathrm{cm}$, respectively. The green shade illustrates the $1\sigma$ confidence interval with the added EDM as the center value and the uncertainty derived from the CRLB for $\sigma = 154 $~fT. }
\end{figure}

The parameters used to generate synthetic data were taken from 9 runs (B76, B78, B84, B87, B89, B90, B91, B92, including 18 sub-runs) of high sensitivity from the 2018 campaign. Since the noise data from real measurements was taken, the background field $B_0$ was set to 3.6~\textmu T instead of 3~\textmu T in order to avoid the correlation between the synthetic Xe precession signal and the used bandstop filters to remove the original precession signals. The electric field pattern included 36 segments of 200~s up to 800~s length, as used in the measurement campaign. The values of other parameters in Eqs.~(\ref{eqn:f_syn}) and (\ref{eqn:SQUID_syn}) are random and uniformly distributed in the ranges listed in Table~\ref{tab:Syn_para}. Three different kinds of noise were separately added to the synthetic data, including two WGN with $\sigma =154$~fT, the standard deviation of the white noise in the real data, and a 5 times lower value of $\sigma = 30.8$~fT, as well as the real SQUID gradiometer noise. The overall EDM values obtained with the GPF method from the 18 synthetic sub-runs for four synthetic EDM input values $d_\mathrm{syn} = (1,2,5~\text{and}~10) \times 10^{-28}~e~\mathrm{cm}$ are plotted in Fig.~\ref{fig:MC}. The averaged overall EDM uncertainty for WGN data with $\sigma =154$~fT is $1.74 \times 10^{-28}~e~\mathrm{cm}$, which is roughly a factor of 5 larger than that obtained from the data with $\sigma = 30.8$~fT and a factor of 1.1 higher than the calculated CRLB for these 18 sub-runs, which is $1.59\times 10^{-28}~e~\mathrm{cm}$. This mainly results from the correlation between the EDM and the parameters of the polynomials in the phase fit. The uncertainty for the real noise is $1.85\times 10^{-28}~e~\mathrm{cm}$, being similar to that for the white noise with $\sigma = 154$ fT. Most of the $1\sigma$ confidence intervals of the derived EDM include the added EDM values $d_\mathrm{syn}$, showing that the GPF method is capable of obtaining $ d_\mathrm{syn} \geq 1 \times 10^{-28}~e~\mathrm{cm}$ in the presence of the chosen noise.
 
\subsection{Anomalous phase drift}
Another topic for the simulation test was the impact of the anomalous phase drift on the derived EDM. The time  for analyzing one sub-run of SQUID data was around 2 minutes, mainly spent on the VP fit to get the phases for each block. To speed up the simulation in order to deal with more cases, the synthetic phase data was directly used, reducing the analysis time per run to several seconds. 

In the synthetic data, white noise was added to the phase. The standard deviation $\sigma$ of the phase noise started with 0.1~mrad and then exponentially increased with time constant $T_2=8000$~s. The total measurement time length was fixed to 38400~s. The electric field pattern was also synthetically generated. No zero voltage segment was included in the simulated pattern. 

\begin{figure}[ht]
    \centerline{\includegraphics[width=.48\columnwidth]{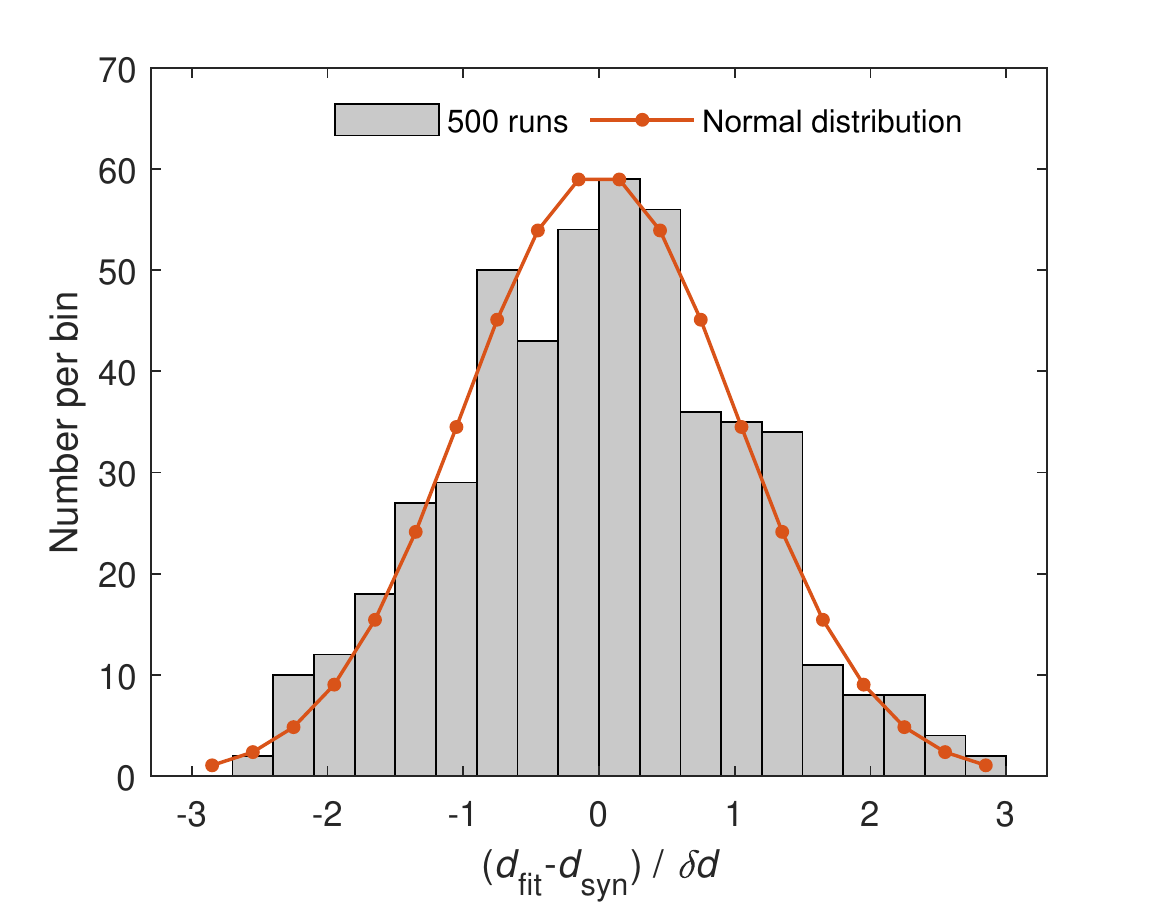} \hspace{1cm}\includegraphics[width=.48\columnwidth]{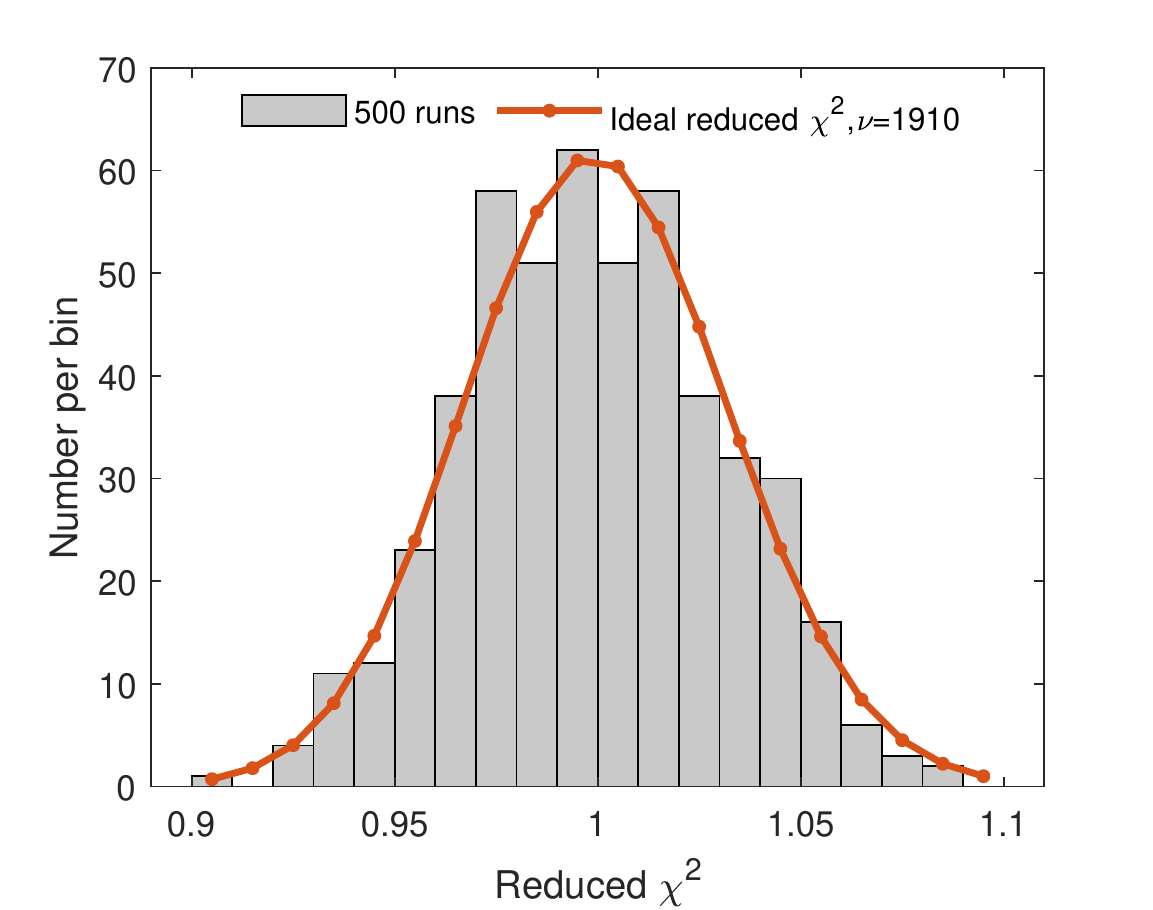}}
    \leftline{ \hspace{3cm}(a) \hspace{8cm} (b)}
    \caption{The fit results of 500 synthetic runs. (a) Histogram of the deviation of the fit EDM result weighted by its uncertainty. (b) Histogram of the reduced $\chi^2$.}
    \label{fig:500_run}
\end{figure}

First a fixed segment number $M=32$ and a fixed EDM value $d_{\text{syn}}= 1 \times 10 ^{-27} e$ cm were used. The results of applying the GPF method to 500 synthetic runs are shown in Fig.~\ref{fig:500_run}. The left plot presents the distribution of the deviation of the fit result weighted with its uncertainty, i.e. $(d_{\text{fit}}-d_{\text{syn}})/\delta d$, which fits the normal distribution well. In the right panel is the distribution of reduced $\chi^2$, which is in good agreement with the theoretical result for 1910 degrees of freedom, given that the total block number was $JM=1920$ and the average number of parameters used in the fit model was around 10. The result shows that the polynomial functions used in the GPF method are able to adequately describe the anomalous comagnetometer phase drift, thus reducing the systematic bias stemming from the analysis algorithm to an indiscernible level.  

\begin{figure}[hbpt]
    \centerline{\includegraphics[width=.49\columnwidth]{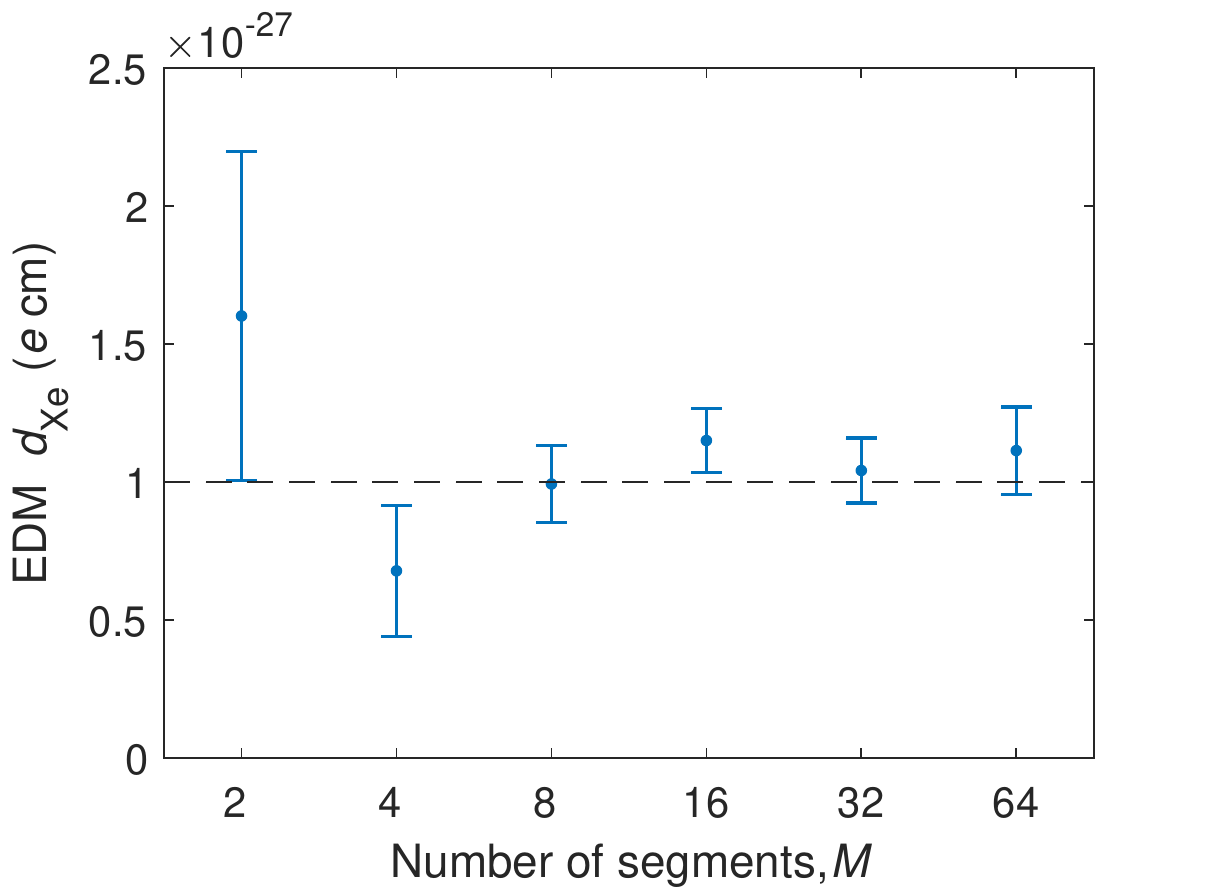} \hspace{1cm}\includegraphics[width=.46\columnwidth]{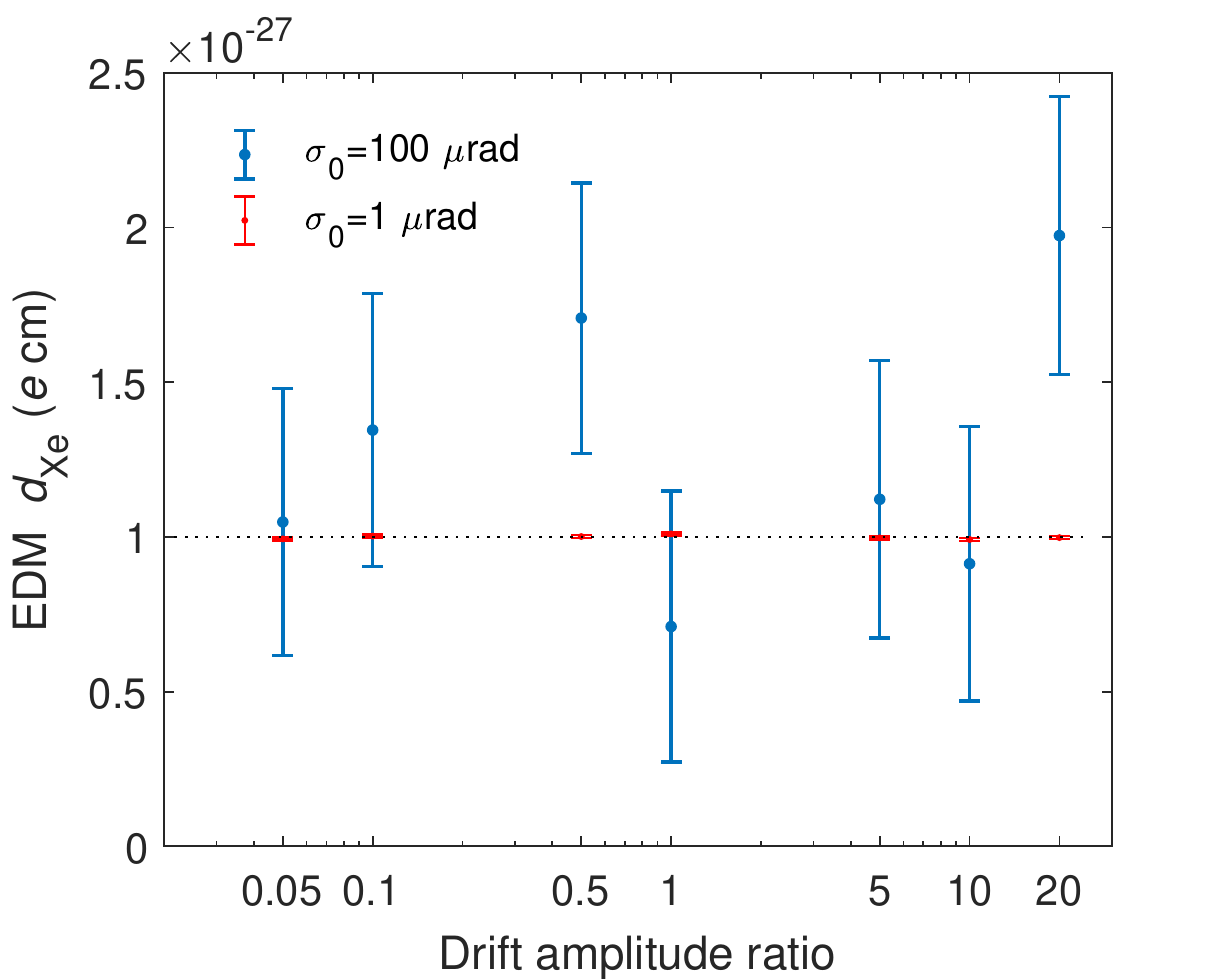}}
    \leftline{ \hspace{3cm}(a) \hspace{8cm} (b)}
\caption{(a) Obtained EDM values as a function of segment numbers $M$ for 100 sub-runs lasting 38400~s. (b) Derived EDM value as a function of the scale ratio of the drift amplitude to the observed drift in two campaigns. The dotted line shows the added EDM value $1 \times 10^{-27} e$~cm. Each result is an average of 40 sub-runs lasting 12800~s and with 32 high voltage segments.}
    \label{fig:Drift_result}
\end{figure}

The impact of the anomalous comagnetometer drift on the derived EDM depends on the segment number $M$ and the drift amplitude. In the following simulations, we separately altered the segment number $M$ and the drift amplitudes $u^\mathrm{Xe/He}$ in Eq.~(\ref{eqn:Ano_f}) for synthetic phase data. Since the real data set included two runs with three segments only, it is necessary to test the GPF method with respect to the number of segments. Fig.~\ref{fig:Drift_result}(a) shows that even for $M=2$, the GPF method is able to obtain the EDM value. The EDM uncertainty is larger for $M=2$ compared to $M=16$, due to the correlation between the EDM function and the used polynomial function accounting for the phase drift. Further increasing the number of segments $M$ leads to an increase in uncertainty, as predicted by CRLB analysis in Eq.~(\ref{eqn:CRLB_EDM_simp}). 

We changed the ranges of the drift amplitudes $u^\mathrm{Xe}$ and $u^\mathrm{He}$ as given in Table~\ref{tab:Syn_para} by factors of 0.05 to 20 in the synthetic phase data. White phase noise was added into the phase data and the noise deviation $\sigma$ increased with time in an exponential way with $T_2=8000$~s. Fig.~\ref{fig:Drift_result}(b) shows the derived EDM values as a function of the scale ratio of the drift amplitudes for two starting noise uncertainties $\sigma_0=100$~\textmu rad, similar to the real data, and $\sigma_0=1$~\textmu rad. No significant dependence between the obtained EDM value and the drift amplitudes could be observed. For the low noise case, the maximum deviation of the central EDM value among 7 results is less than $1 \times 10^{-29} e$~cm. So, we did not assign a model dependent uncertainty for the comagnetometer drift when applying the GPF method, which is also supported by the analysis shown in Fig.~\ref{fig:MC}. 

\chapterimage{chapter_head_1.pdf} 
\chapter{$^{129}$Xe EDM results}

\section{Data properties}

In total, 48 experimental runs were successfully conducted. In order to study the possible correlation between the experimental parameters and the EDM result, various parameters were altered during two campaigns, including the axis of the magnetic field $B_0$, the starting electric field direction, the ramp time of the electric field, the EDM cell, the gas pressure in the cell, the number of segments and the length of a segment. The values of the key parameters for each sub-run are listed in Table~\ref{tab:EDM_result} at the end of this chapter.  

The default number of segments in one sub-run was set to $M=36$, which was optimized with respect to the Pattern Combination method as proposed in Ref.~\cite{Sachdeva2019}. Four sub-runs in the 2017 campaign had 18 segments. To study the idea of the GPF analysis method, we have intendedly performed one sub-run with only three segments but of longer duration time $t_{\text{s}} > 2000$~s in each campaign.   

\subsection{Data anomalies}
\label{sec:data_ano}

Several experimental disturbances during measurements were identified, including discharges, SQUID jumps and incomplete segments. 

\begin{figure}[ht]
    \centerline{\includegraphics[width=.42\columnwidth]{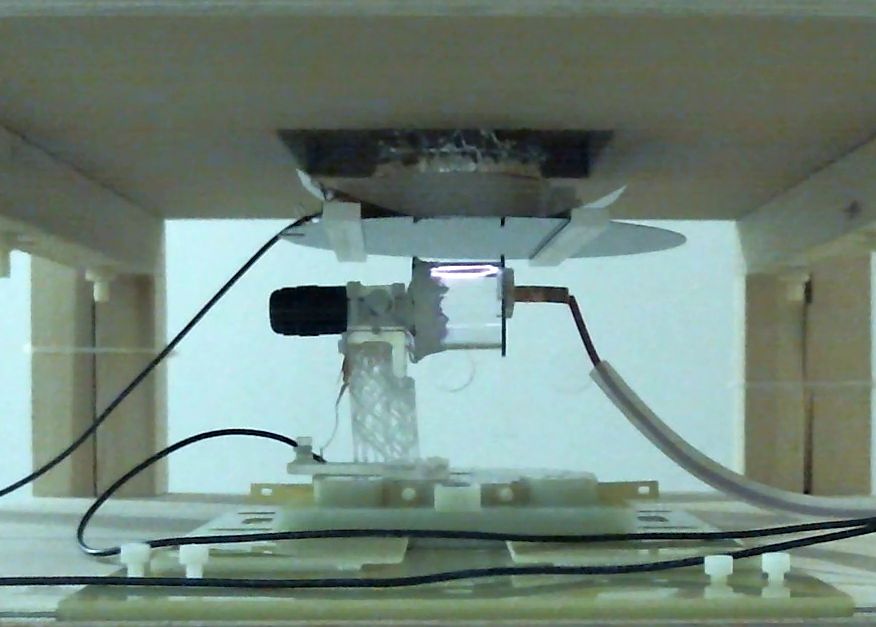} \hspace{1cm}\includegraphics[width=.48\columnwidth]{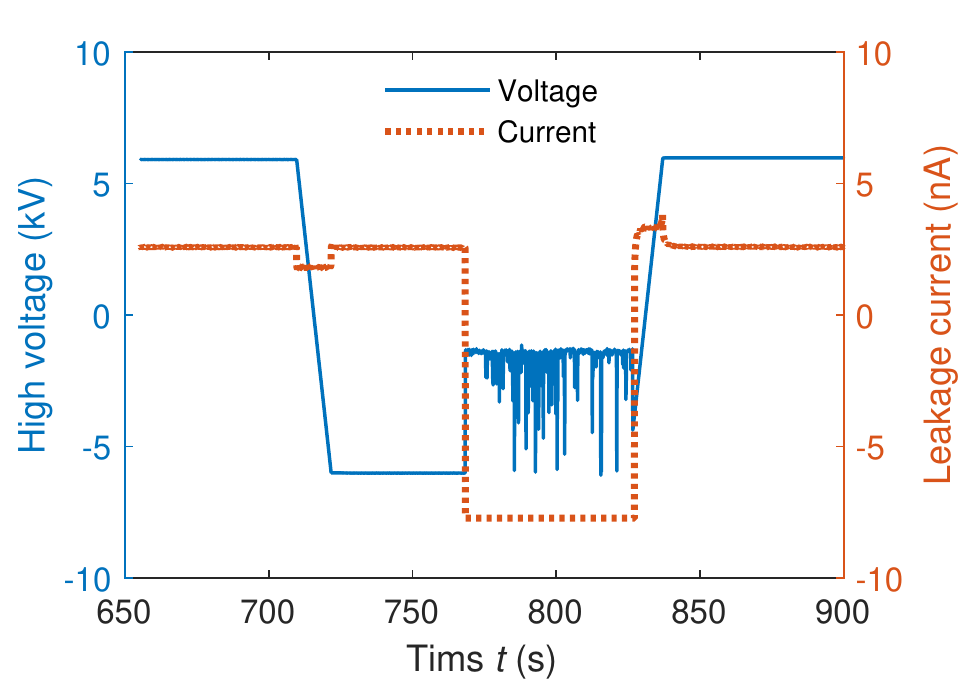}}
    \leftline{ \hspace{3cm}(a) \hspace{8cm} (b)}
\caption{(a) Observed dischargers in a test measurement. The applied high voltage is 6~kV in a low gas pressure $P$=0.43~bar. The light inside the EDM cell is created by the discharge. (b) The measured dischargers happened from 760~s to 820~s during the B64 run.}
    \label{fig:Discharging}
\end{figure}

The high voltage across the EDM cell should be chosen as high as possible, since the EDM sensitivity is linearly dependent to the applied $E$ field. However, several factors limit the high voltage that can be applied to the system. For example, micro discharges must be avoided. Before the measurement, we have tested the maximum usable high voltage. Fig.~\ref{fig:Discharging}(a) shows an observed discharge during a high voltage test. During the EDM measurement, the high voltage was set to a bit lower than the found critical limit. Nevertheless, we had discharges lasting more than 5~s in three runs of the 2018 campaign. The recorded data during the discharging in the B64 run is shown in Fig.~\ref{fig:Discharging}(b). From 760~s to 820~s, the current across the cell reached the limit of the current monitor, and the voltage jumped rapidly. The discharge resulted in a serious spin depolarization destroying the amplitudes of the spin precession. Therefore, two runs (B64, B68) were completely discarded. 

Another anomaly we observed are SQUID jumps, a sudden change of the SQUID offset. This phenomenon happened more frequently than dischargers. The sub-runs with SQUID jumps are listed Table~\ref{tab:SQUID_jump}. Except run B89  (with two sub-runs B891 and B892), all other sub-runs contained not more than 3 SQUID jumps. For a block containing SQUID jumps, the uncertainty of the estimated phase for this block was heavily increased. The SQUID jumps caused a problem to unwrap the phase of the next normal block in Eq.~(\ref{eqn:integer_wrap}). To solve this, the signal frequency of this problematic block is automatically substituted by the average of the frequencies of the adjacent unspoiled blocks. The substitution was done when the sum of the squared residual error values exceed a given threshold level. Owing to the weighted fit in the GPF, the impact of the problematic block on the EDM estimation is suppressed due to its unsubstituted large uncertainty. In the global phase fit step, we just treated the blocks with SQUID jumps as normal blocks. 

\begin{table}[h]
    \centering
    \caption{Number of blocks containing SQUID jumps per sub-run.}
    \begin{tabular}{c | l }
    \toprule
      Number & Sub-run   \\
    \midrule
        1   &  A83,A85,A87,A91,A8,A10,A15,B50,B671,B722,B761,B902   \\
        2   &  A93,B651,B842    \\
        3   &  A82, B631        \\
        >15 &  B891, B892     \\
    \bottomrule
    \end{tabular}
    \label{tab:SQUID_jump}
\end{table}

Last but not least, in four sub-runs the recorded number of nonzero high-voltage segments $M$ was less than the designed number due to experimental problems, such as the loss of data. B782 missed the last two segments; B912 and B942 contained 24 and 25 segments, respectively. However, the GPF method is still able to analyse these incomplete data sets.

\subsection{Anomalous phase drift}

\begin{figure}[!b]
    \centerline{\includegraphics[width=\columnwidth]{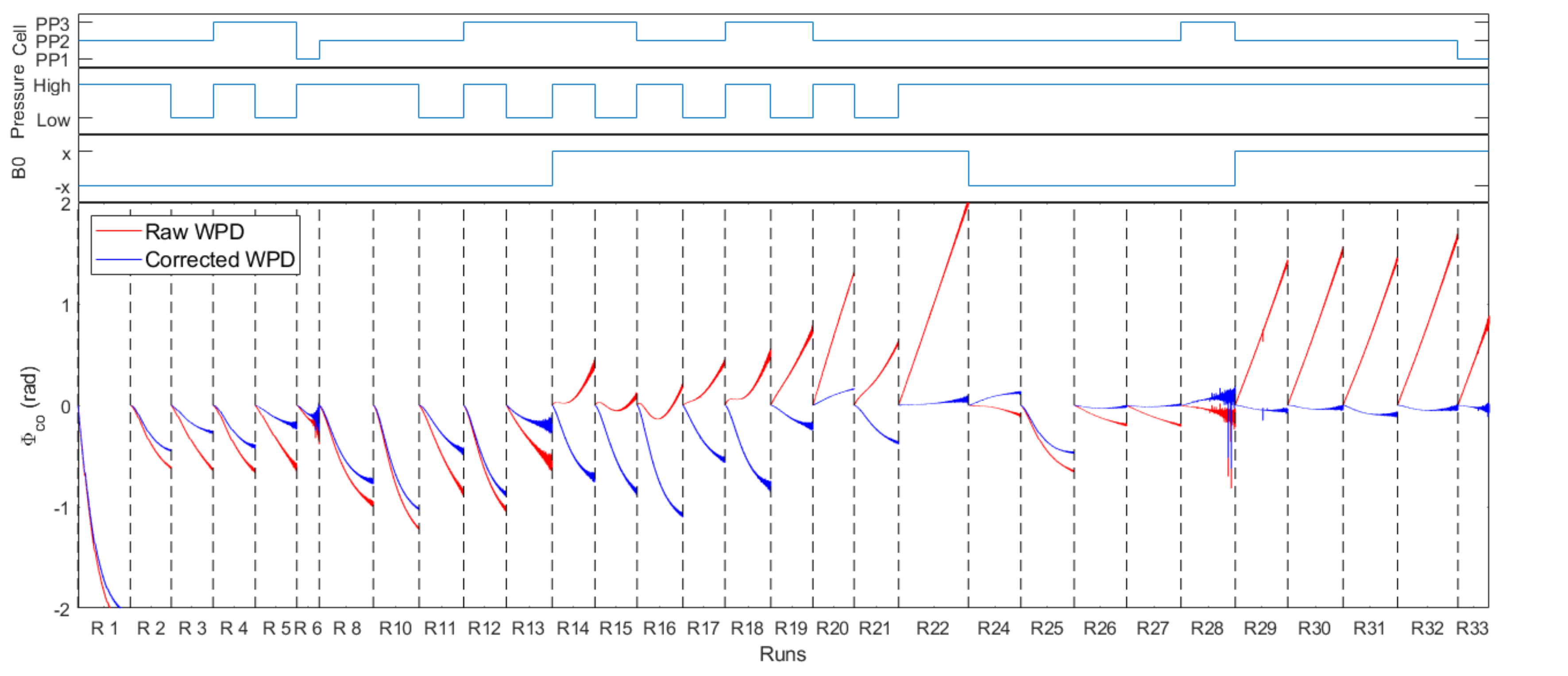} }
    \caption{ The measured weighted phase difference $\Phi_{\text{co}}$ for the 2018 campaign. The corrected phase was obtained via removing the linear phase drift from chemical shift effects as well from Earth's rotation.}   
    \label{fig:WPD_2018}
\end{figure}

The raw comagnetometer phase of each run of the 2018 campaign is plotted in Fig.~\ref{fig:WPD_2018},
together with the corrected comagnetometer phase. Here the correction refers to the cancelling of the linear drift.  It can be seen that in the last week (from run R22 on - with one 'old style' run R25 in between), the anomalous phase after removing the deterministic drifts was significantly smaller than before, owing to an optimized ratio of the starting signal amplitude between $^{129}$Xe and $^{3}$He. The remaining anomalous comagnetometer phase drifts were considered by a polynomial function in the GPF method. 

\section{Data evaluating}

\subsection{Statistical uncertainty}
\begin{figure}[ht]
    \centerline{\includegraphics[width=.7\columnwidth]{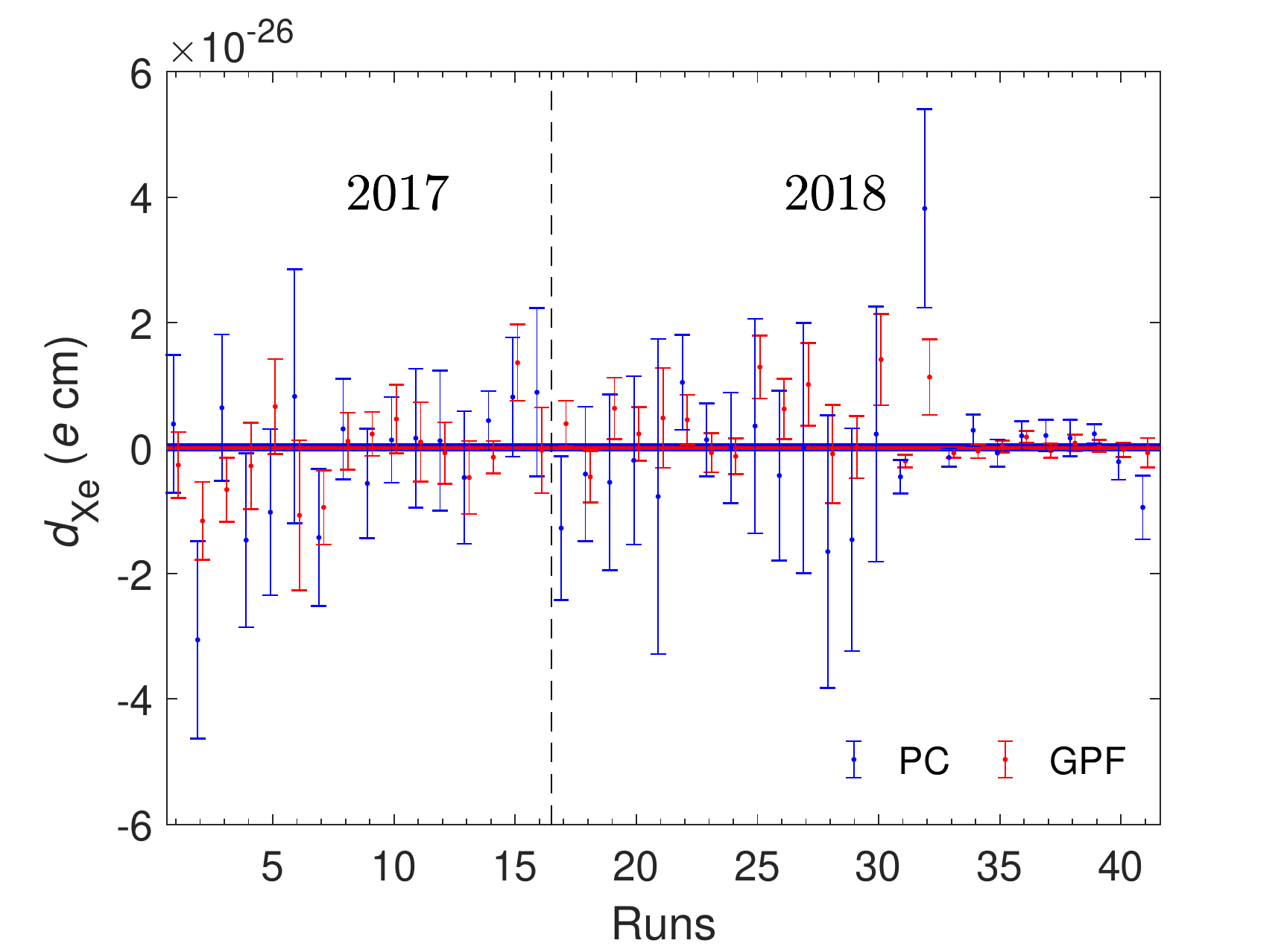} }
    \caption{ Comparison of the GPF method and the PC method per sub-run. The overall averaged result for the PC method is $d_{\text{Xe}}=(1.43 \pm 6.56) \times 10 ^{-28} e$~cm and for the GPF method is $d_{\text{Xe}}=(0.251 \pm 3.06) \times 10 ^{-28} e$~cm.}
    \label{fig:GPF_PC}
\end{figure}

Applying the GPF method to the same data set of 41 runs (80 sub-runs) as analyzed by the PC method \cite{Sachdeva2019a}and using the same analysis parameters, the overall EDM result is in good agreement with the result of the PC method (see Fig.~\ref{fig:GPF_PC}). The statistical uncertainty decreased by a factor of 2.1 from $6.56 \times 10 ^{-28} e$~cm to $3.06 \times 10 ^{-28} e$~cm, as predicted by the theoretical CRLB analysis. 

Due to fewer constraints in the GPF method, runs with $M \neq 4n$ ($n \in \mathbb{N}$) segments  or having SQUID jumps could be included in the data analysis, leading to a total of 45 runs (87 sub-runs). For the analysis, the block length was $t_\mathrm{b}=5$~s, the threshold of the $F$-test was set to $P_\mathrm{min}=0.6$ and the minimum order of the polynomial used in the fit was set to 4 in order to adequately describe the comagnetometer phase drift. The average polynomial order for all sub-runs resulted from the $F$-test procedure to be 6.4 and the maximum order needed was 13. The impact of $P_\mathrm{min}$ on the EDM results is discussed in Sec.~\ref{sec:overall_results}.

\begin{figure}[htpb]
    \centerline{\includegraphics[width=1\columnwidth]{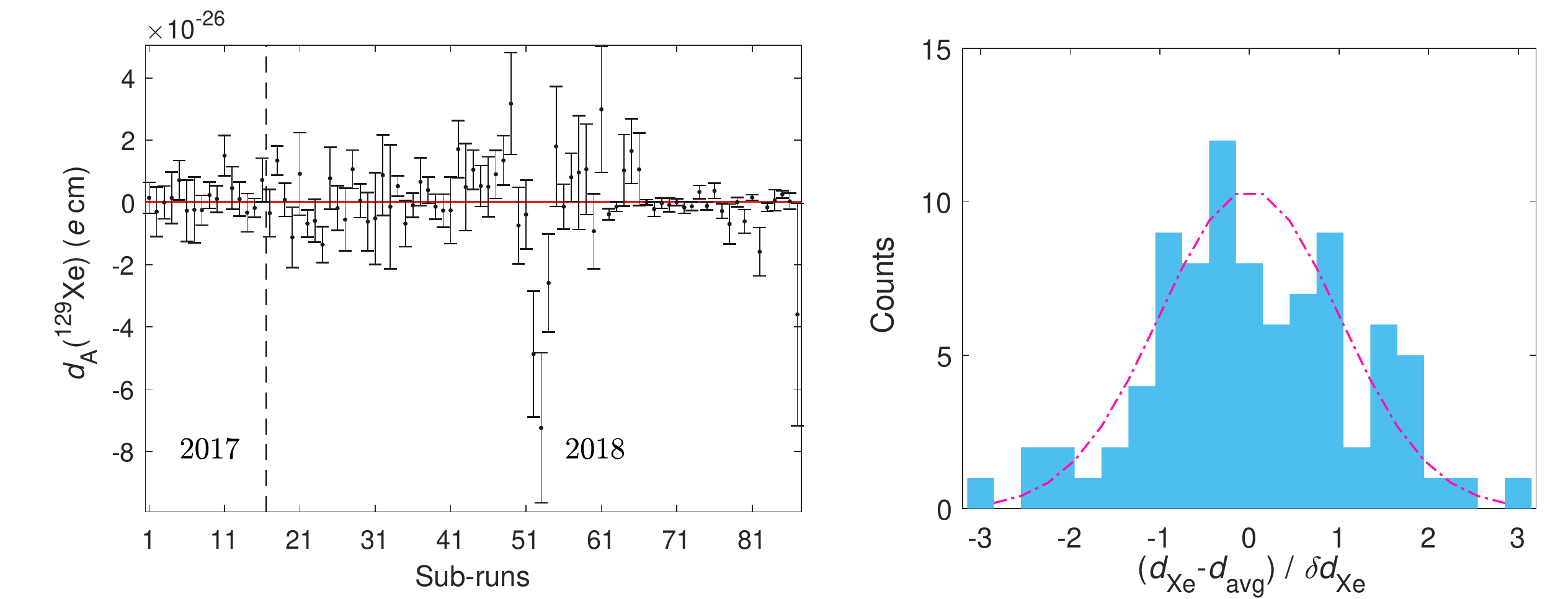}}
    \caption{ (a) EDM results of the 2017 and 2018 campaigns derived with the GPF method by sub-runs. The thin orange bar is the confidence interval of $1\sigma$ around the weighted mean. The reason for a lower uncertainty in the last 20 sub-runs is a change in the experimental parameters as explained in detail in Ref.~\cite{Sachdeva2019}. (b) Histogram of the normalized EDM results $(d_\text{Xe}-d_\text{avg})/\delta d_\text{Xe}$ with $\delta d_\text{Xe}=3.1 \times 10^{-28}~e~\mathrm{cm}$ and the dot-dashed line illustrates the standard normal distribution.}
    \label{fig:Sub_run}
\end{figure}

We have blinded all the data before the GPF analysis as described in Fig.~\ref{fig:Analysis Procedure}. After the extensive calculation of the systematic error, we unblinded the EDM result. After performing the whole analysis we received several suggestions on further testing the analysis method, and the parameters of the GPF method had slightly to be adjusted, which was then applied directly to the unblinded data. Fig.~\ref{fig:Sub_run}(a) and (b) shows the derived EDM results per sub-run and the histogram of the normalized EDM results, respectively. The exact values of the EDM results together with the used polynomial orders for each sub-run are listed in Table~\ref{tab:EDM_result}. The overall averaged result of the 87 sub-runs is $d_\mathrm{A} (^{129}\mathrm{Xe})  = 1.1 \pm 3.1 \times 10^{-28}~e~\mathrm{cm}$  with $\chi^2/\text{\text{dof}}=115.5/86$. As all sub-run measurements were taken with considerable different signal to noise ratio a $\chi^2/\text{dof}  \geq 1$ can be expected. According to the PDG guidelines \cite{Beringer2012} we accounted for these random variations by scaling the statistical uncertainty with the factor $\sqrt{\chi^2/\text{dof}}=1.16$ leading to an uncertainty of $3.6 \times 10^{-28}~e~\mathrm{cm}$. To cross-check the averaged EDM uncertainty, we applied bootstrapping \cite{Efron1982}, yielding an estimate of the statistical uncertainty of $3.14 \times 10^{-28}~e~\mathrm{cm}$ that is consistent with the averaged result. 

\begin{figure}[!ht]
    \centerline{\includegraphics[width=0.7\columnwidth]{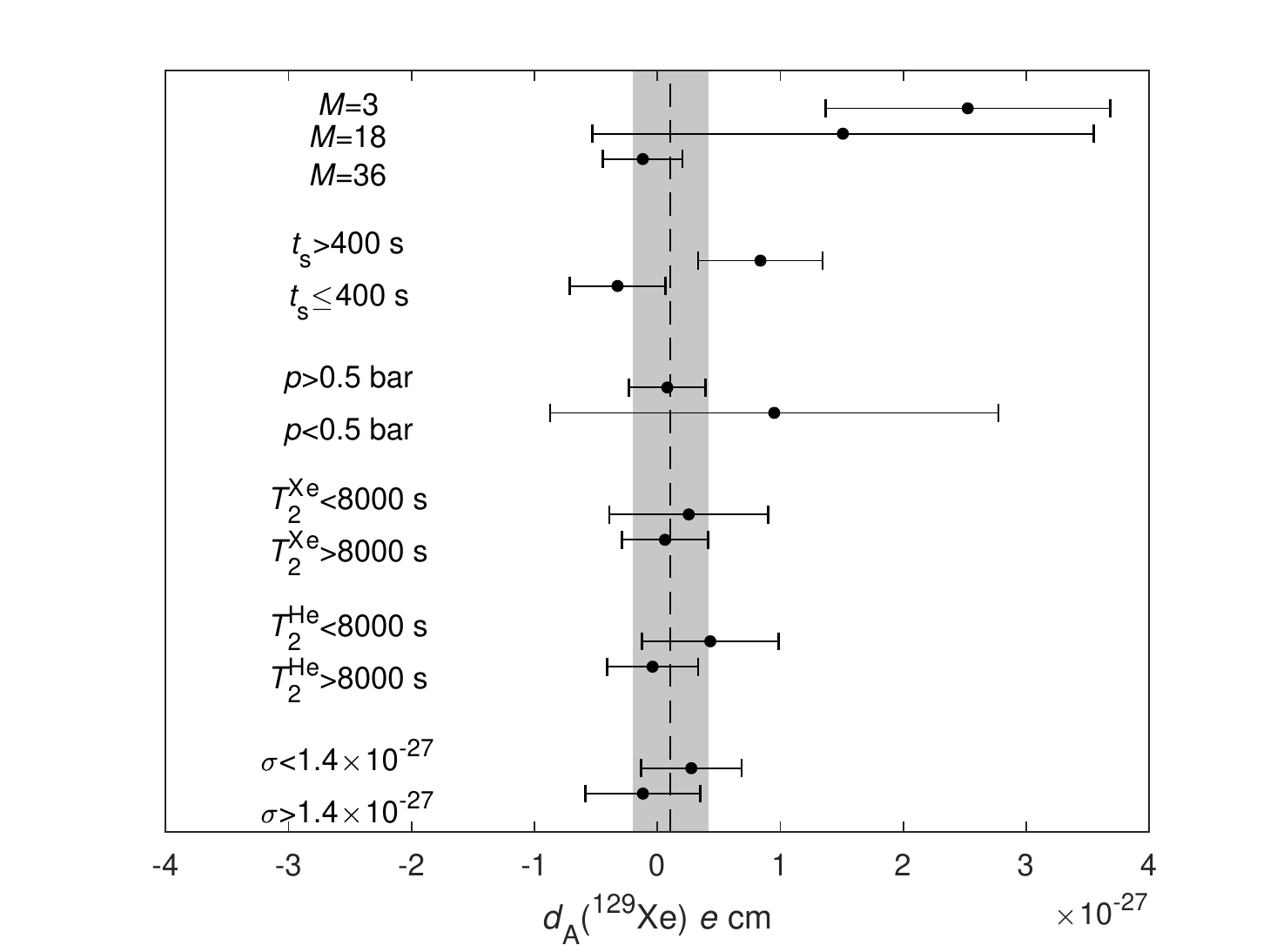}}
    \caption{The EDM results for grouping the data set by number of segments $M$, segment duration $t_s$, gas pressure $p$, $T_2^{Xe}$, $T_2^{He}$, and the statistical uncertainty threshold of  $1.4 \times 10^{-27}~e~\mathrm{cm}$. The dashed line is at $d_\mathrm{A} (^{129}\mathrm{Xe})=1.1 \times 10^{-28}~e~\mathrm{cm}$ and the gray region indicates the confidence interval of $1\sigma$ with the unscaled statistical uncertainty. For clarity of the figure, only a few parameters are plotted here.}
    \label{fig:Correlation}
\end{figure}

\begin{figure}[!ht]
    \centerline{\includegraphics[width=.48\columnwidth]{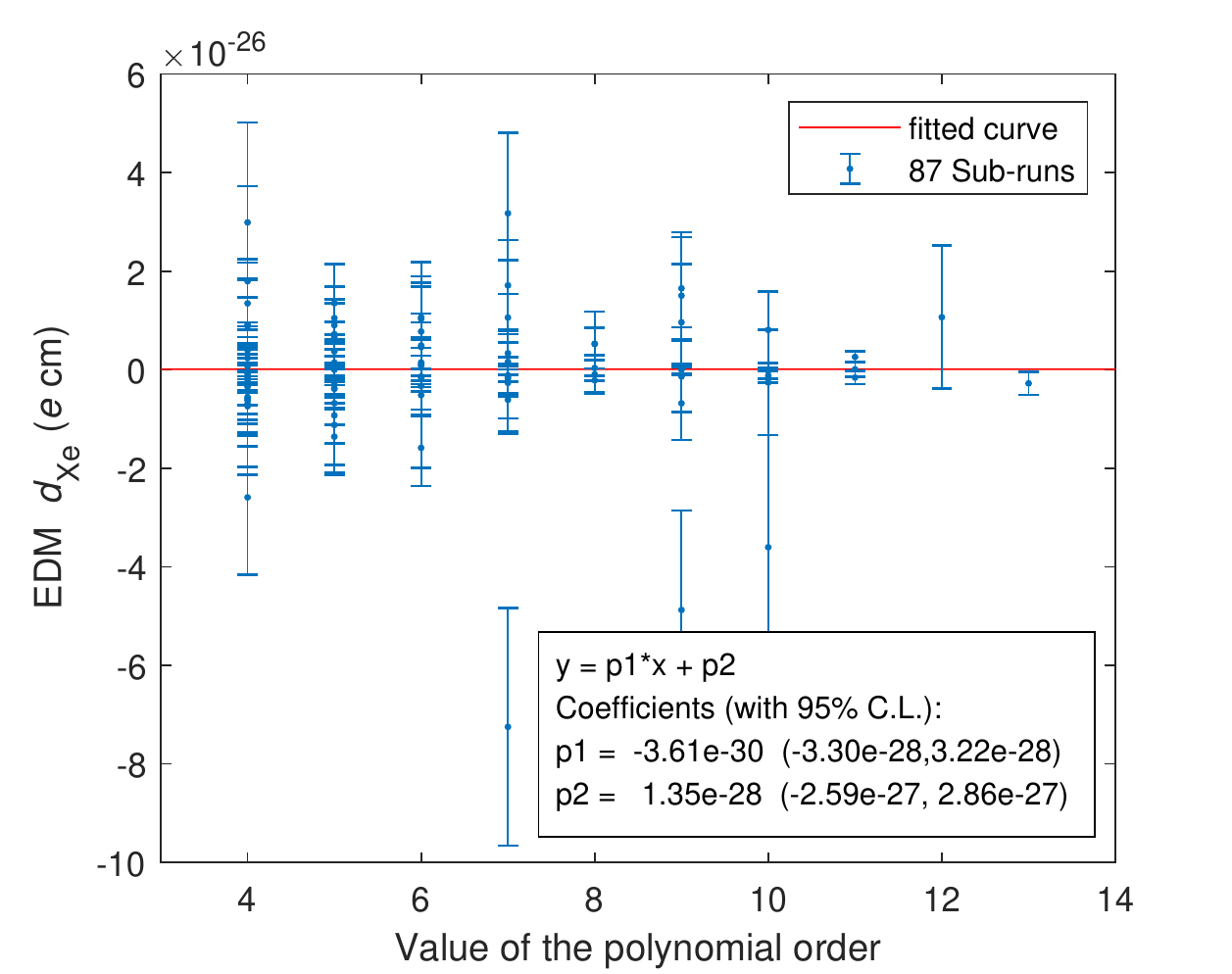} \hspace{1cm}\includegraphics[width=.48\columnwidth]{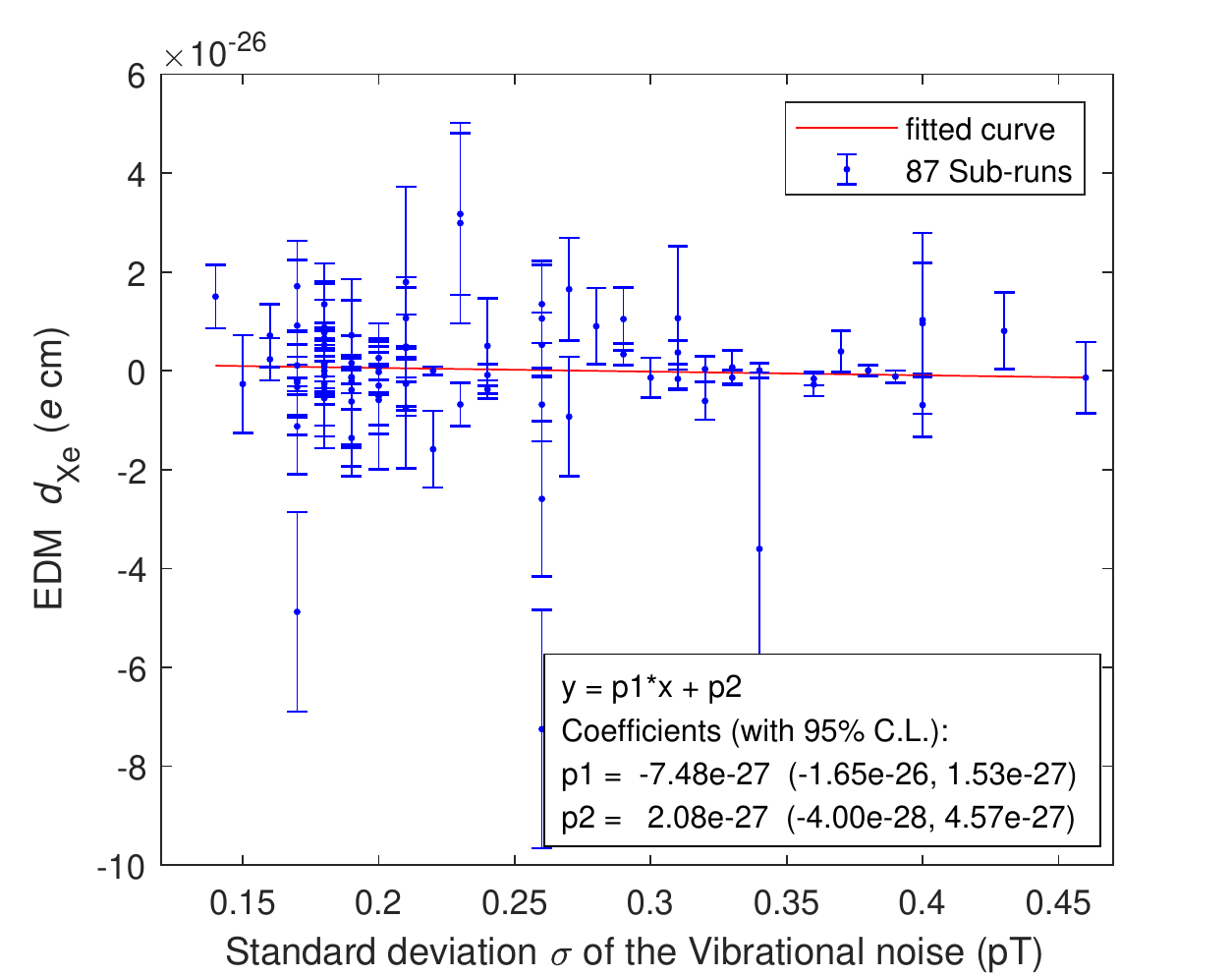}}
    \leftline{ \hspace{3cm}(a) \hspace{8cm} (b)}
\caption{Linear correlation between the derived sub-run Xe EDM values and the $F$-test deduced polynomial order (a) as well as the standard deviation of the vibrational noise (b), which was estimated in the same way as done in Fig.~\ref{fig:fit_quality}. The red lines in both figures are the weighted linear fit to the EDM values from 87 sub-runs, and the text boxs present the fit results. }
    \label{fig:Correlation2}
\end{figure}

Sorting all EDM measurements into groups based on the experimental parameters, such as the cell geometry, $B_0$ field direction, number and duration of segments and the gas pressure, shows no correlation between the deduced EDM value and these parameters, as can be seen in Fig.~\ref{fig:Correlation}. Furthermore, no linear correlation between the derived sub-run EDM values and the $F$-test deduced polynomial order as well as the standard deviation of the vibrational noise was seen, as shown in Fig.~\ref{fig:Correlation2}.

\subsection{Systematic error}
The systematic uncertainties of the two measurement campaigns were extensively studied in Ref.~\cite{Sachdeva2019}. We applied the same analysis to the full data set used here and the derived systematic uncertainties are summarized in Table~\ref{tab:GPF_syst}. The correction to the comagnetometer frequency drift of order higher than one (due to the unweighted average of four segments in the PC method), as it has been done in Ref.~\cite{Sachdeva2019}, becomes obsolete for the GPF method since the model of Eq.~(\ref{eqn:GPF}) considers the higher order drifts implicitly. 

As mentioned above, in contrast to the PC method the GPF used the full data set, including also data during the high voltage ramps. Therefore, the charging current can have an impact on the result in two ways. First the charging current could magnetize parts of the experimental equipment, and change the magnetic field seen by the spins. By this mechanism a false EDM may be generated. This effect has been carefully analyzed in Ref.~\cite{Sachdeva2019} and has been adapted for the data set used for GPF (see Charging current in Table~\ref{tab:GPF_syst}). Secondly, the charging currents will generate magnetic fields that are correlated with the electric field direction, just as the leakage currents. 

\begin{figure}[ht]
    \centerline{\includegraphics[width=.9\columnwidth]{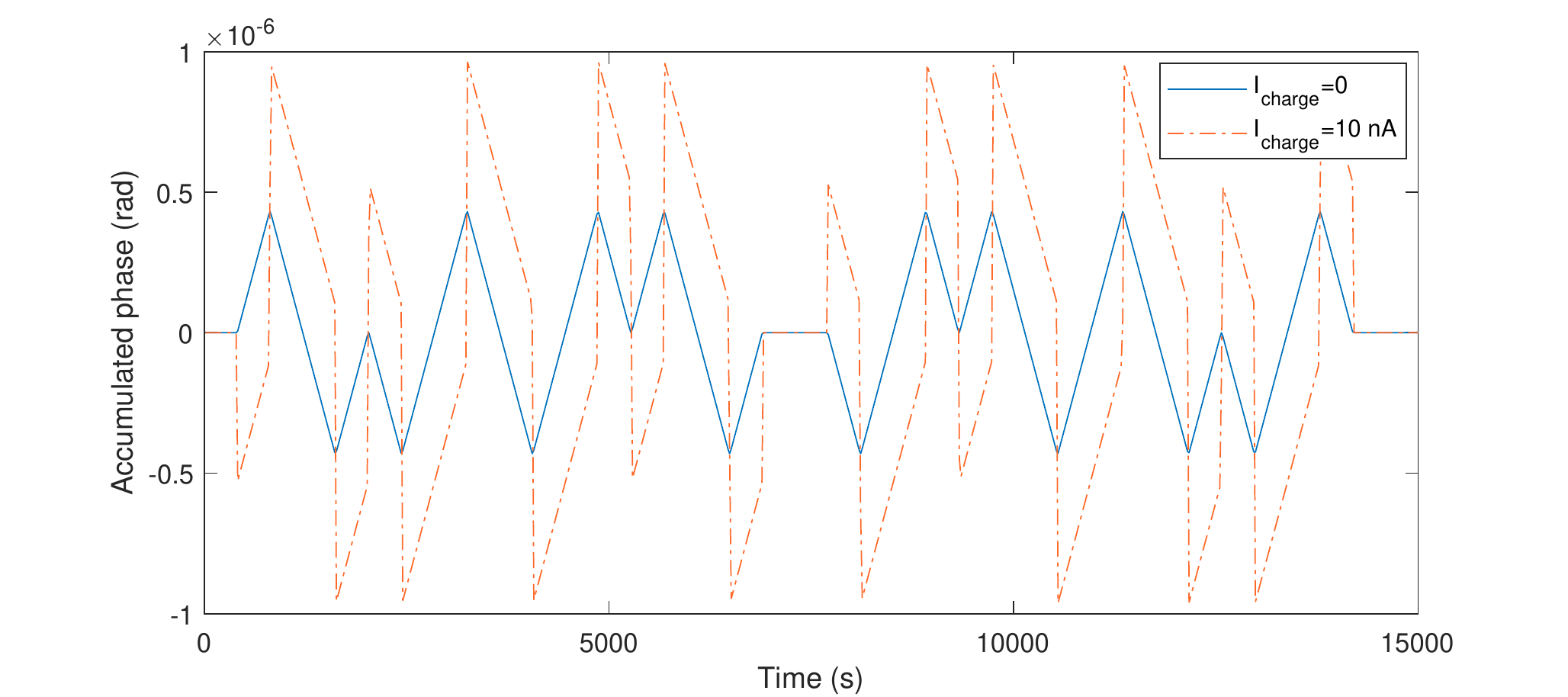} }
    \caption{Accumulated phase caused by the leakage currents over one synthetic sub-run. The used parameters are: $M=36$,$t_s=400$~s, $I_\text{leakage}=0.1$~nA, $C_\text{curr}$=1.7~nHz/nA, $U$=6~kV and the voltage ramp rate is 1~kV/s. The two curves were obtained for different charging currents as given in the legend.}
    \label{fig:Charge_current}
\end{figure}

We used synthetic leakage current data over a whole sub-run to study the impact of charging currents. Both the amplitudes of leakage currents and charging currents were constants and their directions were changed according to the electric field pattern. The maximum conversion factor from leakage current to frequency shift was measured to be $C_\text{curr}$=1.7(0.01)~nHz/nA for the 2017(2018) campaign \cite{Sachdeva2019}. Fig.~\ref{fig:Charge_current} compares the accumulated phase over a sub-run caused by a leakage current with and without considering the charging currents during ramping. The sharp jumps are caused by the charging currents, whose direction is opposite to the leakage current of the following segment. Applying the GPF method to this accumulated phase, we directly obtained the systematic error caused by leakage currents, which are $1.0 \times 10^{-28} e$~cm and $1.2\times 10^{-28} e$~cm, for $I_\text{charge}=10$~nA and $I_\text{charge}=0$~nA respectively. In the course of deriving the EDM error, one had to transfer the frequency shift to the EDM value using the electric field strength, which was set to 2.75~kV/cm, the number of the 2017 campaign as listed in Tab.~\ref{tab:exp_para}. The result shows that charging currents during high voltage ramps slightly reduces the systematic error as compared to the effect of the leakage currents alone. However, this reduction is not significant due to the weak correlation between the accumulated phase caused by pulse-like charging currents and the EDM-induced phase shown in Fig.~\ref{fig:Voltage&Phase}. Applying the same analysis to the sub-run with the maximum systematic error caused by leakage current, the systematic error reduces from $1.20 \times 10^{-28} e$~cm to $1.19 \times 10^{-28} e$~cm for 2017 campaign(A921, $I_\text{leakage}=100$~pA, $I_\text{charge}=11$~nA, $U_0=6$~kV, ramp rate is 2~kV/s, $t_\text{s}=800$~s, $M$ = 18), and from $4.5 \times 10^{-31} e$~cm to $4.4 \times 10^{-31} e$~cm for 2018 campaign (B703, $I_\text{leakage}=73$~pA, $I_\text{charge}=1$~nA, $U_0=7$~kV, ramp rate is 1~kV/s, $t_\text{s}=400$~s, $M$ = 36). The impact of the charging current acting as a leakage current was calculated and turned out to be negligible, compared to the effect of leakage currents as given in Table~\ref{tab:GPF_syst}.

\begin{table}[!h]
    \centering
    \caption{The systematic uncertainties determined as done in Ref.~\cite{Sachdeva2019} based on the data set used for the GPF method.}
    \begin{tabular}{c c  c }     
    \hline\noalign{\smallskip}
      & 2017 ($e$~cm) & 2018($e$~cm) \\
    \noalign{\smallskip}\hline\noalign{\smallskip}
    Leakage current (incl. impact of $I_\text{Charging}$ during ramping)            &  $ 1.2 \times 10^{-28} $    &  $ 4.4 \times 10^{-31} $ \\
    
    Charging current           &  $ 1.7 \times 10^{-29}$     &  $ 1.2 \times 10^{-29} $  \\
    Cell motion (rotation)     &  $ 4.2 \times 10^{-29} $    &  $ 4.0 \times 10^{-29} $  \\    
    Cell motion (translation)  &  $ 2.6 \times 10^{-28} $    &  $ 1.9 \times 10^{-28} $  \\    
     $|E|^2$ effect            &  $ 1.2 \times 10^{-29} $    &  $ 2.2 \times 10^{-30} $  \\    
     $|E|$ uncertainty         &  $ 9.9 \times 10^{-29} $    &  $ 5.7 \times 10^{-30} $   \\    
     Geometric phase           &  $ \leq 2 \times 10^{-31} $ & $ \leq 1 \times 10^{-29} $ \\   
    \noalign{\smallskip}\hline\noalign{\smallskip}
    \textbf{Total systematic uncertainty}    &  $ 3.07 \times 10^{-28}  $    &  $ 1.95 \times 10^{-28} $ \\  
    \textbf{Scaled statistical uncertainty}         &  $ 15.57 \times 10^{-28} $    &  $ 3.67 \times 10^{-28} $ \\
    \noalign{\smallskip}\hline 
    \end{tabular} 
    \label{tab:GPF_syst}
\end{table}

\subsection{Overall results}
\label{sec:overall_results}

The $F$-test threshold $P_\mathrm{min}$ affects the polynomial order used in the GPF method, as listed in Table~\ref{tab:GPF_F_test}. The EDM values for various $P_\mathrm{min}$ are overlapped within the $1\sigma$ statistical uncertainty and are all consistent with zero. Additionally, the upper limit of the $^{129}$Xe EDM is almost insensitive to the threshold. We have chosen 0.6 as the $F$-test threshold as was done also in Ref.~\cite{Sachdeva2019}, from the very beginning also during the blinded analysis and did stick to this decision after unblinding. The highest upper $^{129}$Xe EDM bound shows that this choice even leads to the most conservative result.

\begin{table*}[h]
    \centering
    \caption{ The overall EDM results and upper limit at 95\% C.L. with various $F$-test threshold $P_\mathrm{min}$. }
    \begin{tabular}{c c c c c c c}
    \hline\noalign{\smallskip}
     $P_{\mathrm{min}}$ & Average order & EDM  & Uncertainty & Reduced $\chi^2$ & $P$-value & Upper limit \\
       & & ($10^{-28} e$~cm) &  ($10^{-28} e$~cm) &   &  & ($10^{-28} e$~cm)  \\
    \noalign{\smallskip}\hline\noalign{\smallskip}
    0.4  & 8.2  &  0.08  &  3.22 & 1.32 & 0.03  & 8.2  \\    
    0.5  & 7.3  & -0.36   &  3.20 & 1.24 &0.06  & 8.0  \\
    0.6  & 6.4  &  1.06   &  3.08 & 1.34 &0.02  & 8.3  \\
    0.7  & 6.0  & -0.07  &  3.06 & 1.26 & 0.05  & 7.8  \\    
    0.8  & 5.5  & -0.87  &  3.05 & 1.31 & 0.03  & 8.1  \\    
    
    \noalign{\smallskip}\hline
    \end{tabular}
    \label{tab:GPF_F_test}
\end{table*}

The overall systematic uncertainty is the weighted average of the systematic uncertainties of the two measurement campaigns 2017 and 2018 using the reciprocal of its statistical variance as weights, yielding $2.0  \times 10 ^{-28}~e~ \mathrm{cm}$. The final result from the GPF analysis, separating the statistical and systematic uncertainties, is 
\begin{equation}
    \label{eqn:EDM_result}
    d_\mathrm{A} (^{129}\mathrm{Xe})=(1.1 \pm 3.6_\mathrm{(stat)} \pm 2.0_\mathrm{(syst)})  \times 10 ^{-28} e~\mathrm{cm},
\end{equation}
from which we set an upper limit  $|d_\mathrm{A} (^{129}\mathrm{Xe})| < 8.3 \times 10^{-28}~e~\mathrm{cm}$ at the 95\% C.L. This reanalysis leads to a limit that is a factor of 1.7 smaller compared to the previous result \cite{Sachdeva2019} and a factor of 8.0 compared to the result in the 2001 \cite{Rosenberry2001}. 

\section{Interpretation}\index{Remarks}

This new limit improves constraints on the low energy CP-violating parameters based on $^{129}$Xe EDM calculations, for example, lowering the limit on the Schiffs Moment of $^{129}$Xe by a factor of about 1.8. The detailed values are listed in Table~\ref{tab:Interpre}. At present, the $^{199}$Hg atomic EDM measurement is still the most sensitive, and its upper limit sets constraints to multiple sources of CP-violation \cite{Graner2016}. Considering various potential contributions to an atomic EDM, an improved limit on other systems, like the $^{129}$Xe EDM $d_\mathrm{A} (^{129}\mathrm{Xe})$, will tighten these constraints. The theoretical results for $^{129}$Xe EDM are more accurate and reliable than those obtained for $^{199}$Hg EDM, therefore $^{129}$Xe has the potential to probe new physics \cite{Sakurai2019}.

\begin{table}[h]
    \centering
    \caption{ Upper Limits of low energy parameters based on the obtained $^{129}$Xe EDM in this work (column 3) and the best limits published up to now (column 4). The formula for the calculation of the parameters for this work is given in column 2, mainly from \cite{Allmendinger2019}. No uncertainties were used for the calculations.}
    \begin{tabular}{c c c c}
    \toprule
     \textbf{Para.} &  Formulas & \textbf{Limit(this work)} & \textbf{Best Limit(other work)} \\
    \midrule
    $S_{\text{Xe}}$   & $2.6\times 10^{4} d_{\text{Xe}} \cdot \text{fm}^2$  & $ 2.1 \times 10 ^{-10} e \text{fm}^3 $ &  $ 3.7 \times 10^{-10} e \text{fm}^3$ \cite{Sachdeva2019}\\
	$d_{\text{p}}$    &$S_{\text{Xe}}/(0.125~\text{fm}^2 )$ &  $ 1.7 \times 10 ^{-22} e~\text{cm} $   & $ 1.1 \times 10 ^{-29} e~\text{cm}$ \cite{Andreev2018}  \\
	$d_{\text{n}}$    &$S_{\text{Xe}}/(0.63~\text{fm}^2 )$ &  $ 3.1 \times 10 ^{-23} e~\text{cm} $  & $ 1.6 \times 10 ^{-26} e~\text{cm} $  \cite{Graner2016} \\
    $g^{(0)}_\pi$   &$S_{\text{Xe}}/(0.008~\text{fm}^2 )$ &  $ 2.6 \times 10^{-8} $   & $ 2.3 \times 10^{-12} $ \cite{Graner2016} \\    
	$g^{(1)}_\pi$   &$S_{\text{Xe}}/(0.006~\text{fm}^2 ) $&  $ 3.5 \times 10^{-8} $   & $ 1.1 \times 10^{-12} $  \cite{Graner2016}\\    
	$g^{(2)}_\pi$   &$S_{\text{Xe}}/(0.009~\text{fm}^2 ) $   & $ 2.3 \times 10^{-9} $   & $ 1.1 \times 10^{-12} $  \cite{Graner2016}\\   
	$\bar{\theta}$ & $g^{(0)}_\pi/0.0155 $  &  $ 1.8 \times 10^{-6} $  & $ 1.5 \times 10^{-10} $  \cite{Graner2016}\\  
	$C_T$    & $d_{\text{Xe}}/(5.7 \times 10^{-21}~ e~\text{cm} ) $ & $ 1.5 \times 10^{-6}$ & $ 1.5 \times 10^{-10}$  \cite{Graner2016}\\
	$|g^s_eg^p_N|$ for $m_a<1$~keV & $d_{\text{Xe}}/(1.5 \times 10^{-13}~ e~\text{cm} ) $ & $ 6.4 \times 10^{-15}$ & $ 7.0 \times 10^{-17} $ \cite{Dzuba2018}\\    
    \bottomrule
    \end{tabular}
    \label{tab:Interpre}
\end{table}

\begin{longtable}{c c c c c c c c c c c c} %
    \caption{The experimental parameters and results for each sub-run.}
    \\\hline 
   \bfseries Sub-run &   $A_{\text{Xe}}$  & $A_{\text{He}}$ & $T_{2,\text{Xe}}^*$ & $T_{2,\text{He}}^*$ & $t_{\text{s}}$ & $M$ & $E$ & $d_{\text{Xe}}$ & $\delta d_{\text{Xe}}$& Order \\
       & (pT) & (pT) & (s) &(s) &(s) &  & (kV/cm) & \multicolumn{2}{c}{$(1 \times 10^{-27}  e~\text{cm})$} & \\
    \hline
    \endhead 
 A82& 5.9& 37.3& 6406& 6593& 400& 2.8& 36& 1.47& 5.01& 6 \\
A83& 3.2& 20.8& 7137& 7746& 400& 2.8& 36& -3& 7.98& 4 \\
A84& 4.9& 33.1& 6425& 7581& 400& 2.8& 36& -0.02& 5.26& 5\\
A85& 3.2& 20& 7463& 6924& 400& 2.8& 36& 1.45& 8.24& 5\\
A86& 4.5& 21.8& 5874& 7925& 400& 3.2& 36& 7.13& 6.34&  5\\
A87& 4.5& 20.6& 6019& 5634& 5000& 2.8& 3& -2.67& 9.85& 7 \\
A89& 2.8& 15.4& 6485& 3963& 400& 3.2& 36& -2.38& 10.5& 7\\
A91& 5.4& 28.8& 6557& 4000& 400& 3.2& 36& -2.48& 4.78& 4\\
A92& 3.8& 17.3& 7047& 4040& 800& 3.2& 18& 2.34& 4.24& 4\\
A93& 4.1& 27.6& 6779& 7245& 800& 2.8& 15& 1.09& 4.25& 5\\
A2& 4.4& 32& 4893& 5344& 800& 2.8& 18& 15& 6.44& 9\\
A8& 4.7& 30.5& 4985& 5366& 400& 2.8& 36& 4.6& 6.8& 6\\
A10& 5.3& 26.9& 5783& 3705& 400& 3.2& 36& 1.05& 5.46& 6\\
A12& 4.3& 21.9& 8082& 8579& 400& 2.8& 36& -3.29& 6.17& 6\\
A13& 5.1& 34.4& 7306& 7835& 800& 2.8& 18& -1.79& 3.07& 7\\
A14& 3.5& 21.7& 7282& 7607& 400& 2.8& 36& 7.2& 7.03& 5\\
A15& 3.8& 18.1& 6641& 3858& 400& 3.2& 36& -3.46& 7.59& 4\\
B36& 47.3& 42.6& 10059& 8765& 100& 4.1& 36& 13.5& 4.66& 4 \\
B37& 29& 24.2& 10157& 8776& 150& 4.1& 36& 0.83& 5.31& 6\\
B38& 13.4& 9.9& 10085& 8647& 200& 4.1& 36& -11.2& 9.74& 5\\
B39& 5.7& 3.7& 9916& 8490& 400& 4.1& 36& 9.15& 13.3& 4\\
B40& 40.2& 42.2& 9635& 8382& 120& 4.1& 36& -6.8& 4.39& 5\\
B41& 21.6& 20.7& 9714& 8280& 150& 4.1& 36& -5.9& 6.89& 4\\
B42& 10.2& 8.6& 9581& 8179& 400& 4.1& 36& -13.6& 5.79& 5\\
B43& 25.6& 31.4& 10230& 8658& 100& 3.2& 36& 7.75& 9.92& 6\\
B44& 16.6& 18.8& 10208& 8608& 200& 3.2& 36& -1.84& 7.16& 4\\
B45& 6.5& 6.2& 9978& 8403& 400& 3.2& 36& -5.54& 10.1& 4\\
B47& 41.7& 45& 9075& 6669& 100& 3.7& 36& 10.6& 6.18& 6\\
B48& 25.5& 23& 9348& 6769& 200& 3.7& 36& 0.52& 5.4& 5\\
B49& 10.5& 6.7& 9259& 6694& 400& 3.7& 36& -6.21& 9.39& 4\\
B50& 21.8& 26.1& 9644& 6793& 100& 2.8& 36& -5.16& 14.7& 6\\
B51& 13.8& 13.6& 9645& 6743& 200& 2.8& 36& 8.78& 12.9& 4\\
B52& 6& 4.1& 9382& 6564& 400& 2.8& 36& -1.42& 19.9& 4\\
B60& 38.6& 27.3& 5422& 3503& 400& 3.8& 36& 5.22& 3.3& 8\\
B631& 36.8& 35.2& 9602& 8797& 100& 4.1& 36& -6.83& 7.43& 9\\
B632& 22.9& 20.9& 9661& 8747& 300& 4.1& 36& -0.95& 3.87& 8\\
B633& 6.7& 5.4& 9443& 8462& 500& 4.1& 36& 6.61& 7.77& 5\\
B651& 49.9& 39.8& 8705& 8891& 100& 2.8& 36& 3.92& 4.21& 4\\
B652& 27.8& 22.5& 8829& 8970& 200& 2.8& 36& -1.37& 4.03& 7\\
B653& 10.3& 8.4& 8790& 8856& 400& 2.8& 36& -2.63& 5.31& 5\\
B661& 31.1& 25.5& 9171& 6997& 100& 4.1& 36& -2.59& 10.7&10 \\
B662& 18.4& 12.8& 9275& 7013& 200& 4.1& 36& 17.1& 9.2& 7\\
B663& 7.3& 3.8& 9176& 6920& 400& 4.1& 36& 4.92& 14& 6 \\
B671& 45.5& 33.1& 8480& 6906& 100& 3.2& 36& 10.5& 6.37& 5\\
B672& 26.9& 17.3& 8676& 6945& 200& 3.2& 36& 5.26& 6.55& 8\\
B673& 10.5& 5.4& 8670& 6886& 400& 3.2& 36& 5.03& 9.61& 4\\
B691& 48.7& 29.7& 7235& 5957& 100& 3.2& 36& 9.02& 7.73& 5\\
B692& 26.4& 14.1& 7457& 6067& 200& 3.2& 36& 13.5& 7.89& 5\\
B693& 8.8& 3.7& 7513& 6100& 400& 3.2& 36& 31.7& 16.3& 7\\
B701& 44.5& 21.6& 9160& 6899& 100& 3.2& 36& -7.42& 12.3& 4\\
B702& 27.8& 11.6& 9297& 6939& 200& 3.2& 36& -3.89& 11.1& 5\\
B703& 11.7& 3.6& 9248& 6906& 400& 3.2& 36& -48.8& 20.2& 9\\
B721& 30& 13.5& 8689& 8896& 100& 3.2& 36& -72.5& 24.1& 7\\
B722& 18.3& 8.3& 8772& 8957& 200& 3.2& 36& -25.9& 15.7&4 \\
B723& 7.3& 3.4& 8694& 8762& 400& 3.2& 36& 18& 19.3& 4\\
B731& 51.1& 30.1& 8067& 6403& 100& 4.1& 36& -1.38& 7.2& 9\\
B732& 27.2& 13.6& 8230& 6471& 200& 4.1& 36& 8.06& 7.76& 10\\
B733& 9.3& 3.5& 8227& 6459& 400& 4.1& 36& 9.61& 18.3& 9\\
B741& 24.6& 21.9& 8725& 6514& 100& 3.2& 36& 10.6& 14.5& 12\\
B742& 15& 11.2& 8767& 6494& 200& 3.2& 36& -9.27& 12.1& 5\\
B743& 6& 3.2& 8768& 6434& 400& 3.2& 36& 29.9& 20.3& 4\\
B761& 21.8& 72.5& 8852& 9045& 200& 4.1& 36& -3.76& 1.8& 5\\
B762& 8.6& 29& 8737& 8863& 500& 4.1& 36& -1.37& 1.55& 6\\
B771& 29.8& 19.3& 9680& 9594& 100& 3.2& 36& 10.3& 11.5& 6\\
B772& 18.1& 11.7& 9655& 9486& 200& 3.2& 36& 16.5& 10.4& 9\\
B773& 7.5& 4.7& 9519& 9293& 400& 3.2& 36& 10.6& 11.7& 7\\
B781& 28& 76.7& 9149& 9413& 400& 4.1& 36& 0.03& 0.85& 9\\
B782& 4.7& 13.6& 8973& 9149& 800& 4.1& 34& -2.15& 2.37& 8\\
B841& 18& 60.4& 9583& 9362& 300& 4.1& 36& -0.25& 1.63& 10\\
B842& 5.2& 17& 9481& 9227& 600& 4.1& 36& -0.84& 2.23& 4\\
B85& 41& 68.6& 9602& 9561& 300& 4.1& 36& 0.06& 1.06& 9\\
B86& 11.1& 18.5& 9498& 9347& 600& 4.1& 36& -1.63& 1.87& 6\\
B871& 26.5& 62.2& 8935& 9100& 300& 4.1& 36& -1.26& 1.39&10 \\
B872& 7& 16.9& 8936& 9012& 600& 4.1& 36& 3.34& 2.14& 7\\
B881& 23.9& 60.4& 9718& 9736& 300& 4.1& 36& -1.15& 1.23& 7\\
B882& 6.6& 16.7& 9677& 9616& 600& 4.1& 36& 3.71& 2.42& 5\\
B891& 15.1& 59.9& 7075& 5908& 300& 4.1& 36& -2.77& 2.36& 13\\
B892& 2.7& 7.5& 7648& 6214& 600& 4.1& 36& -6.93& 6.4& 4\\
B901& 26.3& 61.3& 7425& 7720& 300& 4.1& 36& 0.08& 1.52& 11\\
B902& 5.4& 13.2& 7510& 7692& 600& 4.1& 36& -6.11& 3.79& 7\\
B911& 24.8& 55.8& 7929& 8213& 500& 4.1& 36& 1.6& 0.9& 7\\
B912& 2.1& 5.2& 7942& 8089& 600& 4.1& 24& -15.9& 7.78& 6\\
B921& 28.2& 59.9& 7545& 7949& 300& 4.1& 36& -1.6& 1.33& 11\\
B922& 5.4& 12.4& 7604& 7944& 600& 4.1& 36& 0.7& 3.38& 4\\
B93& 25.5& 53.3& 8372& 8616& 12000& 4.1& 3& 2.6& 1.17& 11\\
B941& 20.5& 35& 8458& 4207& 300& 4.9& 36& 0.35& 2.61& 8\\
B942& 5& 2& 8181& 4125& 600& 4.9& 13& -36& 35.7& 10    \\
\hline
    \label{tab:EDM_result}
\end{longtable}

\chapter*{Bibliography}
\bibliography{bibliography}

\begin{thebibliography}{10}
\expandafter\ifx\csname url\endcsname\relax
  \def\url#1{\texttt{#1}}\fi
\expandafter\ifx\csname urlprefix\endcsname\relax\def\urlprefix{URL }\fi
\expandafter\ifx\csname href\endcsname\relax
  \def\href#1#2{#2} \def\path#1{#1}\fi

\bibitem{Sachdeva2019}
N.~Sachdeva, et~al., {New Limit on the Permanent Electric Dipole Moment of
  $^{129}\text{Xe}$ Using $^{3}\text{He}$ Comagnetometry and SQUID Detection},
  Physical Review Letters 123~(14) (2019) 143003.
\newblock \href {http://dx.doi.org/10.1103/PhysRevLett.123.143003}
  {\path{doi:10.1103/PhysRevLett.123.143003}}.

\bibitem{Rosenberry2001}
M.~A. Rosenberry, T.~E. Chupp, {Atomic Electric Dipole Moment Measurement Using
  Spin Exchange Pumped Masers of $^{129}$Xe and $^3$He}, Physical Review
  Letters 86~(1) (2001) 22--25.
\newblock \href {http://dx.doi.org/10.1103/PhysRevLett.86.22}
  {\path{doi:10.1103/PhysRevLett.86.22}}.

\bibitem{Sachdeva2019a}
N.~Sachdeva, {A Measurement of the Permanent Electric Dipole Moment of
  $^{129}$Xe}, Ph.D. thesis, The University of Michigan (2019).

\bibitem{Chupp2019}
T.~E. Chupp, P.~Fierlinger, M.~J. Ramsey-Musolf, J.~T. Singh, {Electric dipole
  moments of atoms, molecules, nuclei, and particles}, Reviews of Modern
  Physics 91~(1) (2019) 015001.
\newblock \href {http://dx.doi.org/10.1103/RevModPhys.91.015001}
  {\path{doi:10.1103/RevModPhys.91.015001}}.

\bibitem{Gemmel2010}
C.~Gemmel, W.~Heil, S.~Karpuk, K.~Lenz, C.~Ludwig, Y.~Sobolev, K.~Tullney,
  M.~Burghoff, W.~Kilian, S.~Knappe-Gr{\"{u}}neberg, W.~M{\"{u}}ller,
  A.~Schnabel, F.~Seifert, L.~Trahms, S.~Baeler, {Ultra-sensitive magnetometry
  based on free precession of nuclear spins}, European Physical Journal D
  57~(3) (2010) 303--320.
\newblock \href {http://dx.doi.org/10.1140/epjd/e2010-00044-5}
  {\path{doi:10.1140/epjd/e2010-00044-5}}.

\bibitem{Allmendinger2019}
F.~Allmendinger, I.~Engin, W.~Heil, S.~Karpuk, H.-J. Krause,
  B.~Niederl{\"{a}}nder, A.~Offenh{\"{a}}usser, M.~Repetto, U.~Schmidt,
  S.~Zimmer, {Measurement of the permanent electric dipole moment of the
  $^{129}\text{Xe}$ atom}, Physical Review A 100~(2) (2019) 022505.
\newblock \href {http://dx.doi.org/10.1103/PhysRevA.100.022505}
  {\path{doi:10.1103/PhysRevA.100.022505}}.

\bibitem{Terrano2019}
W.~A. Terrano, J.~Meinel, N.~Sachdeva, T.~E. Chupp, S.~Degenkolb,
  P.~Fierlinger, F.~Kuchler, J.~T. Singh, {Frequency shifts in noble-gas
  comagnetometers}, Physical Review A 100~(1) (2019) 012502.
\newblock \href {http://dx.doi.org/10.1103/PhysRevA.100.012502}
  {\path{doi:10.1103/PhysRevA.100.012502}}.

\bibitem{Limes2019}
M.~E. Limes, N.~Dural, M.~V. Romalis, E.~L. Foley, T.~W. Kornack, A.~Nelson,
  L.~R. Grisham, J.~Vaara, {Dipolar and scalar $^{3}$He - $^{129}$Xe frequency
  shifts in stemless cells}, Physical Review A 100~(1) (2019) 010501.
\newblock \href {http://dx.doi.org/10.1103/PhysRevA.100.010501}
  {\path{doi:10.1103/PhysRevA.100.010501}}.

\bibitem{Romalis2014}
M.~V. Romalis, D.~Sheng, B.~Saam, T.~G. Walker, {Comment on New Limit on
  Lorentz-Invariance- and CPT-Violating Neutron Spin Interactions Using a
  Free-Spin-Precession $^3$He-$^{129}$Xe Comagnetometer}, Physical Review
  Letters 113~(18) (2014) 188901.
\newblock \href {http://dx.doi.org/10.1103/PhysRevLett.113.188901}
  {\path{doi:10.1103/PhysRevLett.113.188901}}.

\bibitem{Allmendinger2014}
F.~Allmendinger, W.~Heil, S.~Karpuk, W.~Kilian, A.~Scharth, U.~Schmidt,
  A.~Schnabel, Y.~Sobolev, K.~Tullney, {New limit on Lorentz-invariance- and
  CPT-violating neutron spin interactions using a free-spin-precession
  $^{3}\text{He}$ - $^{129}\text{Xe}$ comagnetometer}, Physical Review Letters
  112~(11) (2014) 1--5.
\newblock \href {http://dx.doi.org/10.1103/PhysRevLett.112.110801}
  {\path{doi:10.1103/PhysRevLett.112.110801}}.

\bibitem{PerezGalvan2011}
A.~{P{\'{e}}rez Galv{\'{a}}n}, B.~Plaster, J.~Boissevain, R.~Carr, B.~W.
  Filippone, M.~P. Mendenhall, R.~Schmid, R.~Alarcon, S.~Balascuta, {High
  uniformity magnetic coil for search of neutron electric dipole moment},
  Nuclear Instruments and Methods in Physics Research, Section A: Accelerators,
  Spectrometers, Detectors and Associated Equipment 660~(1) (2011) 147--153.
\newblock \href {http://dx.doi.org/10.1016/j.nima.2011.09.019}
  {\path{doi:10.1016/j.nima.2011.09.019}}.

\bibitem{Abel2019b}
C.~Abel, et~al., {The n2EDM experiment at the Paul Scherrer Institute}, EPJ Web
  of Conferences 219 (2019) 02002.
\newblock \href {http://dx.doi.org/10.1051/epjconf/201921902002}
  {\path{doi:10.1051/epjconf/201921902002}}.

\bibitem{Sakamoto2015}
Y.~Sakamoto, C.~P. Bidinosti, Y.~Ichikawa, T.~Sato, Y.~Ohtomo, S.~Kojima,
  C.~Funayama, T.~Suzuki, M.~Tsuchiya, T.~Furukawa, A.~Yoshimi, T.~Ino,
  H.~Ueno, Y.~Matsuo, T.~Fukuyama, K.~Asahi, {Development of high-homogeneity
  magnetic field coil for $^{129}\text{Xe}$ EDM experiment}, Hyperfine
  Interactions 230~(1-3) (2015) 141--146.
\newblock \href {http://dx.doi.org/10.1007/s10751-014-1109-5}
  {\path{doi:10.1007/s10751-014-1109-5}}.

\bibitem{Abe2018}
M.~Abe, Y.~Murata, H.~Iinuma, T.~Ogitsu, N.~Saito, K.~Sasaki, T.~Mibe,
  H.~Nakayama, {Magnetic design and method of a superconducting magnet for muon
  g-2 /EDM precise measurements in a cylindrical volume with homogeneous
  magnetic fiel}, Nuclear Instruments and Methods in Physics Research Section
  A: Accelerators, Spectrometers, Detectors and Associated Equipment
  890~(October 2017) (2018) 51--63.
\newblock \href {http://dx.doi.org/10.1016/j.nima.2018.01.026}
  {\path{doi:10.1016/j.nima.2018.01.026}}.

\bibitem{Dadisman2018}
J.~R. Dadisman, Magnetic field design to reduce systematic effects in neutron
  electric dipole moment measurements, Theses and dissertations--physics and
  astronomy, University of Kentucky (2018).
\newblock \href {http://dx.doi.org/10.13023/ETD.2018.094}
  {\path{doi:10.13023/ETD.2018.094}}.

\bibitem{Liu2021}
T.~Liu, A.~Schnabel, J.~Voigt, W.~Kilian, Z.~Sun, L.~Li, L.~Trahms, {A built-in
  coil system attached to the inside walls of a magnetically shielded room for
  generating an ultra-high magnetic field homogeneity}, Review of Scientific
  Instruments 92~(2) (2021) 024709.
\newblock \href {http://dx.doi.org/10.1063/5.0027848}
  {\path{doi:10.1063/5.0027848}}.

\bibitem{Sun2020}
Z.~Sun, P.~Fierlinger, J.~Han, L.~Li, T.~Liu, A.~Schnabel, S.~Stuiber,
  J.~Voigt, {Limits of Low Magnetic Field Environments in Magnetic Shields},
  IEEE Transactions on Industrial Electronics 68~(6) (2021) 5385--5395.
\newblock \href {http://dx.doi.org/10.1109/TIE.2020.2987267}
  {\path{doi:10.1109/TIE.2020.2987267}}.

\bibitem{Bork2001}
J.~Bork, H.~Hahlbohm, R.~Klein, A.~Schnabel, {The 8-layered magnetically
  shielded room of the PTB : Design and construction}, in: Proceedings of the
  12th international conference on Biomagnetism, 2001, pp. 970--973.

\bibitem{Gentile2017}
T.~R. Gentile, P.~J. Nacher, B.~Saam, T.~G. Walker, {Optically polarized
  $^3$He}, Reviews of Modern Physics 89~(4) (2017) 045004.
\newblock \href {http://dx.doi.org/10.1103/RevModPhys.89.045004}
  {\path{doi:10.1103/RevModPhys.89.045004}}.

\bibitem{Korchak2013}
S.~E. Korchak, W.~Kilian, L.~Mitschang, {Configuration and Performance of a
  Mobile $^{129}$Xe Polarizer}, Applied Magnetic Resonance 44~(1-2) (2013)
  65--80.
\newblock \href {http://dx.doi.org/10.1007/s00723-012-0425-7}
  {\path{doi:10.1007/s00723-012-0425-7}}.

\bibitem{Walker1997}
T.~G. Walker, W.~Happer, {Spin-exchange optical pumping of noble-gas nuclei},
  Reviews of Modern Physics 69~(2) (1997) 629--642.
\newblock \href {http://dx.doi.org/10.1103/RevModPhys.69.629}
  {\path{doi:10.1103/RevModPhys.69.629}}.

\bibitem{Stefan2018}
Z.~Stefan, {Search for a Permanent Electric Dipole Moment of $^{129}$Xe with a
  He/Xe Clock-Comparison Experiment}, Ph.D. thesis, Johannes Gutenberg
  University of Mainz (2018).

\bibitem{Cates1988}
G.~D. Cates, S.~R. Schaefer, W.~Happer, {Relaxation of spins due to field
  inhomogeneities in gaseous samples at low magnetic fields and low pressures},
  Physical Review A 37~(8) (1988) 2877--2885.
\newblock \href {http://dx.doi.org/10.1103/PhysRevA.37.2877}
  {\path{doi:10.1103/PhysRevA.37.2877}}.

\bibitem{Allmendinger2017}
F.~Allmendinger, P.~Bl{\"{u}}mler, M.~Doll, O.~Grasdijk, W.~Heil, K.~Jungmann,
  S.~Karpuk, H.-J. Krause, A.~Offenh{\"{a}}usser, M.~Repetto, U.~Schmidt,
  Y.~Sobolev, K.~Tullney, L.~Willmann, S.~Zimmer, {Precise measurement of
  magnetic field gradients from free spin precession signals of $^3$He and
  $^{129}$Xe magnetometers}, The European Physical Journal D 71~(4) (2017) 98.
\newblock \href {http://dx.doi.org/10.1140/epjd/e2017-70505-4}
  {\path{doi:10.1140/epjd/e2017-70505-4}}.

\bibitem{Storm2017}
J.-H. Storm, P.~H{\"{o}}mmen, D.~Drung, R.~K{\"{o}}rber, {An ultra-sensitive
  and wideband magnetometer based on a superconducting quantum interference
  device}, Applied Physics Letters 110~(7) (2017) 072603.
\newblock \href {http://dx.doi.org/10.1063/1.4976823}
  {\path{doi:10.1063/1.4976823}}.

\bibitem{Golub2003}
G.~Golub, V.~Pereyra, {Separable nonlinear least squares: the variable
  projection method and its applications}, Inverse Problems 19~(2) (2003)
  R1--R26.
\newblock \href {http://dx.doi.org/10.1088/0266-5611/19/2/201}
  {\path{doi:10.1088/0266-5611/19/2/201}}.

\bibitem{Fan2016}
I.~Fan, S.~Knappe-Gr{\"{u}}neberg, J.~Voigt, W.~Kilian, M.~Burghoff,
  D.~Stollfuss, A.~Schnabel, G.~W{\"{u}}bbeler, O.~Bodner, C.~Elster,
  F.~Seifert, L.~Trahms, {Direct measurement of the $\gamma_\text{He}$ /
  $\gamma_\text{Xe}$ ratio at ultralow magnetic field}, Journal of Physics:
  Conference Series 723~(1) (2016) 012045.
\newblock \href {http://dx.doi.org/10.1088/1742-6596/723/1/012045}
  {\path{doi:10.1088/1742-6596/723/1/012045}}.

\bibitem{Refaat2009}
R.~E. Attar, {Legendre Polynomials And Functions}, CreateSpace Independent
  Publishing Platform, South Carolina, 2009.

\bibitem{Bevington1992}
P.~R. Bevington, D.~K. Robinson, {Data Reduction and Error Analysis for the
  Physical Sciences}, Vol.~7, McGraw-Hill, 1992.
\newblock \href {http://dx.doi.org/10.1063/1.4823194}
  {\path{doi:10.1063/1.4823194}}.

\bibitem{Allan1981}
D.~Allan, J.~Barnes, {A Modified "Allan Variance" with Increased Oscillator
  Characterization Ability}, IEEE, Philadelphia, Pennsylvania, USA, 1981, pp.
  470--475.
\newblock \href {http://dx.doi.org/10.1109/freq.1981.200514}
  {\path{doi:10.1109/freq.1981.200514}}.

\bibitem{Venzon1988}
D.~J. Venzon, S.~H. Moolgavkar, {A Method for Computing
  Profile-Likelihood-Based Confidence Intervals}, Applied Statistics 37~(1)
  (1988) 87.
\newblock \href {http://dx.doi.org/10.2307/2347496}
  {\path{doi:10.2307/2347496}}.

\bibitem{KAY1993}
S.~M. KAY, {Fundamentals of Statistical Signal Processing, Volume I: Estimation
  Theory.}, Prentice Hall PTR, New Jersey, 1993.

\bibitem{Thrasher2019}
D.~A. Thrasher, S.~S. Sorensen, J.~Weber, M.~Bulatowicz, A.~Korver, M.~Larsen,
  T.~G. Walker, {Continuous comagnetometry using transversely polarized Xe
  isotopes}, Physical Review A 100~(6) (2019) 061403.
\newblock \href {http://dx.doi.org/10.1103/PhysRevA.100.061403}
  {\path{doi:10.1103/PhysRevA.100.061403}}.

\bibitem{Beringer2012}
J.~Beringer, et~al., {Review of Particle Physics}, Physical Review D 86~(1)
  (2012) 010001.
\newblock \href {http://dx.doi.org/10.1103/PhysRevD.86.010001}
  {\path{doi:10.1103/PhysRevD.86.010001}}.

\bibitem{Efron1982}
B.~Efron, {The jackknife, the bootstrap, and other resampling plans}, Society
  for Industrial and Applied Mathematics, Philadelphia, 1982.

\bibitem{Graner2016}
B.~Graner, Y.~Chen, E.~G. Lindahl, B.~R. Heckel, {Reduced Limit on the
  Permanent Electric Dipole Moment of $^{199}$Hg}, Physical Review Letters
  116~(16) (2016) 1--5.
\newblock \href {http://dx.doi.org/10.1103/PhysRevLett.116.161601}
  {\path{doi:10.1103/PhysRevLett.116.161601}}.

\bibitem{Sakurai2019}
A.~Sakurai, B.~K. Sahoo, K.~Asahi, B.~P. Das, {Relativistic many-body theory of
  the electric dipole moment of $^{129}$Xe and its implications for probing new
  physics beyond the standard model}, Physical Review A 100~(2) (2019) 020502.
\newblock \href {http://dx.doi.org/10.1103/PhysRevA.100.020502}
  {\path{doi:10.1103/PhysRevA.100.020502}}.

\bibitem{Andreev2018}
V.~Andreev, D.~G. Ang, D.~DeMille, J.~M. Doyle, G.~Gabrielse, J.~Haefner, N.~R.
  Hutzler, Z.~Lasner, C.~Meisenhelder, B.~R. O'Leary, C.~D. Panda, A.~D. West,
  E.~P. West, X.~Wu, {Improved limit on the electric dipole moment of the
  electron}, Nature 562~(7727) (2018) 355--360.
\newblock \href {http://dx.doi.org/10.1038/s41586-018-0599-8}
  {\path{doi:10.1038/s41586-018-0599-8}}.

\bibitem{Dzuba2018}
V.~A. Dzuba, V.~V. Flambaum, I.~B. Samsonov, Y.~V. Stadnik, {New constraints on
  axion-mediated P, T -violating interaction from electric dipole moments of
  diamagnetic atoms}, Physical Review D 98~(3) (2018) 035048.
\newblock \href {http://dx.doi.org/10.1103/PhysRevD.98.035048}
  {\path{doi:10.1103/PhysRevD.98.035048}}.

\end{thebibliography}

\end{document}